\documentclass[a4paper,11pt]{article}
\pdfoutput=1 

\usepackage{jcappub} 

\usepackage[T1]{fontenc} 

\usepackage{amsfonts}
\usepackage{amsmath}
\usepackage{booktabs}
\usepackage{siunitx}
\usepackage{lineno}

\title{\boldmath Downward ultra-high-energy neutrino detection in the air with radio antennas at ground-based observatories}

\author[a,1]{Baobiao Yue,\note{Corresponding author.}}
\author[a]{, Karl-Heinz Kampert}
\author[a]{, and Julian Rautenberg}

\affiliation[a]{Bergische Universität Wuppertal, Department of Physics, Gaußstraße 20, Wuppertal, Germany}

\emailAdd{bayue@uni-wuppertal.de}

\abstract{
Ultra-high-energy (UHE) neutrinos are unique cosmic messengers that can traverse cosmological distances unattenuated, providing direct insight into the most energetic processes in the universe. 
Radio detection offers significant advantages for detecting highly inclined air showers induced by UHE neutrinos. This is due to a larger exposure range compared to particle detectors, which is a result of minimal atmospheric attenuation of radio signals combined with good reconstruction precision. Furthermore, this technique improves the air shower longitudinal reconstruction, which can be used to identify neutrinos with their first interaction far below the top of the atmosphere.
In this work, we present a method for identifying UHE neutrinos using ground-based radio antennas. 
A reconstruction algorithm is introduced based on the radio emission maximum ($X^{\text{radio}}_{\text{max}}$), which demonstrates its power in distinguishing deeply developing neutrino-induced showers from background cosmic rays. 
Using simulations of $\nu_e$-CC-induced air showers, we evaluate the trigger efficiency, reconstruction performance, and resulting effective area and aperture prediction for a reference array.
Our results show that radio detection significantly enhances the sensitivity to very inclined showers above 1~EeV, complementing traditional surface detectors. This technique is highly scalable and applicable to future radio observatories, such as GRAND. 
The proposed reconstruction and identification strategy provides a pathway toward achieving the sensitivity required to detect UHE neutrinos.
}

\begin{document}
\maketitle
\flushbottom

\newpage
\section{Introduction}

Ultra-high-energy (UHE) neutrinos, with energies exceeding $10^{17}$~eV, are important messengers for uncovering the origins of the most energetic cosmic rays in the universe \cite{Halzen:2002pg}. Due to their weak interactions and lack of electric charge, neutrinos can propagate over cosmological distances without deflection or significant energy loss, carrying the information of the direction and the radiant energy from their sources. 

A growing global effort is underway to detect these elusive particles. The IceCube Neutrino Observatory has pioneered neutrino astronomy by observing a diffuse astrophysical neutrino flux in the TeV–PeV range \cite{IceCube:2014stg}
Most recently, the KM3NeT collaboration reported the detection of a $\sim$220~PeV neutrino candidate~\cite{KM3NeT:2025npi}, the most energetic such event to date. 
Its origin remains unknown, which has increased the interest in UHE neutrino detection.

Detecting UHE neutrinos requires enormous effective volumes and advanced background rejection. 
A promising approach involves large-area radio antenna arrays that are deployed on the ground or embedded in ice. 
Ground based radio detection exploits the coherent radio emission generated by the geomagnetic deflection of charged particles or the charge excess (Askaryan effect) \cite{Askaryan:1961pfb} in an extensive air shower (EAS). 
Radio signals are minimally attenuated in the atmosphere, offering nanosecond timing information that enables a precise reconstruction of the geometry and energy of the shower. 

\begin{figure}[bt]
    \centering
    \includegraphics[width=0.8\linewidth]{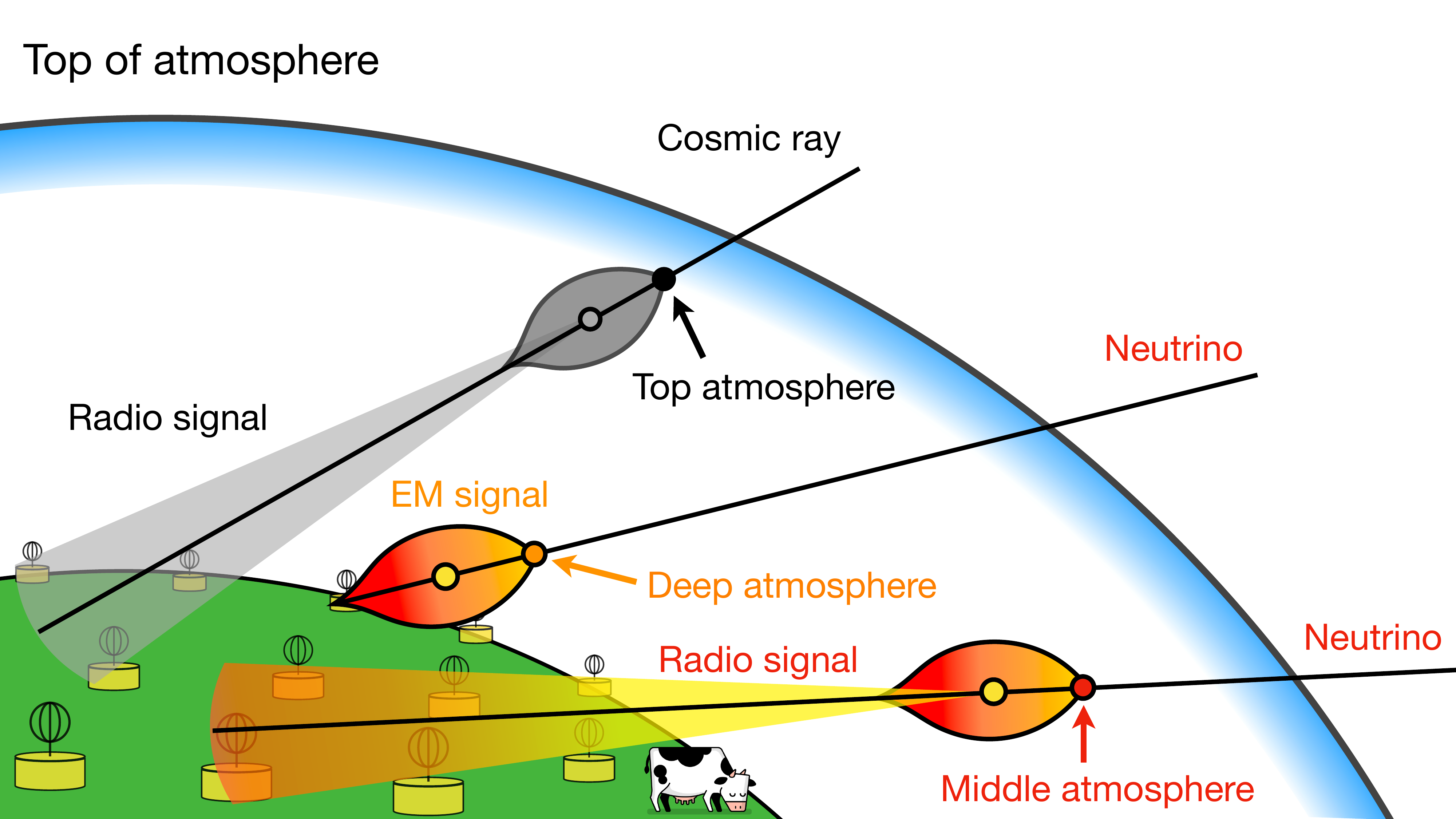}
    \caption{The neutrino detection enhancement with radio antennas. Three shows are shown here for the illustration. The first one is the cosmic ray interacting the top of the atmosphere, which can be measured by radio antenna. The second one is about the neutrino induced young shower with significant electromagnetic component measured and identified by particle detector. The third one is the neutrino interaction in the middle of the atmosphere and its detection with radio antenna. Since the radio signal can propagate much longer than particles in the atmosphere, the neutrino detection enhancement with radio antenna is feasible, especially for the neutrino interaction in the middle and the top of the atmosphere.}
    \label{fig:sketch}
\end{figure}

In this study, we examine the potential of radio detection to enhance UHE neutrino searches at ground-based observatories. 
Using a reference observatory that is similar in design to the Pierre Auger Observatory \cite{PierreAuger:2015eyc}, except for its area, we evaluate the performance of radio-based triggering and reconstruction for inclined showers.
This reference observatory covers an area of 10,000 km$^2$ at an altitude of 1.4 km with an antenna spacing of 1.5\,km.
The geomagnetic field $\vec{B}$ is assumed to be the same as that at the Pierre Auger Observatory.
Figure\,\ref{fig:sketch} illustrates how radio antennas can enhance the detection of downward-going neutrinos w.r.t.\ particle detectors. 
The identification of neutrino-induced showers is based on measurements of the slant depth at which the radio-emission reaches its maximum, $X_{\mathrm{max}}^{\mathrm{radio}}$, in the atmosphere. We also estimate the sensitivity that can be achieved with a standalone radio array. Our findings demonstrate that radio detection can complement existing techniques and expand the capabilities of neutrino observatories to the EeV regime. 

\section{Radio detection for neutrino-induced air showers}
Downward-going neutrinos can interact with nuclei in the atmosphere, producing extensive air showers (EAS). Due to their small cross sections, these interactions can occur at much greater slant depths than those of cosmic rays. The key parameter that distinguishes neutrinos from hadronic cosmic rays is the first interaction point, $X_1$, or the position of maximum radio emission, $X^{\text{radio}}_{\text{max}}$ \cite{PierreAuger:2023rgk}. 
Since radio radiation is coherent, we can reconstruct the position of maximum radio emission, $X^{\text{radio}}_{\text{max}}$, by analyzing the timing of the maximum pulse in the radio antennas. This allows us to differentiate steeply downward-going neutrinos that interact deeply within the atmosphere from UHECR-induced EASs, which reach their shower maximum at much shallower depths. Traditional particle detector arrays work on the same principle, but can identify neutrino-induced EASs only when the showers develop sufficiently close to ground to still contain a significant electromagnetic component (\textit{c.f.} Fig.\,\ref{fig:sketch}). The time profile of this component of electrons, positrons, and photons is much broader than that of high-energy muons reaching the detectors from shallow interaction depths. This method of discrimating between ``old'' from ``young'' showers has been pioneered using the water Cherenkov detectors of the Pierre Auger Observatory \cite{PierreAuger:2015eyc,PierreAuger:2007vvh}. We will use this as a benchmark for our studies.

In addition, Earth-skimming tau neutrinos can initiate tau-decay-induced air showers near the ground from directions a few degrees below the horizon. If the air shower is close to the observatory array, the particle detectors can again identify it as a young shower because of its distinct electromagnetic signal. 
Using radio interferometry \cite{LOPES:2005ipv,Schoorlemmer:2020low}, the shower axis can be reconstructed with high precision. This enables the identification of upward-going neutrinos traversing the Earth, as well as downward-going tau neutrinos emerging from interactions within mountains.

In this work, we focus on detecting downward-going $\nu_e$-CC interactions, which is the dominant channel for downward-going neutrino detections in ground based observatories and demonstrate the potential of radio antennas for neutrino detection using ground-based radio arrays.


\section{Simulations of the signal and noise for air showers induced by neutrinos}
We use CoREAS \cite{Huege:2013vt} simulations of neutrino $\nu_e$ charged current (CC) interactions to study the radio detection sensitivity for downward-going EAS. 
These simulations include the time-dependent electric field vector from air showers at specific antenna locations. 
The CoREAS simulations use a 1.5 km antenna spacing to match the configuration of the radio antenna array (RD) at the Pierre Auger Observatory \cite{Pont:2025wdg}. 
To mimic the frequency response of the radio antenna and the analog filter-amplifier chain, the simulations employ a 30-80\,MHz bandpass filter, similar to those used in common radio detectors, such as those at the Pierre Auger Observatory.
In addition, we adopted an approximate model of the radio antenna response pattern to simulate the real response of the Auger RD. Details of the radio antenna response mode model will be presented below.

We generate an uncertainty of $\sigma^\text{GPS}_t=5$\,ns for the local time of the GPS receiver in each individual antenna without external synchronization signals.
Additionally, we simulate an unpolarized white noise with a root mean square (RMS) of approximately 25\,\textmu V/m, denoted as RMS$_\mathrm{noise}$, in the radio signal traces. This mimics the radio background from the Milky Way, which is one of main sources of noise for air showers, especially in the 30–80\,MHz frequency range.


\subsection{Modeling the radio antenna response pattern}
\begin{figure}
    \centering
    \includegraphics[width=0.8\linewidth]{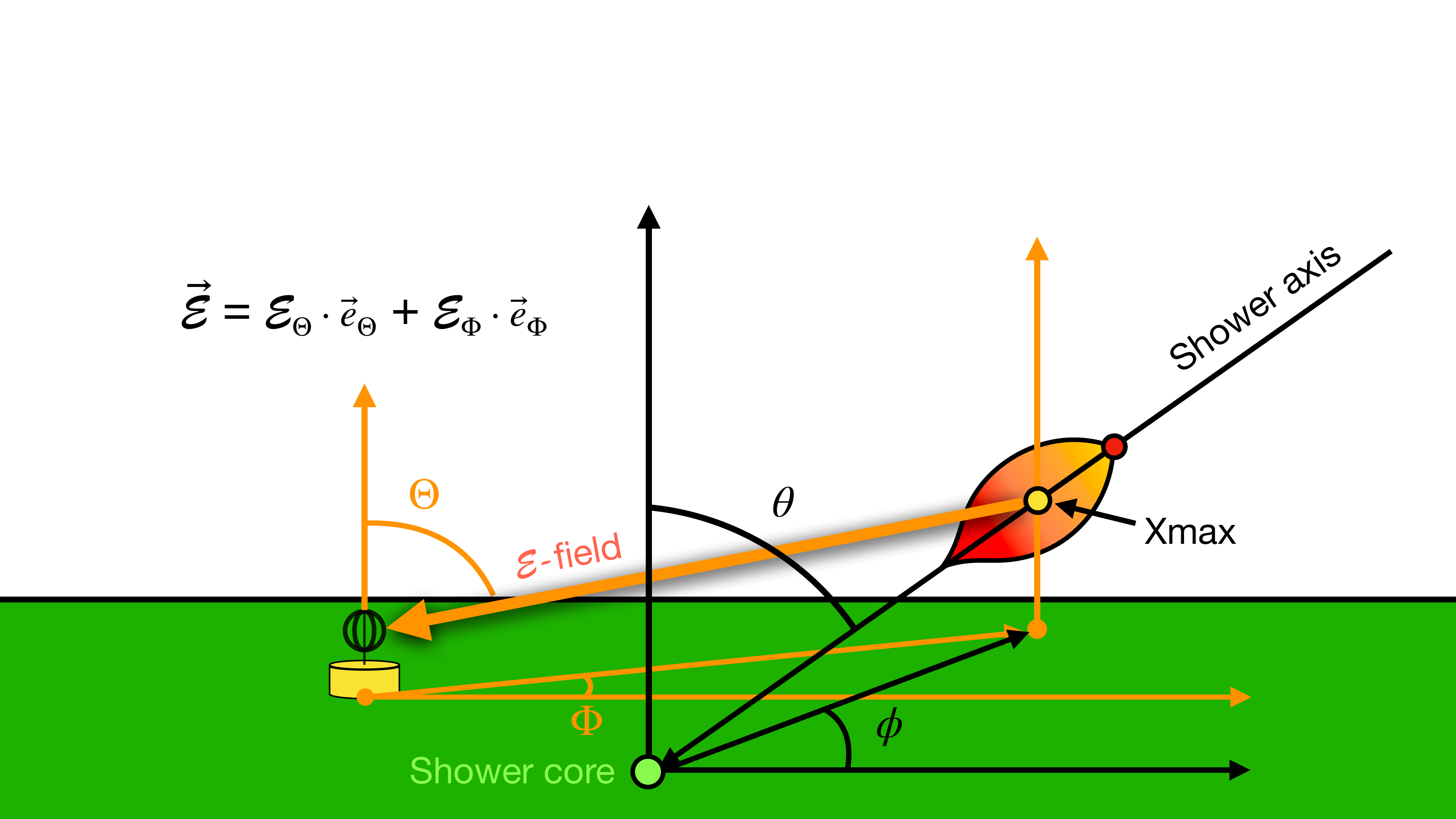}
    \caption{The sketch for the electric-field propagation from the shower maximum to the radio antenna. Zenith $\Theta$ and azimuth $\Phi$ are for electric field geometry and zenith $\theta$ and azimuth $\phi$ are for shower axis.}
    \label{fig:geometry_sketch}
\end{figure}
The radio array is configured to measure the electric-field trace released by air showers. 
It can be extracted by unfolding the induced voltage from the radio antenna response.
Therefore, accurate knowledge of the antenna response is crucial for the unfolding processes.
The response of a radio antenna can be described by the Vector Effective Length (VEL) $\vec{H}$, which depends on the frequency and direction of the incoming electric field.
To distinguish the zenith and azimuthal angles of the electric field more easily from the zenith and azimuth of the shower axis, the electric field geometry is referred to as $\Theta$ and $\Phi$, and the shower axis is referred to as $\theta$ and $\phi$, as illustrated in Fig.~\ref{fig:geometry_sketch}.
In the frequency domain, the induced voltage $V(f)$ is given by
\begin{equation}
    V(f) = \vec{H}(\Theta, \Phi, f)\cdot \vec{\mathcal{E}}(f)\,,
\end{equation}
where $\vec{\mathcal{E}}(f)$ denotes the trace of the electric field in the frequency domain.
If the shower development is far away from the antennas, the radio emission position of the air shower can be treated as one point for the antennas inside its footprint at the ground.
Therefore, the coherent summation of the electric field does not have a component in the direction of propagation.
In spherical coordinates, the electric field vector, $\vec{\mathcal{E}}(f)$, can be decomposed into
\begin{equation}
    \vec{\mathcal{E}}(f) = {\mathcal{E}}_\Theta(f) \cdot \vec{e}_\Theta + {\mathcal{E}}_\Phi(f) \cdot \vec{e}_\Phi\,,
\end{equation}
where $\vec{e}_\Theta$ and $\vec{e}_\Phi$ are the unit vectors.
Similarly, $\vec{H}(\Theta, \Phi, f)$ can be decomposed using $\vec{e}_\Theta$ and $\vec{e}_\Phi$.
The induced voltage $V(f)$ can be rewritten as
\begin{equation}
    V(f) = {H}_\Theta(\Theta, \Phi, f)\cdot{\mathcal{E}}_\Theta(f) +  {H}_\Phi(\Theta, \Phi, f)\cdot{\mathcal{E}}_\Phi(f) \,.
\end{equation}
In the frequency domain, ${H}_\Theta$ and ${H}_\Phi$ are complex values.
They can be decomposed with real and imaginary parts
\begin{align}
    {H}_\Theta(f) &= {H}^{real}_\Theta(f) + i  {H}^{imag}_\Theta(f)  \\
    {H}_\Phi(f) &= {H}^{real}_\Phi(f) + i  {H}^{imag}_\Phi(f) \,.
\end{align}

To unfold the electric field trace induced by air showers, two antennas with orthogonal polarization can be used.
The induced voltage at each antenna arm (EW and NS) can be expressed as follows:
    \begin{align}
        V_{EW}(f) & = {H}^{EW}_\Theta(\Theta, \Phi, f)\cdot {\mathcal{E}}_\Theta(f) +  {H}^{EW}_\Phi(\Theta, \Phi, f)\cdot{\mathcal{E}}_\Phi(f) \\
        V_{NS}(f) & = {H}^{NS}_\Theta(\Theta, \Phi, f)\cdot{\mathcal{E}}_\Theta(f) +  {H}^{NS}_\Phi(\Theta, \Phi, f)\cdot{\mathcal{E}}_\Phi(f) \,.
    \end{align}

For this study, we created a simplified response model inspired by the SALLA response \cite{PierreAuger:2023zdi}. 
More details can be found in Appendix \ref{AntennaModeling}.
We assume that there is no uncertainty from the antenna response model.

\subsection{Galactic noise}
\begin{figure}
    \centering
    \includegraphics[width=0.49\linewidth]{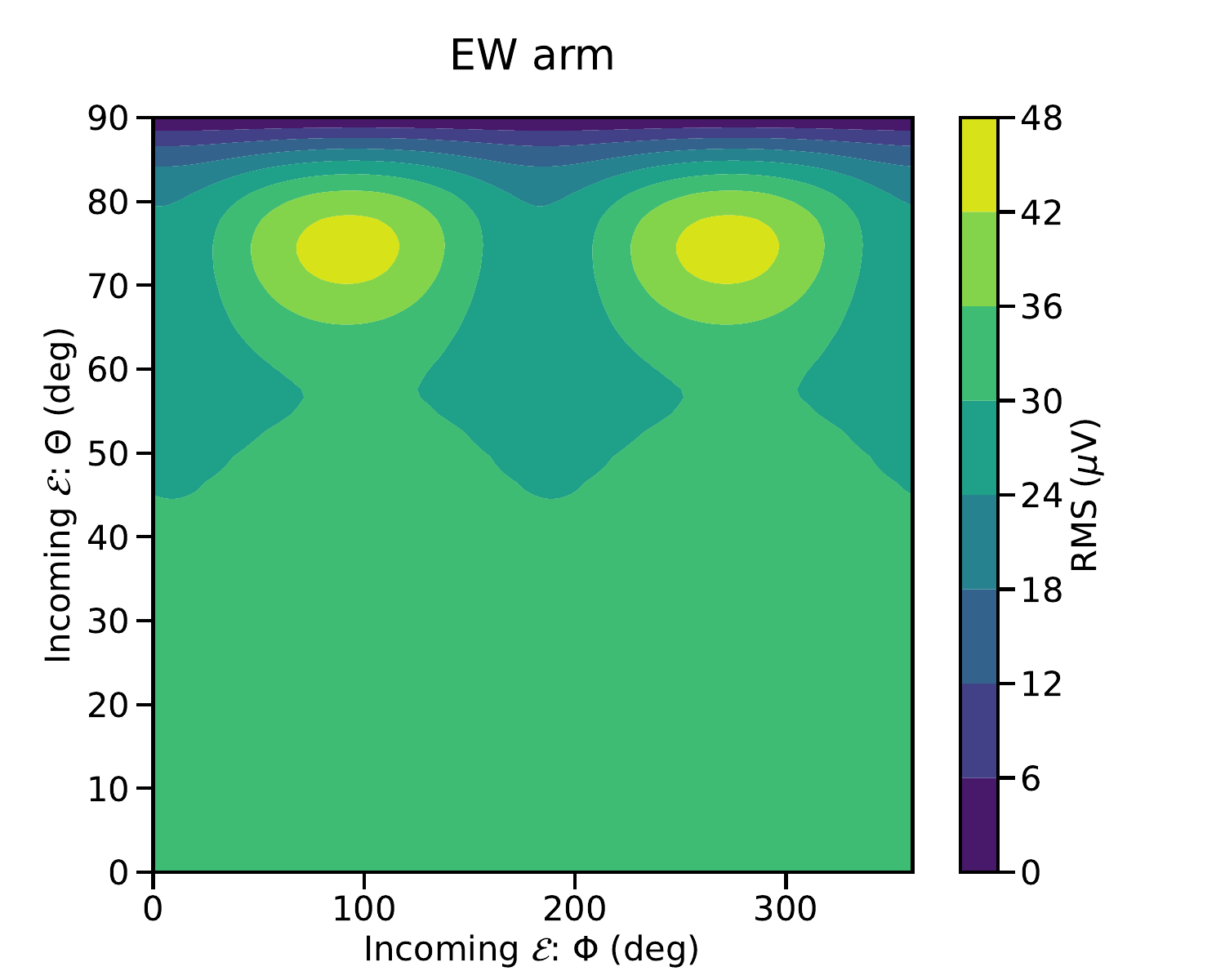}
    \includegraphics[width=0.49\linewidth]{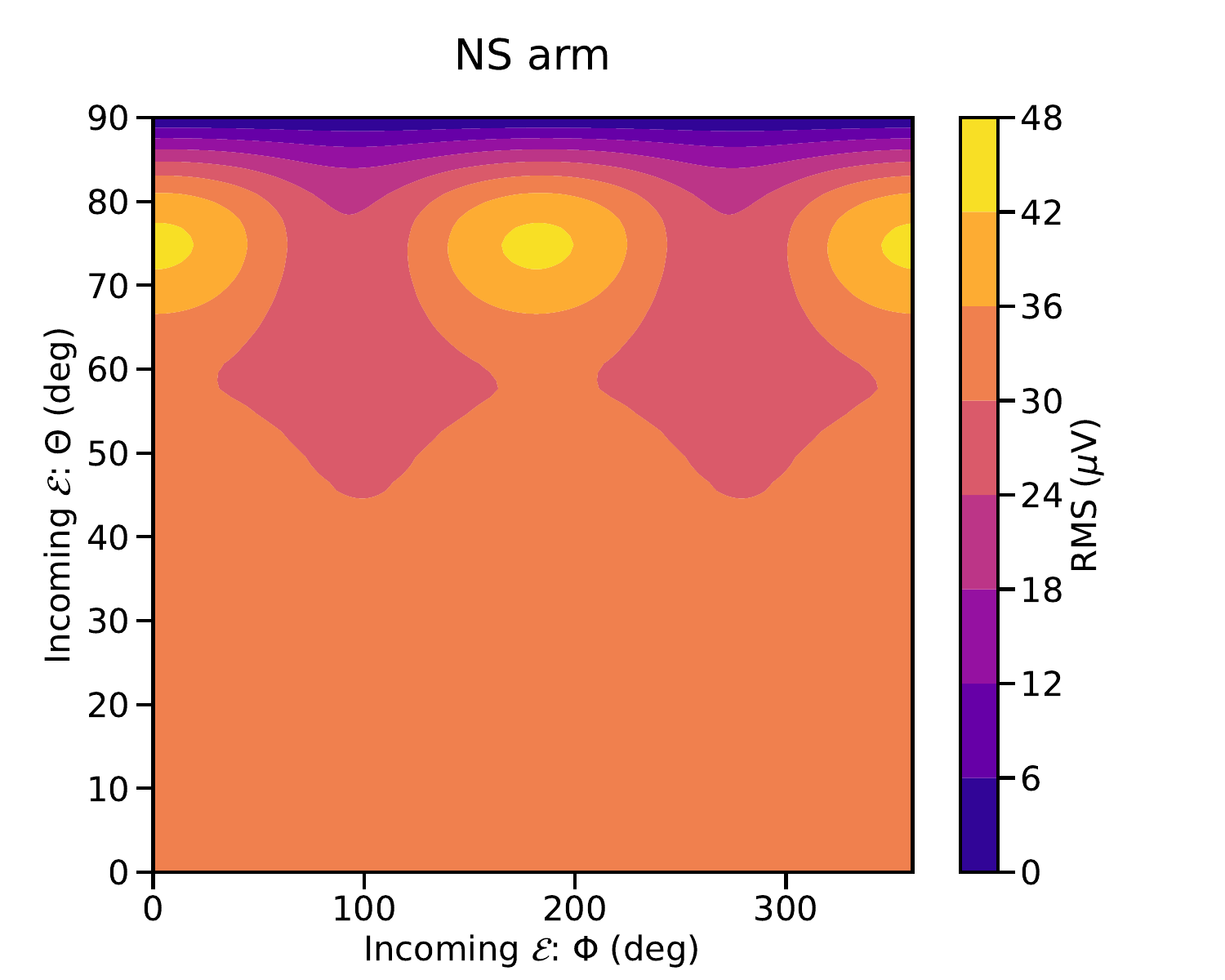}
    \caption{RMS of the induced voltages $V_{\rm EW}$ (left) and $V_{\rm NS}$ (right) for the response of the EW and NS arm antennas, respectively, to Galactic noise.}
    \label{fig:AntennaResponse}
\end{figure}

As mentioned, one of the main sources of external noise is the Galactic plane. It generates an RMS of about 25\,\textmu V/m at frequencies between 30 and 80 MHz.
The elevation angle of the Galactic plane radio emission relative to the radio array depends on local sidereal time.
For this study, we assume that the Galactic radio emission originates from a single point in the sky.
Figure~\ref{fig:AntennaResponse} presents the antenna response in units of induced voltage for a given direction of the Galactic radio emission point. 
The figure shows that the EW and NS arms complement each other in azimuth angles.
This type of antenna has a maximum response at around 75$^\circ$ in zenith angle.

Since the antenna response to Galactic noise depends on the direction from which the electric field arrives, 
the Galactic noise in the antenna varies over time \cite{PierreAuger:2025hoe}. 
In this study, we make the simplified and conservative assumption that the Galactic emission point is fixed at $\Theta\approx75^\circ$ and $\Phi \approx 45^\circ$.
Therefore, the Galactic noise will be maximum, and the EW and NS arms will have similar noise levels.

\subsection{Radio signal simulation}

\begin{figure}
    \centering
    \includegraphics[width=0.8\linewidth]{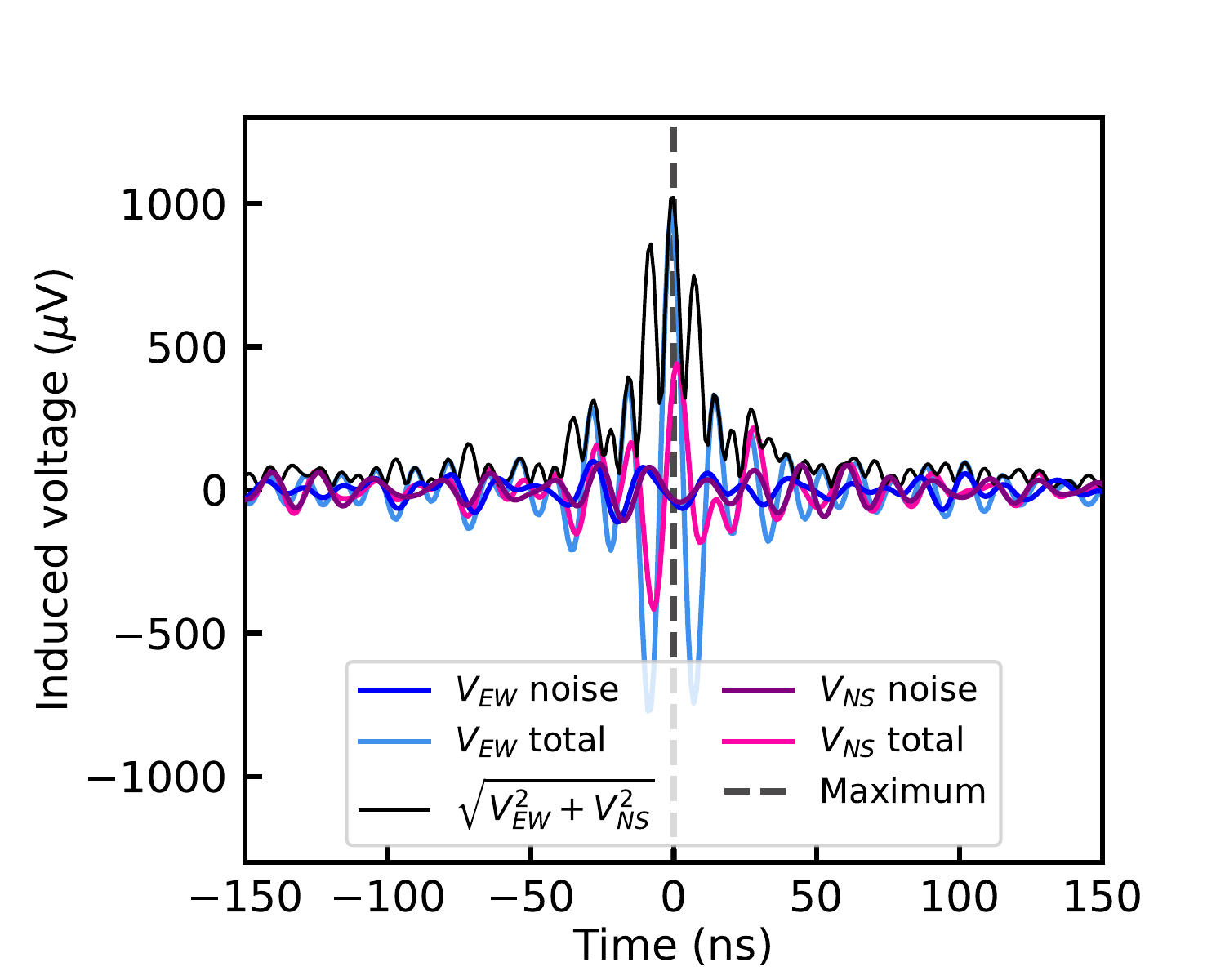}
    \caption{Electric field simulation with the antenna response in 30-80 MHz bandpass and the Galactic noise. 
    }
    \label{fig:RD_signal}
\end{figure}

We obtained the electric-field simulations for each antenna from CoREAS simulations.
Using the radio antenna response at the EW and NS arms, we simulated the induced voltage at each arm of the registered antenna.
Figure\,\ref{fig:RD_signal} presents the induced voltage strengths of the EW and NS arms of one radio antenna from the neutrino simulation and the Galactic noise.








\section{Neutrino detection scenario}


Neutrino detection is very challenging due to its small interaction cross-section, low flux intensity, and high UHECR background.
In this section, we study the trigger efficiency,  reconstruction, and neutrino identification, as well as the enhancement of the effective area, with the goal of complementing particle detector arrays in the detection of downward-going neutrinos.


\subsection{Trigger efficiency of $\nu_e$-CC induced air showers}

We define the trigger threshold, denoted $T_\mathrm{trigger}^\mathrm{SNR}$, in units of the signal-to-noise ratio (SNR):
\begin{equation}
    \mathrm{SNR} = \left(\frac{\sqrt{V_{EW}^2+V_{NS}^2}}{\mathrm{RMS}_\mathrm{noise}}\right)^2\,.
\end{equation}
If the SNR is greater than $T_\mathrm{trigger}^\mathrm{SNR}$, the antenna is tagged as triggered.
We define three cases: an ideal case with $T_\mathrm{trigger}^\mathrm{SNR}=10$, one intermediate case with $T_\mathrm{trigger}^\mathrm{SNR}=30$, and one conservative case $T_\mathrm{trigger}^\mathrm{SNR}=100$.


The latter case is intended to account for surrounding noise from the horizon, as discussed in \cite{AbdulHalim:2024FH}.
The vertical black dashed line in Fig.\,\ref{fig:RD_signal} represents the maximum of the absolute electric field and indicates the timing, $t_i$, when the radio signal is received from the position of the shower maximum $X^{\text{radio}}_{\text{max}}$.
\begin{figure}
    \centering
    \includegraphics[width=0.8\linewidth]{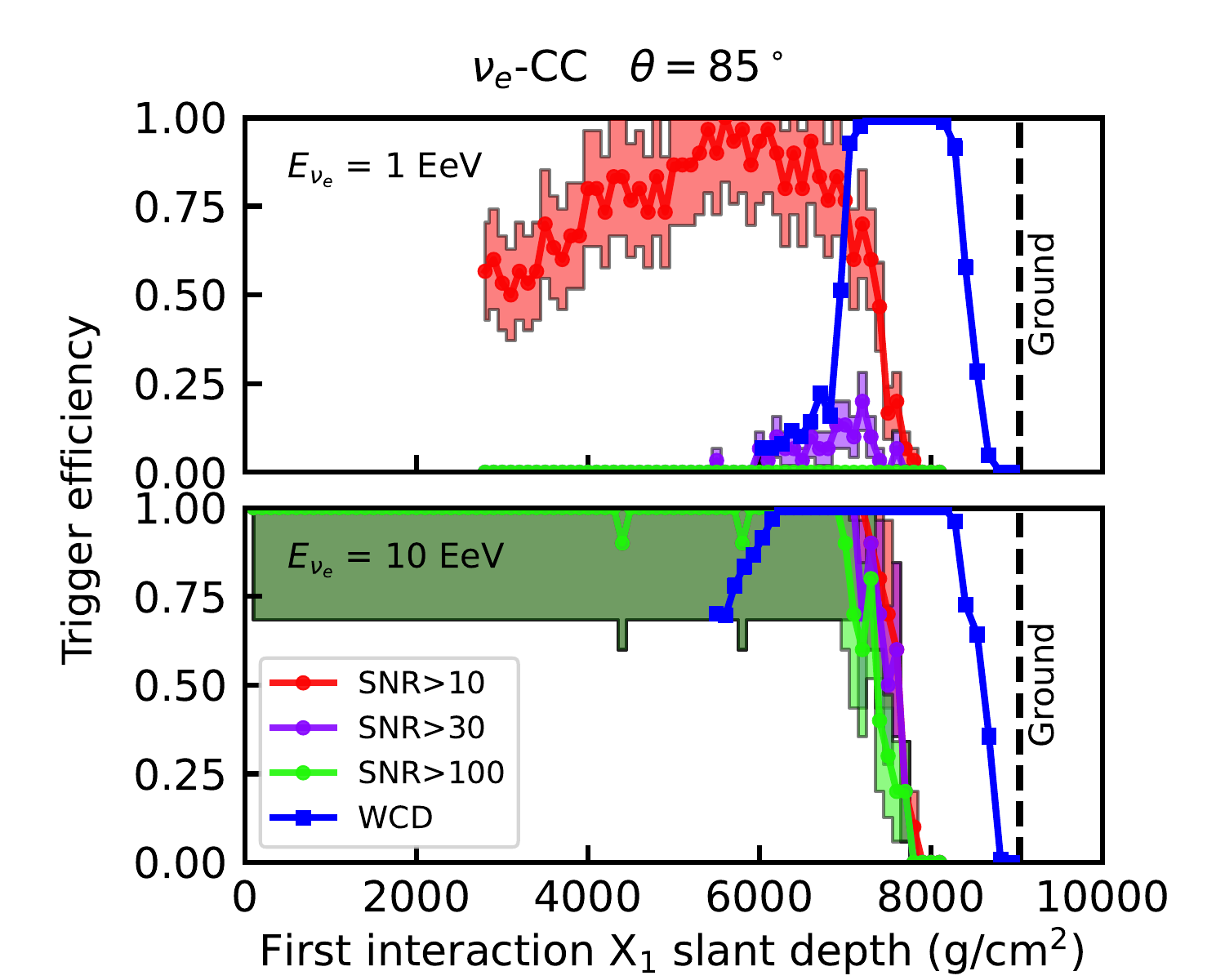}
    \caption{
    $\nu_e$-CC trigger efficiency for radio antennas at different thresholds. It is compared to the trigger efficiency of water Cherenkov detectors (\textit{c.f.} Fig.\ 7.1 and Fig.\ 7.3 in \cite{Tiffenberg:2011thesis}). $X_1=0$ represents the top of the atmosphere. Since the efficiency cannot exceed 1, values above 1 arising from statistical fluctuations were truncated to 1. There are no simulations for $X_1$ below about 3000 g cm$^{-2}$ for the upper plot because it was not expected to trigger the detector with the realistic trigger conditions.}
    \label{fig:RD_trigger_eff}
\end{figure}

An event accepted by the data acquisition (DAQ) is assumed to require at least three triggered antenna stations, i.e.\ $N_\mathrm{trigger} \ge 3$.
Figure~\ref{fig:RD_trigger_eff} shows the radio trigger efficiencies compared with the water Cherenkov detector (WCD) at the Pierre Auger Observatory \cite{Tiffenberg:2011thesis} for energies of 1 and 10\,EeV at a zenith angle of $\theta = 85^\circ$.
The abscissa is the $E_{\nu_e}^\mathrm{CC}$ interaction injection position $X_1$ in units of the slant depth, which denotes the integrated air density. It is approximated as follows:
\begin{equation}
\label{eq:X1}
    X_{1} = \int_{l_{1}}^{l_{\text{top}}(\theta)} \rho(l,\theta) \, dl\,,
\end{equation}
where $l$ represents the distance from the shower core on the ground to a point in the air, $l_{1}$ and $l_{\text{top}}(\theta)$ are the distances to the first interaction point and the top of atmosphere, respectively.
The curvature of the atmosphere has been considered in the air density $\rho(l,\theta)$ and the length $l_{\text{top}}(\theta)$.


At $E_{\nu_e} = 1$\,EeV, the trigger efficiency is highly sensitive to the value of  $T_\mathrm{trigger}^\mathrm{SNR}$. 
For $T_\mathrm{trigger}^\mathrm{SNR}=10$, the trigger efficiency complements the WCD trigger efficiency at the middle of the atmosphere.
However, for $T_\mathrm{trigger}^\mathrm{SNR}\geq 30$, the trigger efficiency is lower than that of the WCD at any point of the first interaction.
At higher energies of $E_{\nu_e}^\mathrm{CC}=10$\,EeV, the trigger efficiency exceeds that of the WCD for a wide range of shower starting points, $X_1$, regardless of the trigger threshold. 
However, triggering the radio array is difficult for very young showers starting close to the ground because the showers must develop sufficiently to generate a detectable radio signal. Additionally, the cone-shaped development of the footprint must be large enough to trigger a sufficient number of antennas spaced 1.5\,km apart.
Therefore, as shown in Fig.\,\ref{fig:RD_trigger_eff}, the radio trigger efficiency is lower than that of a particle detector array when the showers are young.
On the other hand, radio signals in the 30-80\,MHz band can propagate much farther than particles can trigger their respective detectors, making radio antenna arrays sensitive to older showers that are invisible to particle detector arrays.
Thus, we expect that the exposure of a complementary radio antenna array will surpass that of a particle detector array at energies higher than $E \gtrsim 10^{18}$\,eV.

An observatory with a denser array of radio antennas than the Pierre Auger Observatory (which has 1.5 km spacing) would have a greater complementary radio trigger exposure, especially at lower energies. This is because more antennas would fall within a typical EAS footprint. This increases the probability of meeting multi-station trigger and reconstruction requirements.
Furthermore, neutrino detection is more effective at increasingly inclined zenith angles, because the radio footprint grows rapidly with the zenith angle, illuminating multiple stations. Additionally, inclined trajectories provide larger slant depths, which increases the probability of neutrino interactions. Therefore, the following analysis will focus on zenith angles greater than $75^\circ$.
In addition, the neutrino trigger efficiency exhibits interesting behavior, which is discussed in Appendix~\ref{sec:TriggerEff}. 

\subsection{The maximum radio emission point $\vec{r}_{\rm max}$}
\label{sec:RadioPositionRecon}
The position and emission time of the radio emission maximum, $X^{\text{radio}}_{\text{max}}$, are represented by the vectors $\vec{r}_{\rm max}$ and $t_{\rm max}$, respectively.
The maximum radio pulse at the antenna corresponds to the maximum emission point, $X^{\text{radio}}_{\text{max}}$.
Assuming that the maximum radio wavefront is spherical \cite{Corstanje:2014waa}, we posit that the wavefront propagates to the antenna at a velocity of $\gamma_{n} \cdot c$ in the air. 
%
%
The parameter $\gamma_n$ is introduced as an effective correction factor in the timing reconstruction. 
It accounts, in a phenomenological way, for deviations from straight-line propagation at the vacuum speed of light, including the effects of the atmospheric refractive index, refraction, and the non-spherical geometry of the true radio wavefront. 
These effects become more relevant for very inclined air showers, for which the actual propagation path of the radio signal can be longer than the geometric straight-line distance between the effective emission region and an antenna. 
In this sense, $\gamma_n$ can be interpreted as an effective propagation parameter that absorbs these delays \cite{Fliescher:2011rxa}.

The reconstruction itself is a simple geometrical fit based on the expected arrival times of the radio pulse at the different antenna positions. 
For each triggered antenna $i$, we define $V_i^{\rm max}$ as the maximum amplitude of $\sqrt{V_{\rm EW}^2+V_{\rm NS}^2}$, and denote by $t_i$ the time at which this maximum occurs. 
If the pulse originates from an effective source position $\vec{r}_\text{max}$ at time $t_\text{max}$, the expected arrival time at antenna position $\vec r_i$ is $t_\text{max}+\frac{|\vec r_\text{max}-\vec r_i|}{\gamma_n c}.$

The source position $\vec r_\text{max}$ is then obtained by minimizing the timing residuals over all triggered antennas using the $\chi^2$ function
\begin{equation}
\chi^2=\sum_i \left( \frac{t_i- \left(t_\text{max}+\frac{|\vec{r}_\text{max}-\vec{r}_i|}{\gamma_{n} \cdot c}\right)}  {\sigma^\text{GPS}_t} \right)^2
+ \left( \frac{1-\gamma_{n}}{\sigma_{\gamma_{n}}}\right)^2 \: .
\end{equation}
The first term represents the agreement between the measured pulse times and the arrival times expected from the geometrical model. 
The second term acts as a penalty term on $\gamma_n$: it constrains the fit to remain close to the physically expected value $\gamma_n$=1, while still allowing for small deviations required by refractive and wavefront effects. 
Here we use a Gaussian penalty around $\gamma_n=1$ in the form $(1-\gamma_n)^2/\sigma_{\gamma_n}^2$.
Without such a constraint, $\gamma_n$ could absorb statistical timing fluctuations and become weakly constrained, which would in turn bias the reconstructed source position $r_\text{max}$.

The width of this constraint is parameterized by $\sigma_{\gamma_n}=\frac{c}{\Delta s}\sqrt{2}\,\sigma_t^\text{GPS}$, where $\Delta s$=1.5 km is taken as the characteristic spacing between two antennas and $\sigma_t^\text{GPS}$ is the GPS timing uncertainty. 
This expression can be understood as translating the timing uncertainty between two stations separated by a typical baseline $\Delta s$ into an uncertainty on the effective propagation parameter $\gamma_n$. 
In this way, the allowed variation of $\gamma_n$ is set by the intrinsic timing resolution of the detector array.

To obtain a reliable estimate of the emission point $\vec{r}_\text{max}$, a quality cut of $\chi^2/ndf<10$ is applied. 
In Appendix~\ref{sec:Wavefront}, we compare different wavefront models. Among them, the spherical radio wavefront with the $\gamma_{n}$-factor is the best model, especially for the horizontal showers considered here.

\subsection{The radio footprint}
\label{sec:Footprint}
\begin{figure}[ht]
    \centering    
   \includegraphics[width=0.8\linewidth]{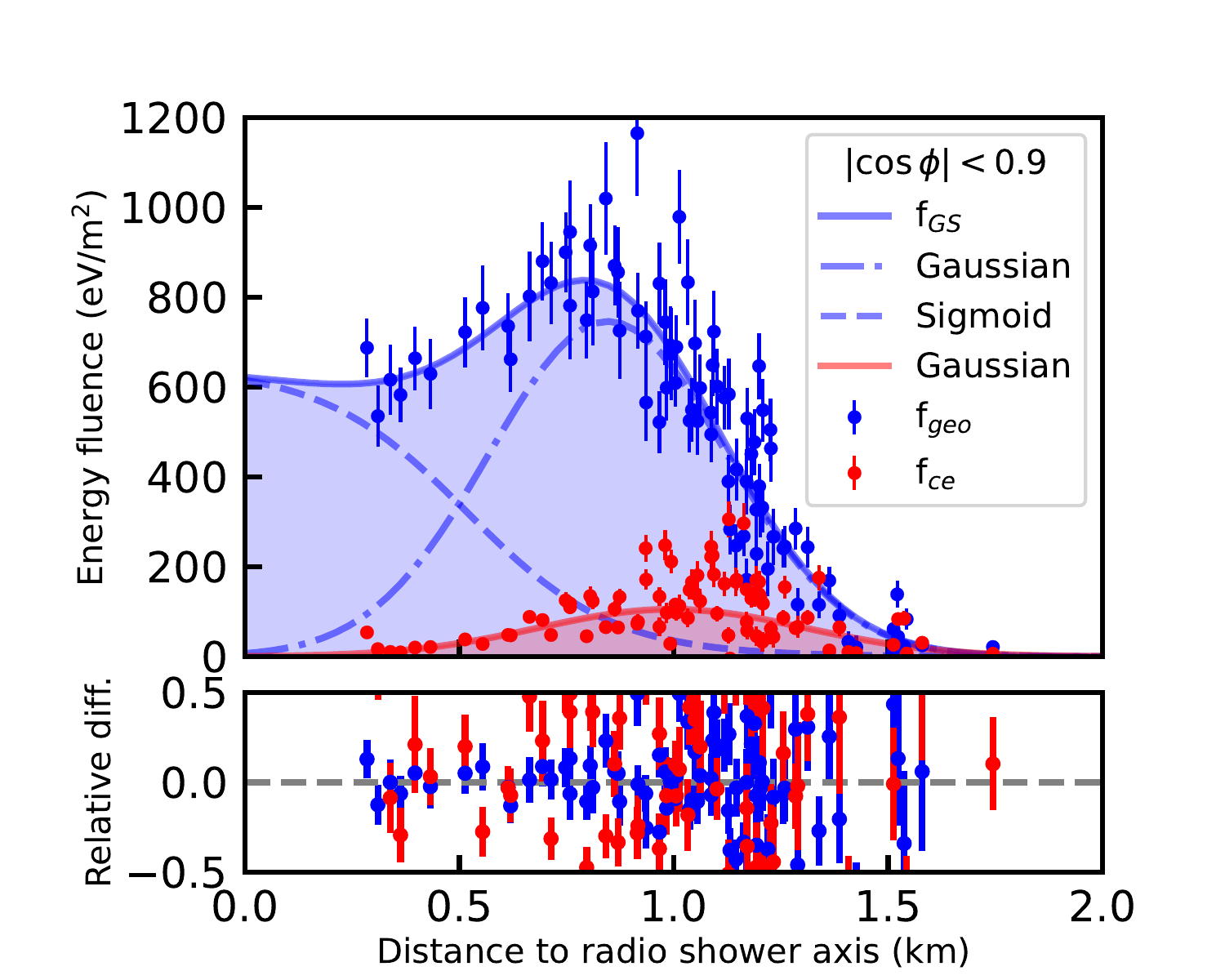}
    \caption{The performance of the footprint fits for a $\nu_e$-CC simulation of $E_{\nu_e}=30$ EeV, $\theta=87^\circ$, and $X_1=8000$\,g\,cm$^{-2}$ with 30-80\,MHz bandpass, white noise, antenna response and GPS timing uncertainty (5 ns). 
    $\phi$ represents the angle in the shower plane between the antenna position and the $\vec{v}\times\vec{B}$ axis ($\phi=0$: $\vec{v}\times\vec{B}$, $\phi=90^\circ$: $\vec{v}\times(\vec{v}\times\vec{B})$).
    }
    \label{fig:Footprint}
\end{figure}

The geomagnetic energy fluence is radially symmetric in the shower plane ($\vec{v} \times \vec{B}$,  $\vec{v} \times (\vec{v} \times \vec{B})$), thereby enabling the determination of the position of the shower core, $\vec{r}_\text{core}$, at ground level. 
The energy fluence of the signal at each polarization ($\vec{v} \times \vec{B}$,  $\vec{v} \times (\vec{v} \times \vec{B})$) on the shower plane can be obtained via the noise subtraction method (\cite{PierreAuger:2025aik})
\begin{equation}
\label{eq:energyfluence}
    f = \epsilon_0 \, c \left( \int_{t_1}^{t_2}  \mathcal{E} ^2 (t) {\rm d}t - \frac{t_2-t_1}{t_4-t_3} \cdot \int_{t_3}^{t_4}  \mathcal{E} ^2 (t) {\rm d}t\right) \,,
\end{equation}
where $\epsilon_0$ is the vacuum permittivity, $\mathcal{E}$ is the electric field at each polarization on the shower plane, $t_1$ and $t_2$ represent the signal window, and $t_3$ and $t_4$ the noise window.
As delineated in \cite{Schluter:2022mhq}, the reconstruction of the shower core can be achieved through the utilization of the geomagnetic energy fluence via
\begin{equation}
\label{eq:fgeo1}
    f_\text{geo} = \left(\sqrt{f_{\vec{v}\times\vec{B}}} - \frac{\cos\phi}{|\sin\phi|}\cdot \sqrt{f_{\vec{v}\times(\vec{v}\times\vec{B})}} \right)^2\,,  \quad  
\end{equation}
where $\vec{B}=0.24$ G is the magnetic field with an inclination of approximately $-36^\circ$ at the reference observatory site, and $\vec{v}$ is the direction of the shower axis. 
The reconstructed position, $\vec{r}_\text{max}$, of $X^{\text{radio}}_{\text{max}}$ is an input for the shower axis $\vec{v}=\vec{r}_\text{core}-\vec{r}_\text{max}$. 
After correcting for the so called ``early-late'' effect and subtracting the charge excess component, the geomagnetic energy fluence can be expressed as a combination of Gaussian and sigmoid functions:
\begin{equation}
\label{eq:fgeo2}
    f_{GS} = f_0 \Bigg( \underbrace{  \exp\left( - \left(\frac{r-r_0}{\sigma}\right)^{p(r)}\right) }_\text{Gaussian} + \underbrace{\frac{a_\text{rel}}{1+\exp(s\cdot\left(r/r_0 - r_{02}\right))} }_\text{Sigmoid} \Bigg) \,,
\end{equation}
where $f_0$, $r_0$, $\sigma$, $p(r)$, $a_\text{rel}$, $s$, and $r_{02}$ are free parameters. 
In this study, we fix $s=5$ as done in Ref.\,\cite{Schluter:2022mhq}.
Furthermore, $p(r)=2$ when $r\lq r_0$ and $p(r)=(r_0/r)^{b/1000}$ when $r>r_0$.
The charge excess energy fluence can be calculated using $ f_\text{ce} = \frac{1}{\sin^2\phi}\cdot f_{\vec{v}\times(\vec{v}\times\vec{B})}$. 
We found that the Gaussian part of Eq.~\ref{eq:fgeo2} can also adequately describe the charge excess energy fluence:
\begin{equation}
\label{eq:fce}
    f_{G} = f_0  \cdot \exp\left( - \left(\frac{r-r_0}{\sigma}\right)^{p(r)}\right) \,.
\end{equation}
Figure \ref{fig:Footprint} illustrates the performance of the footprint fits to a simulated event for both the geomagnetic and charge excess components. 

Since geomagnetic radiation is the dominant component, we only use the geomagnetic radiation fit to estimate $\vec{r}_\text{core}$. 
A quality cut of $\chi^2/ndf<10$ is used to select good fits. 

\subsection{The shower axis}
\label{sec:ShowerAxisRecon}
There exists a multitude of methodologies for the reconstruction of the shower axis through the utilization of radio antennas.
In the following, we will present some methodologies for the reconstruction of the shower axis.
\begin{enumerate}
    \item Interferometric mapping \cite{LOPES:2005ipv,Schoorlemmer:2020low}.
    \item The shower maximum $\vec{r}_{\rm max}$ and the radio footprint core position $\vec{r}_{\rm core}$.
    \item The shower maximum $\vec{r}_{\rm max}$ and the barycenter position $\vec{r}_{b}$.
    \item The shower maximum $\vec{r}_{\rm max}$ and the particle footprint core position $\vec{r}_{p}$.
\end{enumerate}

In this study, the second method mentioned above will be adopted as the default option for the reconstruction of the shower axis.
The precise estimation of the radio footprint core is imperative for the reconstruction of the shower axis.
However, in scenarios where the number of triggered antennas is limited, the reconstruction of the radio footprint core becomes a formidable challenge.
The alternative options for the radio cores could be the barycenter position or the particle footprint core reconstructed by particle detectors.
In circumstances where atmospheric refraction is minimal for radio propagation, the particle footprint core will be in close proximity to the radio footprint core.
The measurement of the particle footprint core is facilitated by the particle detectors.
It is hypothesized that for less inclined air showers or for neutrino-induced showers situated in proximity to the ground, the core of the particle footprint and the core of the radio footprint will demonstrate a high degree of similarity. This assertion is supported by the findings of \cite{Schluter:2020tdz,Corstanje:2017djm}.

The position of the barycenter $\vec{r}_{b}$ is intrinsically different from the radio footprint core.
However, when the shower maximum is far from the ground, it is an acceptable approximation to treat the barycenter position $\vec{r}_{b}$ as the radio footprint core.
Alternative options will be adopted only when the radio footprint core reconstruction is poor.
In this study, the alternative option for the footprint core is the barycenter position $\vec{r}_{b}$.

\subsubsection{Reconstruction methods}
\paragraph{The radio emission maximum $\vec{r}_{\rm max}$ and the radio footprint core $\vec{r}_{\rm core}$}
The radio emission position $\vec{r}_{\rm max}$ is reconstructed as illustrated in Sec.\,\ref{sec:RadioPositionRecon}.
The radio footprint core $\vec{r}_{\rm core}$ can be found from the footprint fit as shown in Sec.\,\ref{sec:Footprint}.
Linking these two points, the shower axis can be found:
\begin{equation}
    \vec{v} = \vec{r}_{\rm core}  - \vec{r}_{\rm max}\,.
\end{equation}

\paragraph{The radio emission maximum $\vec{r}_{\rm max}$ and the barycenter position $\vec{r}_{b}$}
In this study, we use barycenter position $\vec{r}_{b}$ of the radio antennas as the alternative shower core on the ground.
The barycenter position, $\vec{r}_{b}$, can be found using the following equation:
\begin{equation}
    \sum_i^N V_i^{\rm max}\cdot \vec{d}_i = 0 \,,
\end{equation}
where $i$ is the antenna index, $V_i^{\rm max}$ is the maximum amplitude of $\sqrt{V_{\rm EW}^2+V_{\rm NS}^2}$, and $\vec{d}_i=\vec{r}_i-\vec{r}_{b}$ is the distance from the antenna position to the barycenter position.
Then, we obtain the shower axis by
\begin{equation}
    \vec{v} = \vec{r}_{b}  - \vec{r}_{\rm max} \, .
\end{equation}

\begin{figure}
    \centering
    \includegraphics[width=0.9\linewidth]{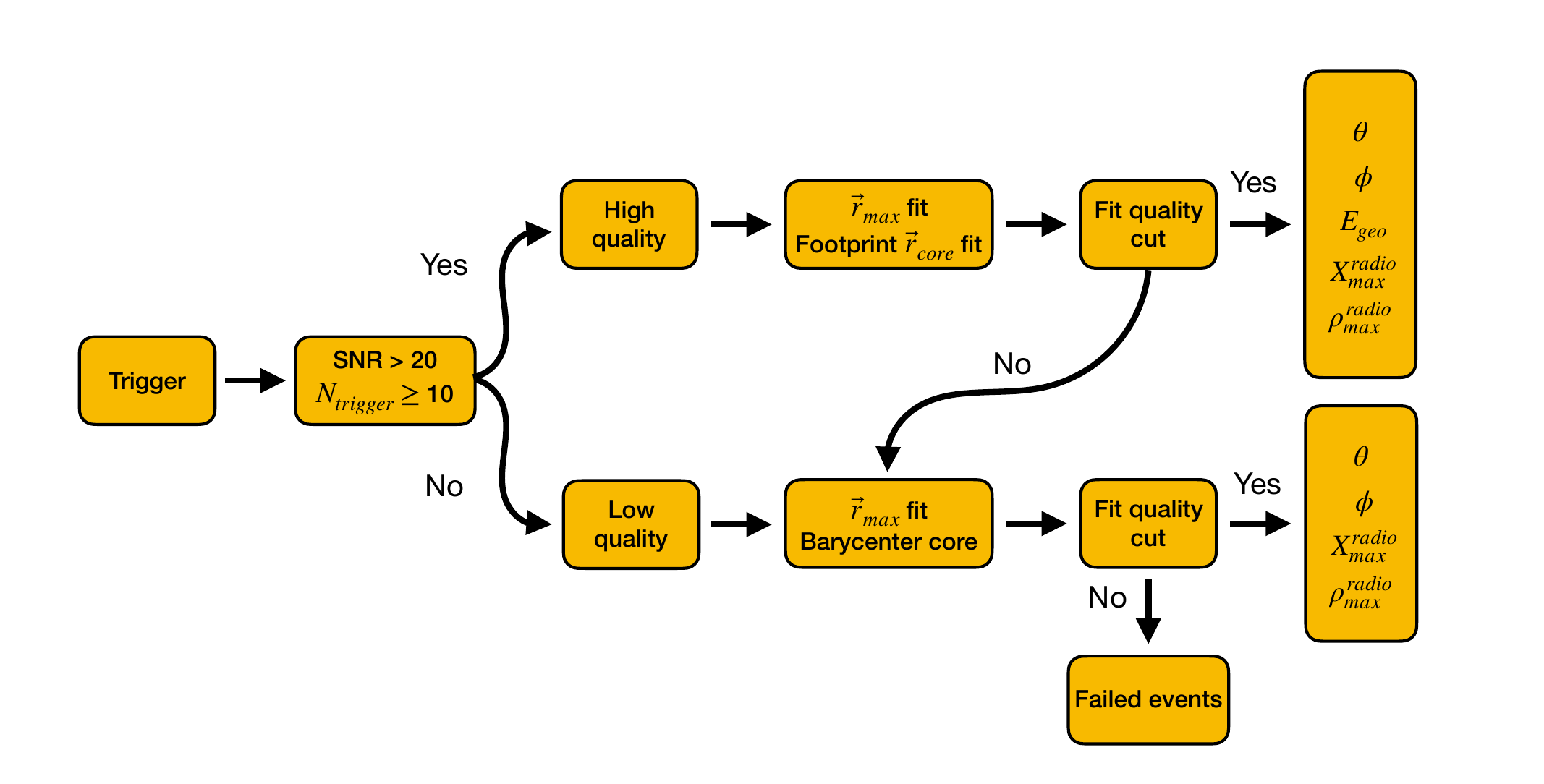}
    \caption{The logic diagram of the reconstruction process.}
    \label{fig:LogicDiagram}
\end{figure}

\subsubsection{Reconstruction performance}
\paragraph{Reconstruction strategy}
We categorize triggered events as either high quality (HQ) or low quality (LQ).
To demonstrate our reconstruction process, we created a logic diagram, which is shown in Fig.~\ref{fig:LogicDiagram}.
Since LQ events have limited capability to reconstruct the radio footprint, we use the alternative option for the shower core: the barycenter position.
For this analysis, we set the trigger conditions to $\mathrm{SNR}>10$ and $N_{\rm trigger}\ge3$.

\begin{figure}[ht]
    \centering
    \includegraphics[width=0.48\linewidth]{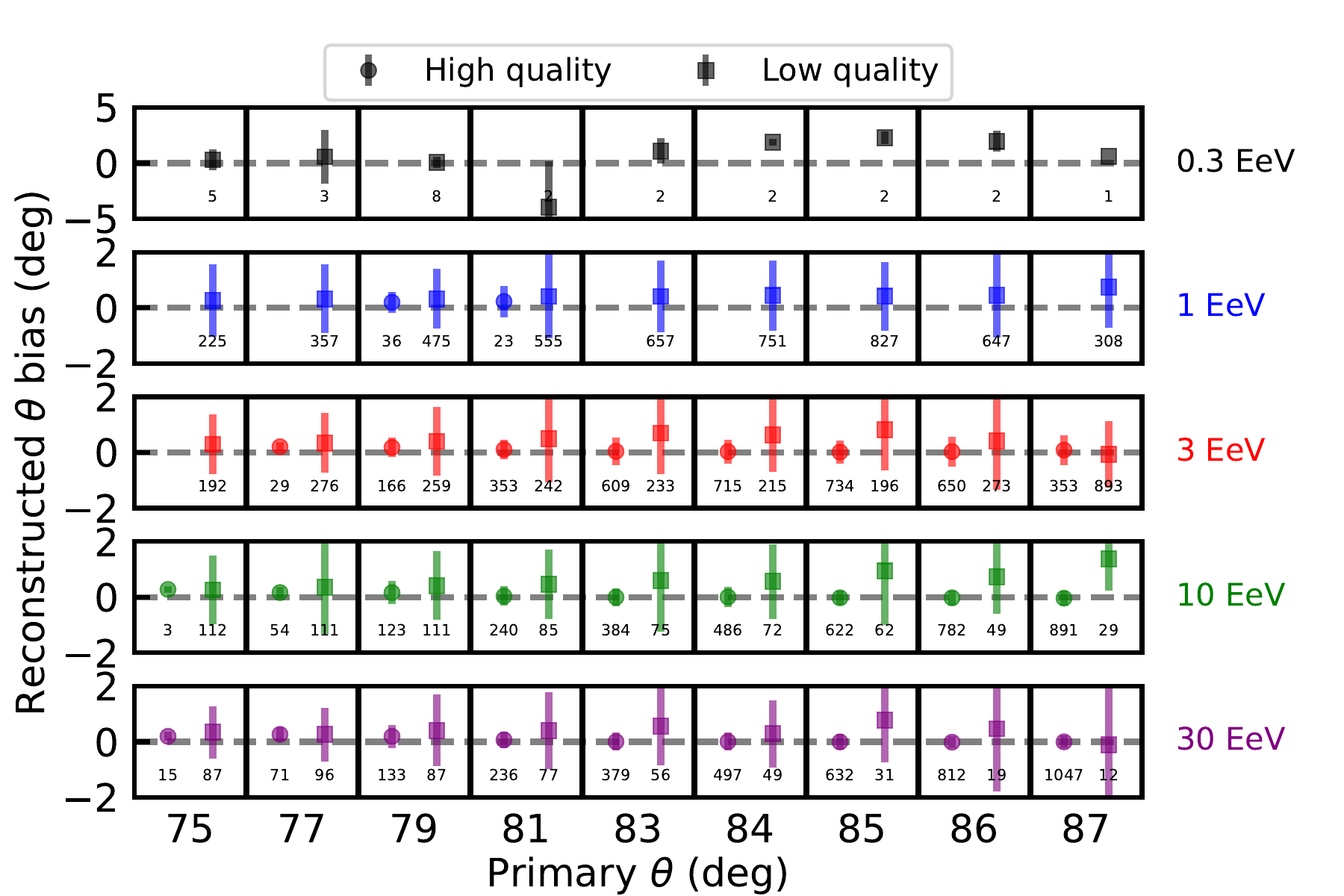}
    \includegraphics[width=0.48\linewidth]{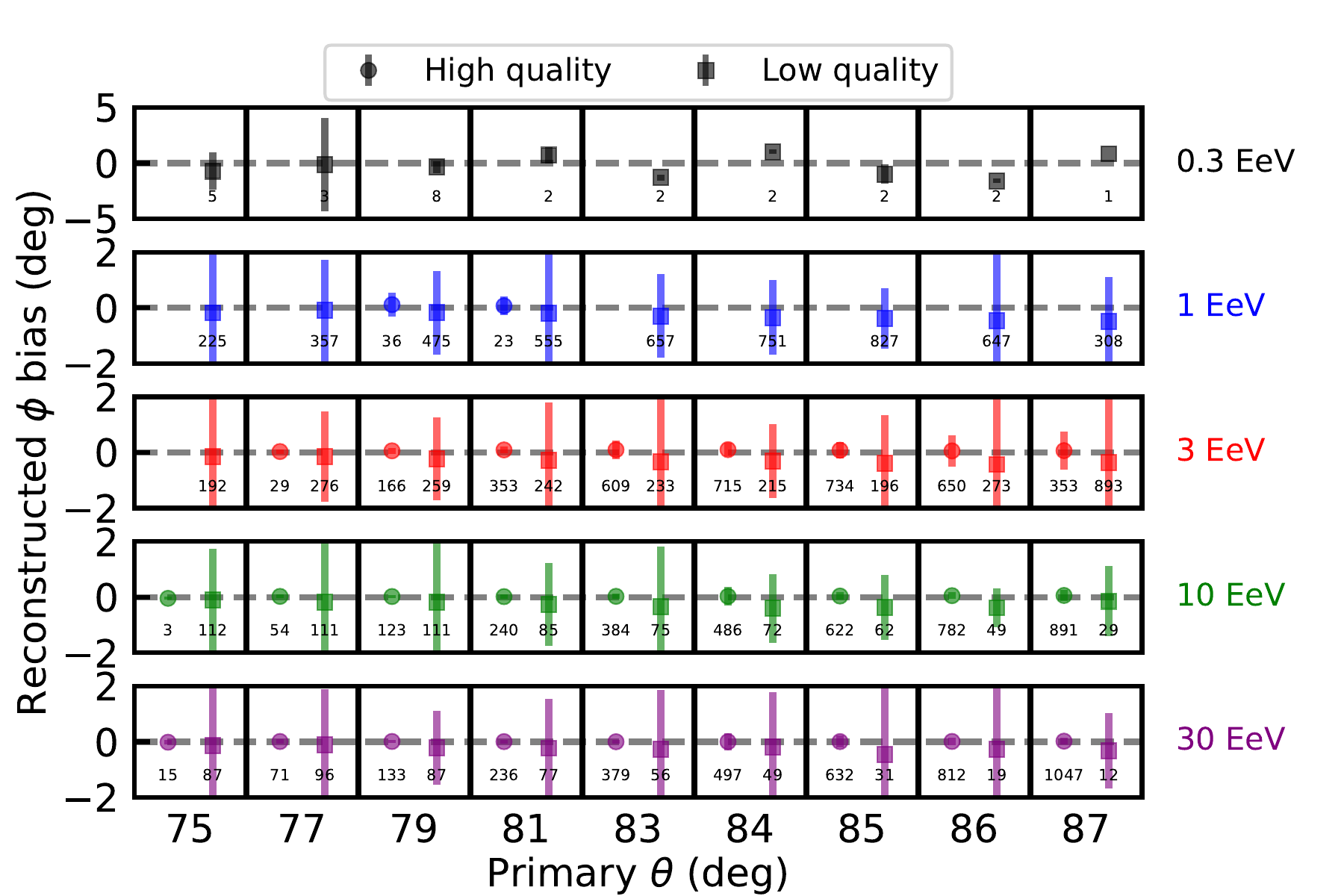}
    \caption{Left: bias of the reconstructed zenith angle $\theta$ as a function of primary $\theta$ for different primary energies.
    Right: same, but for the reconstructed azimuth angle $\phi$ of the shower axis.
    The numbers inside the boxes indicate the number of simulated events.}
    \label{fig:ShowerAxisRecon}
\end{figure}

\paragraph{High quality events}
The HQ event conditions are $\mathrm{SNR}>20$ and $N_{\rm trigger}\ge 10$.
To properly estimate the uncertainty of the shower axis, we must consider the propagation of errors from the reconstructed values of $\vec{r}_\text{max}$ and $\vec{r}_\text{core}$.
Using two quality cuts for both the $\vec{r}_{\rm max}$ fit and the radio footprint fit, we demonstrate the reconstruction performance using circle markers in both figures in Fig.\,\ref{fig:ShowerAxisRecon}.
The figure on the left (right) shows the bias of the reconstructed angle $\theta$ ($\phi$) as a function of the primary angle $\theta$ for different primary neutrino energies.
At a given primary zenith angle, $\theta$, the higher the energy, the lower the uncertainty of the reconstructed angles.


\paragraph{Low quality events}
The reconstruction of the LQ events is shown by the square markers in Fig.~\ref{fig:ShowerAxisRecon}.
The bias and uncertainty are larger for LQ events than for HQ events.
Additionally, a few outliers with significant bias appear in the reconstruction of $\phi$ for $N_{\rm trigger}=3$. 
These outliers were caused by the limited number of triggered stations and have been removed from Fig.\,\ref{fig:ShowerAxisRecon}. 

More details about the performance of the shower axis reconstruction can be found in Appendix~\ref{sec:ShowerAxisReconWithDifferentCores} with three different shower core assumptions.






\subsection{The radio emission depth $X_{max}^{radio}$}
Once the positions of $r_{\text{max}}$ and the shower axis are known, $X^{\text{radio}}_{\text{max}}$ can be obtained by integrating the air density from the top of the atmosphere to the maximum emission point $\vec{r}_\text{max}$. 
We considered the propagation of errors from the reconstructed shower axis ($\theta$) and the maximum emission point $\vec{r}_\text{max}$ to the reconstructed $X^{\text{radio}}_{\text{max}}$.
This depth is similar to Eq.~\eqref{eq:X1} and is written as 
\begin{equation}
\label{eq:Xradio}
    X^{\text{radio}}_{\text{max}} = \int_{l_{\text{max}}}^{l_{\text{top}}(\theta)} \rho(l,\theta) \, dl\,,
\end{equation}
where $l_{\text{max}}$ is the distance from the top of the atmosphere to the position of the maximum radio emission.

\subsection{Energy reconstruction}
Once neutrino events are registered by the observatory, it is crucial to reconstruct the energy to understand the processes from which the neutrinos originate.
Radio techniques have the potential to precisely reconstruct the electromagnetic energy of air showers, even far from the shower maximum. This makes radio techniques very attractive for neutrino astrophysics.
However, we must first investigate the correlation between radio radiation energy and the electromagnetic energy of EAS induced by neutrinos.

In this subsection, we analyze simulations of $\nu_e$-CC induced EAS to derive the conversion between radio radiation and electromagnetic shower energy.
Based on these simulations, we then derive a conversion model for $\nu_e$-CC interactions.
Once an EAS induced by a $\nu_e$-CC interaction is identified, the derived model can be used to estimate the primary electromagnetic energy. Next, we discuss the performance of electromagnetic energy reconstruction. Lastly, we briefly discuss the primary neutrino energy reconstruction.


\subsubsection{Radio radiation energy from $\nu_e$-CC interactions}
Radio emission from extensive air showers contains transverse current radiation (which we refer to as ``geomagnetic radiation'' in this work), charge excess radiation, and geosynchrotron radiation \cite{Chiche:2024yos}. 
The magnetic field strength of 0.24 G at the Auger site does not significantly contribute to geosynchrotron emission for the inclined showers discussed here. Since its decoupling is still unknown, we ignore this contribution in the present study.
Studying the contribution of this component to radio emission polarization is beyond the scope of this paper, but we note its importance in discriminating neutrinos from cosmic rays \cite{Chiche:2024yos}.

In this section, we study simulations $\nu_e$-CC interactions without considering any detector effects besides a rectangle 30-80 MHz bandpass filter. 
We call these simulations with only a rectangle 30-80 MHz bandpass filter as ``signal-only'' in this study.

\paragraph{Geomagnetic radiation energy to the primary electromagnetic energy}
The geomagnetic radiation energy can be obtained by integrating Eq.\,\eqref{eq:fgeo2}:
\begin{equation}
    E_{\rm geo} = 2\pi\int_0^\infty r\cdot f_{GS}\,(r)dr \,.
\end{equation}
References \cite{Glaser:2016qso,Schluter:2022mhq} derive the geomagnetic radiation fraction relative to a specific air density at the shower maximum for a given electromagnetic energy of showers.
Here, we adopt the convention from Ref.~\cite{Schluter:2022mhq} to convert geomagnetic emission energy to electromagnetic energy from $\nu_e$-CC interactions.
However, since this study focuses on radio array observations, we use the air density $\rho$ at the radio emission maximum instead of that at the particle or $dE/dX$ maximum.
Modeling the radiation energy with the particle maximum is presented in Appendix~\ref{sec:ParticleMaxModeling}.
Given a specific amount of electromagnetic energy from a shower, the geomagnetic emission can be described as a function of the air density at the radio emission maximum:
\begin{equation}
    \label{eq:EgeoFraction}
    \frac{E_{\rm geo}}{\sin^2\alpha} = S_{\rm geo} \cdot \left(1-p_0 + p_0\cdot \exp{\left( p_1\cdot(\rho - \rho_{ref}) \right)} \right)\, .
\end{equation}
Here, $\alpha$ is the angle between the shower axis $\vec{v}$ and the magnetic field $\vec{B}$, and $S_{\rm geo}$ is the reference radiation energy from the shower, with the radio emission maximum at the reference air density $\rho_{\rm ref}$ when $\alpha=90^\circ$.
The geomagnetic radiation energy $S_{\rm geo}$ can be correlated with the electromagnetic energy of showers.
For the inclined EAS discussed here, we adopt the reference air density $\rho_{\rm ref}$ at the average air density $\langle\rho\rangle=0.3$ kg\,m$^{-3}$, as given by Ref.\,\cite{Schluter:2022mhq}.
Therefore, we can directly compare the parameters of UHECR and $\nu_e$-CC interactions.

\begin{figure}
    \centering
    \includegraphics[width=0.48\linewidth]{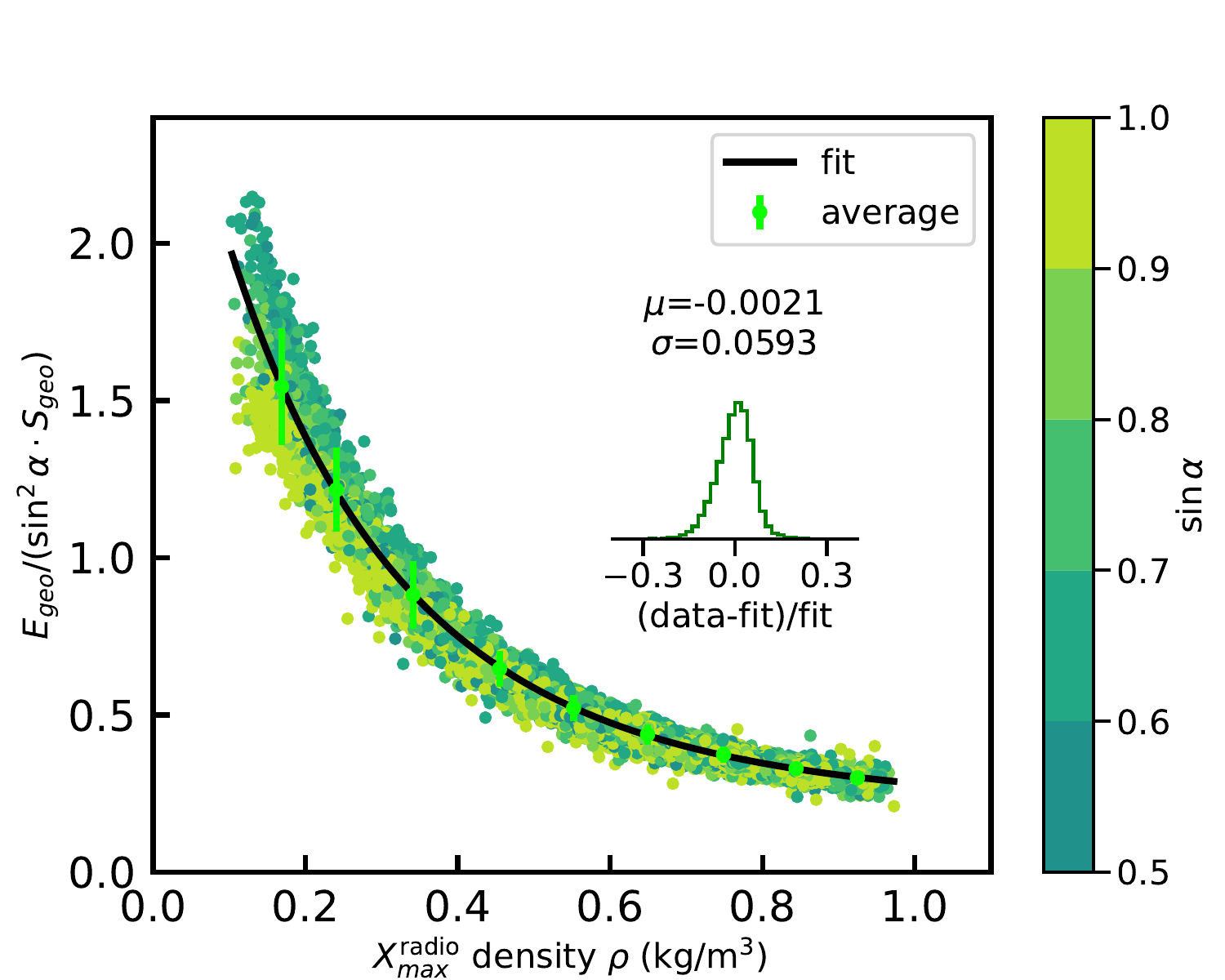}
    \includegraphics[width=0.48\linewidth]{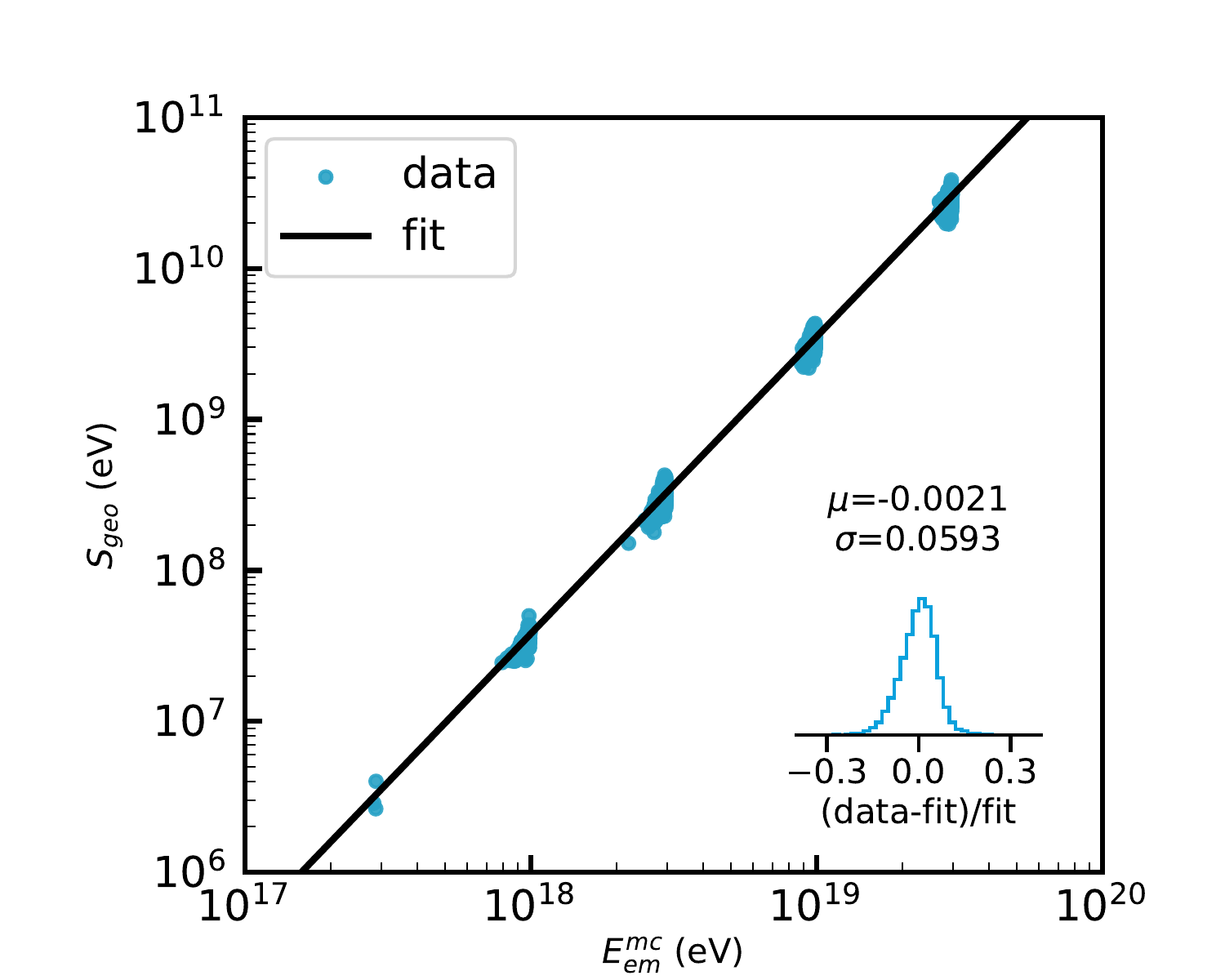}
    \caption{Left: geomagnetic radiation energy as a function of the air density at the position of the radio emission maximum for ``signal-only'' simulations with Sibyll-2.3d model. 
    Right: conversion between the reference radiation energy and the electromagnetic energy for ``signal-only'' simulations with Sibyll-2.3d model. 
    }
    \label{fig:EgeoRadiationEnergyModeling}
\end{figure}


The reference energy of geomagnetic radiation $S_{\rm geo}$ can then be correlated with the electromagnetic energy $E_{\rm em}$ by using a power law:
\begin{equation}
\label{eq:EgeoEem}
    E_{\rm em} = 10\,\text{EeV} \left(\frac{S_{\rm geo}}{S_{19}}\right)^{1/\gamma}\, .
\end{equation}
Here, $S_{19}$ represents the reference geomagnetic radiation energy from a shower with a radio emission maximum at the reference air density $\rho_{\rm ref}$ when the electromagnetic energy is 10\,EeV. $\gamma$ represents the power-law index.

Using Eq.\eqref{eq:EgeoFraction} and \eqref{eq:EgeoEem}, we can correlate the geomagnetic radio radiation $E_{\rm geo}$ with the electromagnetic shower energy $E_{\rm em}$.
We employed two datasets for $\nu_e$-CC interactions that used different hadronic interaction models: Sibyll-2.3d \cite{Riehn:2019jet} and EPOS-LHC \cite{Pierog:2013ria}.
The simulations use the length factor ``STEPFC'' of 1 for electron multiple scattering.
To correct the geomagnetic radiation energy in the simulations, we scale $E_{\rm geo}$ by 111\,\% based on Ref.\,\cite{Gottowik:2017wio}.
We then performed a combined fit to determine the parameters in Eq.\,\eqref{eq:EgeoFraction} and \eqref{eq:EgeoEem} for these two datasets.
They are presented in Table~\ref{tab:EgeoEem} for both hadronic models.

\begin{table*}[!ht]
\centering
{\small
\begin{tabular}{ lcc }
 \toprule
 $\nu_e$-CC  & Sibyll-2.3d & EPOS-LHC \\
\midrule\midrule
$p_0$                   & 0.5442 $\pm$ 0.0275  & 0.5482 $\pm$ 0.0245 \\
$p_1$  (m$^{3}$\,kg$^{-1}$)     & -2.8168 $\pm$ 0.2986 & -2.7795 $\pm$ 0.2746 \\
$S_{19}$ (GeV)          & 3.5384 $\pm$ 0.0995 &  3.5327 $\pm$ 0.0982 \\
$\gamma$                & 1.9691 $\pm$ 0.0109 & 1.9689 $\pm$ 0.0092 \\
\bottomrule
\end{tabular}
}
\caption{The combined fits based on Eq.\eqref{eq:EgeoFraction} and \eqref{eq:EgeoEem}. The reference air density $\rho_{\rm ref}$ is 0.3\,kg\,m$^{-3}$, which is $\langle\rho\rangle$ of the inclined showers in Ref.\,\cite{Schluter:2022mhq}. }
\label{tab:EgeoEem}
\end{table*}

The left plot in Fig.~\ref{fig:EgeoRadiationEnergyModeling} shows the relative strength of the geomagnetic radiation energy as a function of the air density at the radio emission maximum.
It shows that the transverse current decreases as the air density increases because electrons and positrons scatter more in denser air and lose coherence.
The right plot of Fig.~\ref{fig:EgeoRadiationEnergyModeling} presents the energy scale from the geomagnetic radiation $S_{\rm geo}$ to the primary electromagnetic energy $E_{\rm em}$.
The model based on Eq.\eqref{eq:EgeoFraction} and \eqref{eq:EgeoEem} has an intrinsic uncertainty of about $5.9\,\%$ in total.

\paragraph{Charge excess radiation energy}

\begin{figure}
    \centering
    \includegraphics[width=0.6\linewidth]{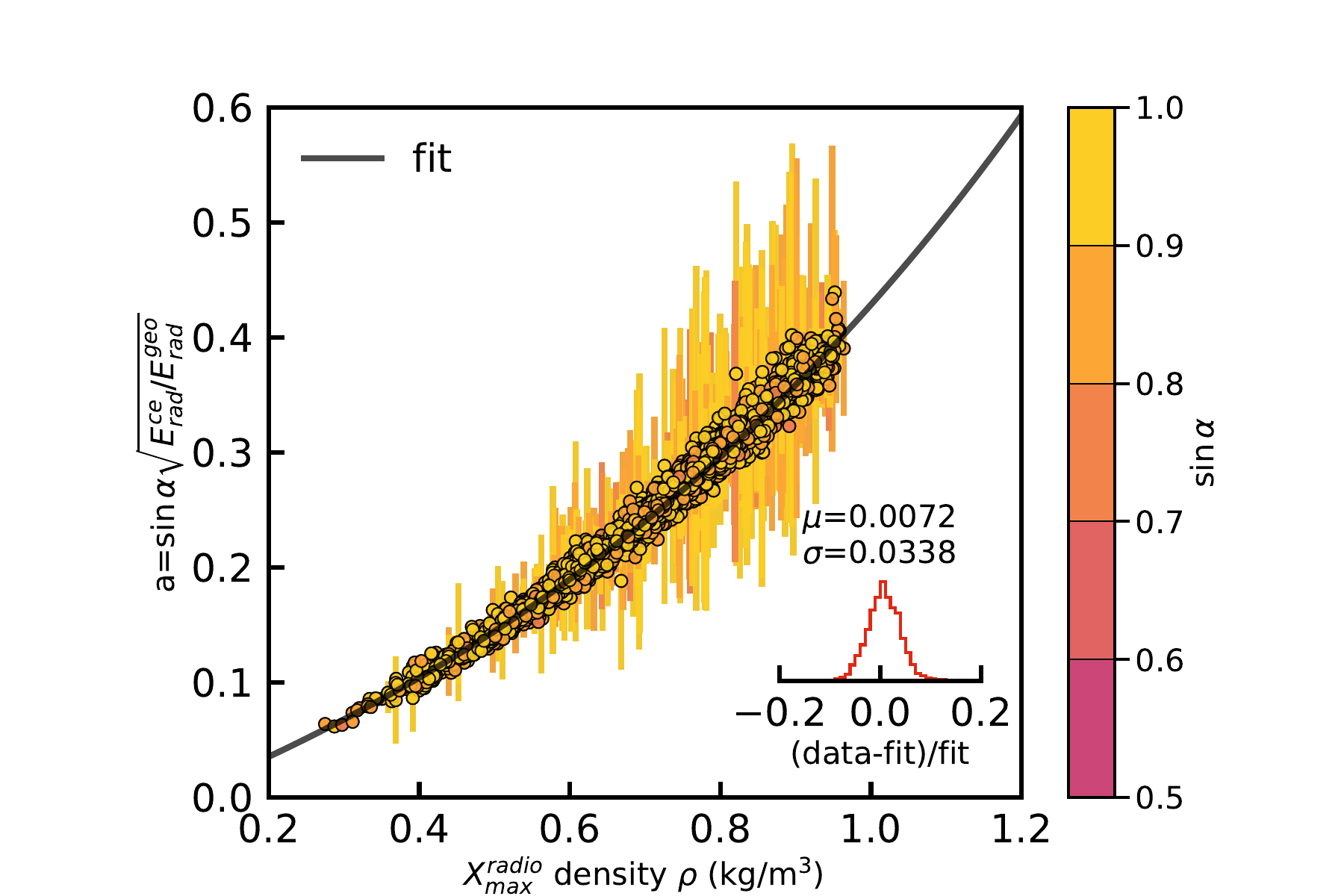}
    \caption{Charge excess fraction as a function of the air density at the radio emission maximum for ``signal-only'' simulations with Sibyll-2.3d model.}
    \label{fig:charge_excess_fraction}
\end{figure}


The charge excess radiation energy can be obtained by integrating Eq.\,\eqref{eq:fce}. This equation describes the charge excess energy fluence, as shown in Fig.~\ref{fig:Footprint}.
The charge excess radiation energy $E_{\rm ce}$ can be expressed as:
\begin{equation}
    E_{\rm ce} = 2\pi\int_0^\infty r\cdot f_{G}\,(r)dr \,.
\end{equation}
The ratio of the charge excess radiation energy to the geomagnetic radiation energy is defined as 
\begin{equation}
    a = \sin\alpha \cdot \sqrt{\frac{E_{\rm ce}}{E_{\rm geo}}} \,.
\end{equation}
It can be modeled as proposed in Ref.\,\cite{Glaser:2016qso}:
\begin{equation}
\label{eq:charge_excess_a}
    a=q_2 + q_0 \exp ( q_1( \rho - \rho_{\rm ref}) )\,.
\end{equation}
We adopt the reference air density, $\rho_{\rm ref}$, at the average air density $\langle\rho\rangle=0.3$\,kg\,m$^{-3}$ of the inclined air showers.
The length factor “STEPFC” is again set to 1 in the simulations.
However, the correction for the charge excess radiation energy remains unknown. 
Therefore, we do not scale the geomagnetic or charge excess radiation energies.
Further investigation is needed to determine these corrections.

As shown in Fig.~\ref{fig:charge_excess_fraction}, the charge excess fraction increases as the atmospheric density increases because more electrons can be accumulated in the shower front in the denser media.
Table~\ref{tab:EceFraction} presents the fitting results for the charge excess radiation energy.

\begin{table*}[!ht]
\centering
{\small
\begin{tabular}{ lcc }
 \toprule
 $\nu_e$-CC  & Sibyll-2.3d & EPOS-LHC \\
\midrule\midrule
$q_0$                   & 0.31475 $\pm$ 0.00388  & 0.30926 $\pm$ 0.00358 \\
$q_1$  (m$^{3}$\,kg$^{-1}$)     & 1.09051 $\pm$ 0.00995 & 1.10638 $\pm$ 0.00952 \\
$q_2$                   & -0.24679 $\pm$ 0.00404 & -0.24164 $\pm$ 0.00371 \\
\bottomrule
\end{tabular}
}
\caption{Fit result based on Eq.\,\eqref{eq:charge_excess_a} for the charge excess radiation energy fraction. The reference air density, $\rho_{\rm ref}$, is $\langle\rho\rangle=0.3$ kg\,m$^{-3}$.}
\label{tab:EceFraction}
\end{table*}

\begin{figure}[!ht]
    \centering
    \includegraphics[width=0.5\linewidth]{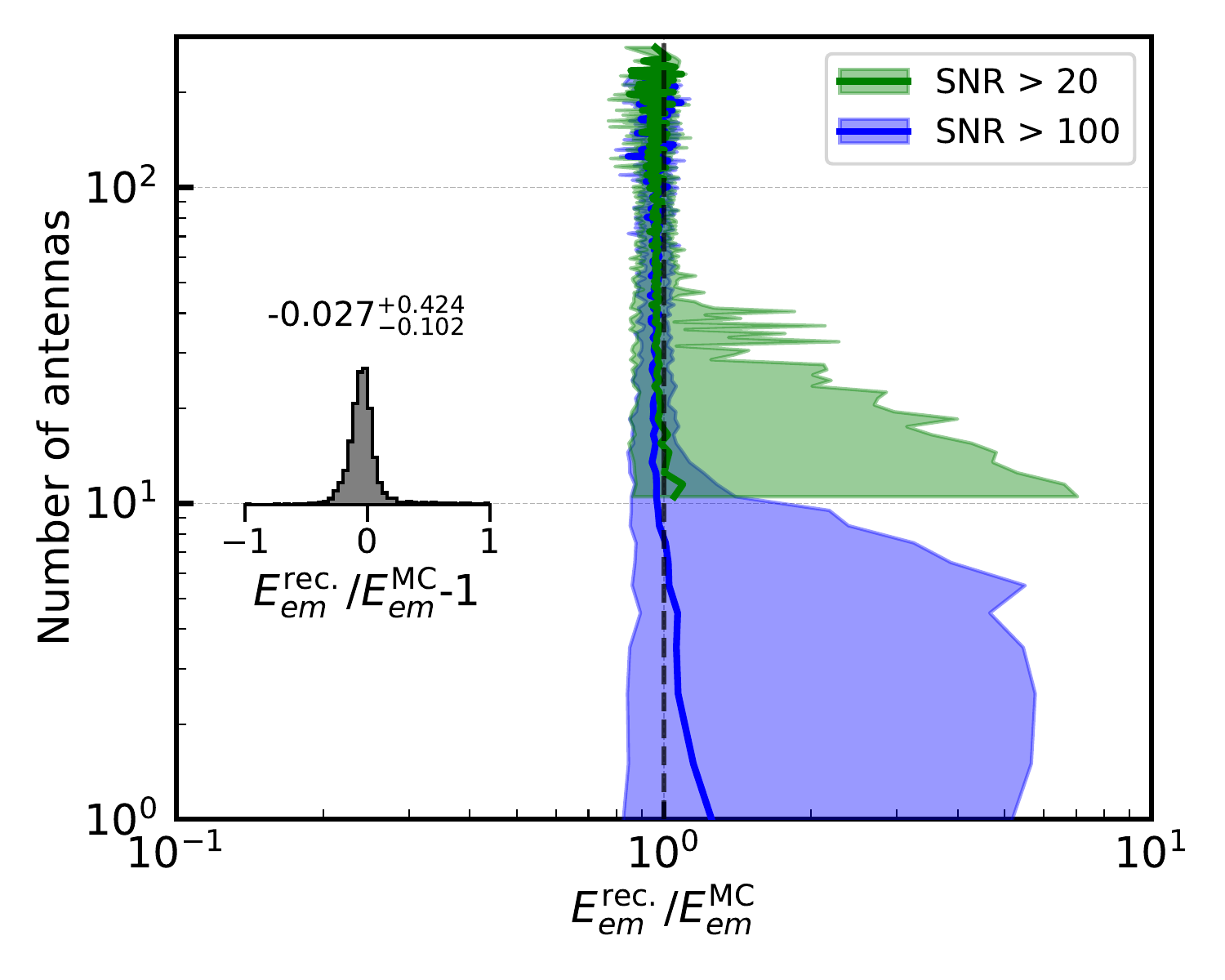}
    \caption{Distribution of the electromagnetic energy bias, $E_{\mathrm{em}}^{\mathrm{rec.}}/E_{\mathrm{em}}^{\mathrm{MC}}$, for the same set of HQ events (Sibyll-2.3d), with median and 1$\sigma$ bands.
    The three colored bands do not correspond to different event samples, but to the same HQ sample evaluated with different SNR definitions and with different subsets of triggered antennas. 
    This comparison is intended to identify which SNR regime and antenna multiplicity are the dominant sources of the bias in the electromagnetic energy reconstruction.
    For the case of SNR > 20, the seriously big uncertainty (1$\sigma$ band) can be removed if the corresponding number of antennas is greater than 40.
    For case of SNR > 100, these serious uncertainties only occurs if the corresponding number of antennas is less than 10.
    The histogram inside of this figure shows the total distribution of the electromagnetic energy bias for the HQ events.
    It shows a small bias, $-2.7$\%, in total.
    }
    \label{fig:EemRecon}
\end{figure}

\begin{figure}[!ht]
    \centering
    \includegraphics[width=1\linewidth]{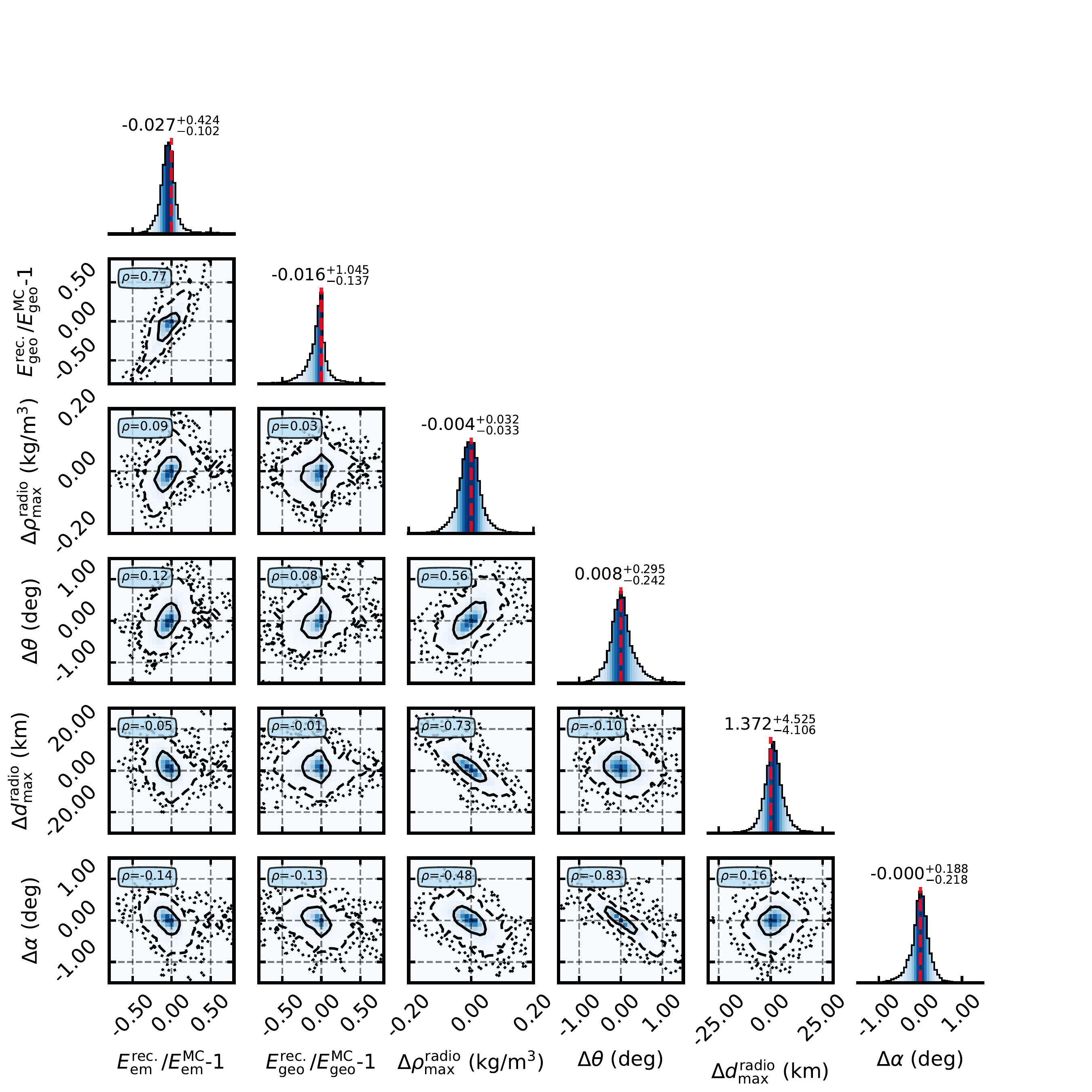}
    \caption{Corner plot showing the distributions and correlations of the reconstructed electromagnetic energy bias, $E_{\rm em}^{\rm rec.}/E_{\rm em}^{\rm MC}-1$, with the relative bias in the geomagnetic radiation energy, $E_{\rm geo}^{\rm rec.}/E_{\rm geo}^{\rm MC}-1$, and with the reconstruction differences ($\Delta$) in $\rho_{\rm max}^{\rm radio}$, $\theta$, and $\alpha$ from their truths for the HQ events (Sibyll-2.3d). The diagonal panels show the marginalized distributions, with the median and the 15.87\% and 84.13\% percentiles indicated, while the lower panels display the corresponding pairwise correlations with 1, 2, and 3 $\sigma$ contours and the corresponding coefficients.
    }
    \label{fig:EemReconCorrelation}
\end{figure}

\subsubsection{Performance of the electromagnetic energy reconstruction}

To evaluate the performance of the electromagnetic energy reconstruction for the HQ events selected as shown in the diagram of Fig.~\ref{fig:LogicDiagram}, we employ the Sibyll-2.3d model, as shown in Table~\ref{tab:EgeoEem}, and present the results in Fig.~\ref{fig:EemRecon}. We set all the footprint model parameters free with good initialization values.
The reconstruction performance is strongly affected by the signal quality. 
For this study, the HQ events are not divided into different subsamples; instead, the same events are displayed with different tags according to the number of antennas fulfilling a given SNR condition (SNR>20 or SNR>100). 
The corresponding bias distributions therefore show how the electromagnetic energy $E_{\rm em}$ reconstruction depends on the SNR regime and on the number of antennas with sufficiently strong signals in the same event sample. 
The bias is reduced when more antennas satisfy higher SNR requirements, demonstrating the importance of strong-signal antenna measurements for an accurate reconstruction.

Figure \ref{fig:EemRecon} shows a bias of $-2.7$\% in the reconstructed electromagnetic energy. 
To investigate this bias, we compare multiple reconstructed observables with their true values.
The observables include the geomagnetic radiation energy, $E_{\rm geo}$, the air density at the radio emission maximum, $\rho_{\rm max}^{\rm radio}$, the zenith angle $\theta$ of the shower axis, the angle between the shower axis and the geomagnetic field, $\alpha$, and the distance between the radio shower core and the radio emission maximum $d_{\rm max}^{\rm radio}$.
The true values of $E_{\rm em}$, $\theta$, and $\alpha$ come from the CORSIKA 7 simulations, while the true values of $E_{\rm geo}$, $\rho_{\rm max}^{\rm radio}$, and $d_{\rm max}^{\rm radio}$ come from the reconstructions of the ``signal-only'' simulations that adopt a 30-80 MHz rectangular bandpass filter.

The multi-parameter comparison is shown in Fig.~\ref{fig:EemReconCorrelation}. 
All of these parameters can cause considerable bias in the geomagnetic radiation energy $E_{\rm geo}$.
In particular, the reconstruction biases in $E_{\rm geo}$, $\rho_{\rm max}^{\rm radio}$, and $\alpha$ can propagate to $E_{\rm em}$ via Eqs.\,\eqref{eq:EgeoFraction} and \eqref{eq:EgeoEem}.
The reconstruction bias of the geometry reconstruction, $\theta$ and $d_{\rm max}^{\rm radio}$, can directly influence the reconstruction of $\rho_{\rm max}^{\rm radio}$, and indirectly impact the reconstruction of $E_{\rm em}$. 
Further investigations about the reconstruction bias can be found in Appendix \ref{sec:EemRecon}.


\subsubsection{From the electromagnetic energy to the primary neutrino energy}
Once the electromagnetic energy is determined, the primary neutrino energy can be inferred through neutrino interaction simulations based on assumptions about the primary neutrino flux.
However, this topic is beyond the scope of this work, so it will not be discussed further here.
If the flavor of the primary neutrino and the type of neutrino interaction can be identified, the primary neutrino energy can be estimated more accurately using neutrino simulations with known neutrino flavors and interactions.
In Appendix~\ref{sec:EemOverEnu}, we provide the electromagnetic energy fraction, which can be used to estimate the neutrino energy if a $\nu_e$-CC induced shower is identified.

\begin{figure}[ht]
    \centering    
\includegraphics[width=0.48\linewidth]{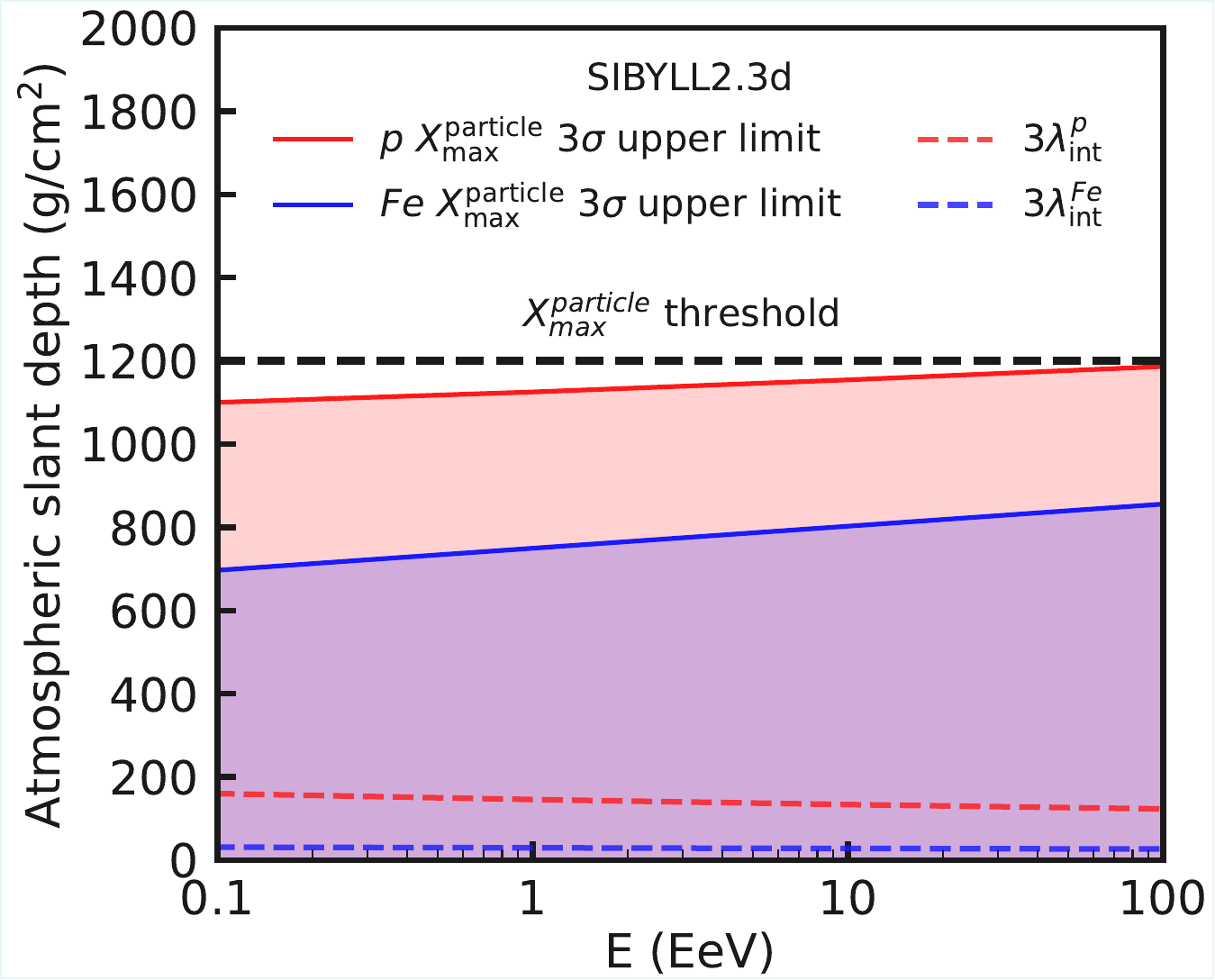}
\includegraphics[width=0.48\linewidth]{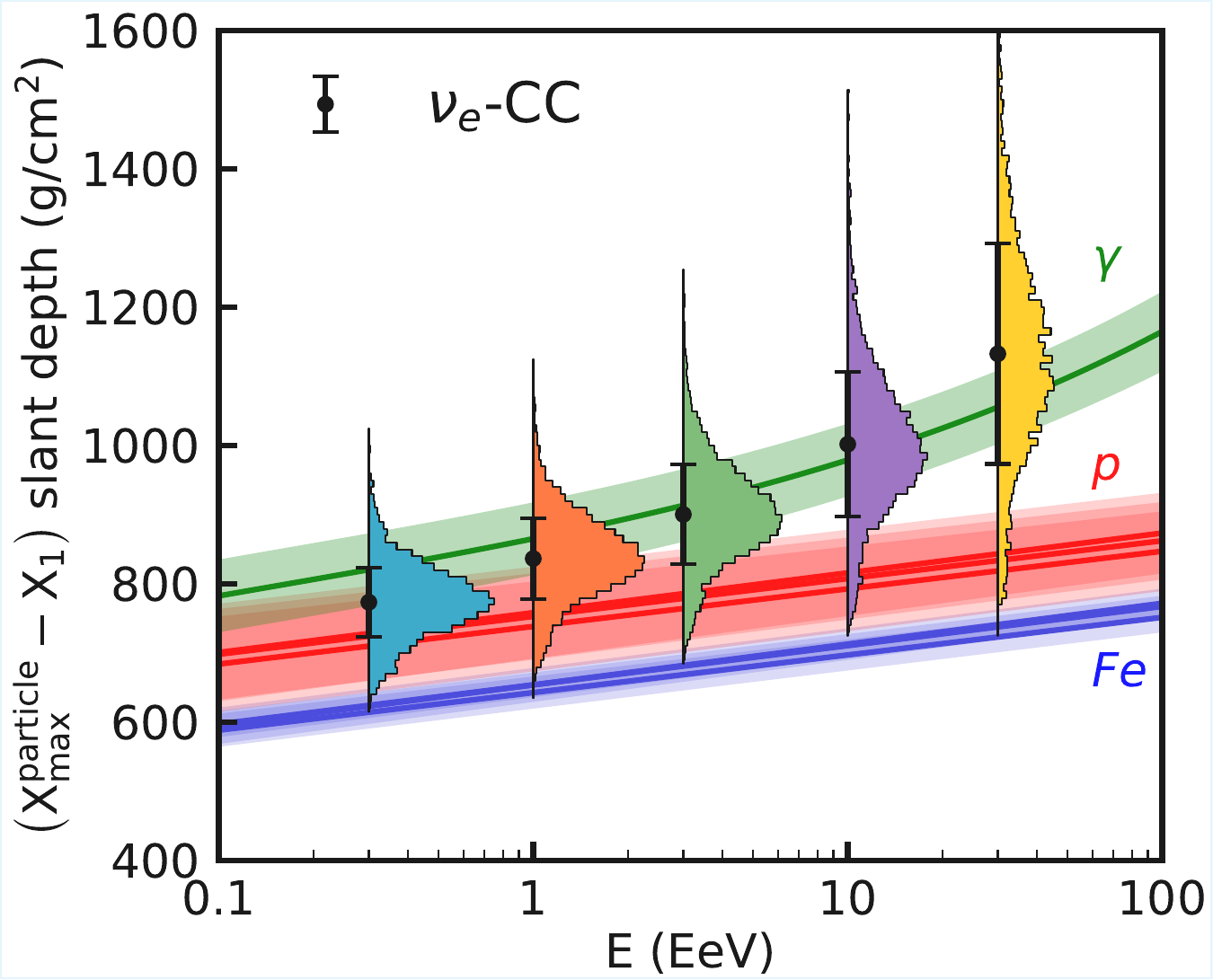}
\caption{
Left: upper limit of the atmospheric slant depth at which the first interaction occurs for proton and iron primaries (red and blue dashed lines), using Sibyll-2.3d as hadronic interaction model.  The colored bands mark the $3\sigma$ upper range in which the particle shower maximum occurs.
The black dashed line represents the threshold applied to discriminate hadron-induced air showers.
Right: distributions of the relative particle maximum ($X^{\text{particle}}_{\text{max}}-X_1$) for proton, iron (EPOS-LHC, QGSJetII-04 \cite{Ostapchenko:2010vb}, and Sibyll-2.3d), and photon ($\lambda=1$ in \cite{DeDomenico:2013wwa}) primaries, compared to $\nu_e$ (CC) primaries. The distribution of $\nu_e$-CC induced showers are obtained from simulations of showers at very high inclinations ($\theta > 75^\circ$).
}
\label{fig:ParticleXmax}
\end{figure}

\subsection{Neutrino identification}
\label{sec:NeutrinoID}

In this work, we propose using the shower depth at which the maximum radio emission is observed, $X_{\mathrm{max}}^{\mathrm{radio}}$, to differentiate between neutrino and UHECR induced EAS. 

Because of the large interaction cross-sections of UHECR (neutrons, charged CRs, and photons), the depth of the first interaction, $X_1$, is expected to be less than 200\,g\,cm$^{-2}$ at the $3\sigma$ confidence level (three interaction lengths, $\lambda_{\rm int}$) in the UHE domain, as shown on the left side of Fig.\,\ref{fig:ParticleXmax}.
The plot on the right of Fig.\,\ref{fig:ParticleXmax} compares the maxima and $1\sigma$ fluctuations of the particle maxima, $\Delta X^{\text{particle}}_{\text{max}} = (X^{\text{particle}}_{\text{max}}-X_1)$ for $\nu_e$-induced showers (considering CC interactions only), compared to what is expected for photon, proton, and iron primaries, using different hadronic interaction models. 
The horizontal histograms show the full distributions for different $\nu_e$-energies, from which the mean values and standard deviations of particle maxima (represented by the black symbols) are obtained.
The colored bands and lines illustrate the same for proton, iron, and photon primaries.

First, we will discuss the methods for identifying neutrinos. Then, we will discuss the possibility of determining their flavor and interaction type.

\subsubsection{Constant $X^{\text{radio}}_{\text{max}}$ threshold}
\label{sec:constant_threshold}
To determine a conservative threshold for $X^{\text{radio}}_{\text{max}}$ that excludes cosmic rays, we calculate the limit step by step.
Since protons have a longer interaction length and a deeper shower maximum in air than heavier nuclei, we start by defining a proton threshold that imposes an even stronger constraint on heavier nuclei.
The average distance that a proton travels before interacting with a nucleus in the atmosphere, $X_1^p$, is given by its mean free path $\lambda_{\rm int}^p = \frac{A_{\rm air}}{N_A \sigma_{p\text{-air}}}$, with 
$A_{\rm air}=14.5$\, g\,mol$^{-1}$ being the average mass number of air and $\sigma_{p\text{-air}}$ the interaction cross section.
The particle shower maximum relative to the first interaction is denoted by $\Delta X^{\text{particle}}_{\text{max}}$.
The absolute position of the particle shower maximum is then given by $X^{\text{particle}}_{\text{max}} = \Delta X^{\text{particle}}_{\text{max}} + X_1^p$.
$\Delta X^{\text{particle}}_{\text{max}}$ and can be approximated by a Gaussian distribution.
Thus, the distribution of $X^{\text{particle}}_{\text{max}}$ can be obtained by convolving a Gaussian distribution with and an exponential function.
We define a constant threshold, $X^{\text{radio}}_{\text{max}}$, at the $3\sigma$ upper limit maximum value for protons below 100 EeV.
The left plot in Fig.~\ref{fig:ParticleXmax} shows these $3\sigma$ upper limits for protons and iron based on Sibyll-2.3d simulations.
The upper limit for iron is much smaller than that for protons, so that the upper for protons can be considered as conservative assumption.
Finally, we set a constant threshold, $X_{\text{threshold}}$, at $1200$\,g\,cm$^{-2}$, which excludes all cosmic rays  below 100 EeV at the $3\sigma$ confidence level.


The particle shower maximum position $X^{\text{particle}}_{\text{max}}$ for $\nu_e$-induced showers will be required to be  greater than $X_{\text{threshold}} = 1200$\,g\,cm$^{-2}$. 
Furthermore, as discussed in \cite{Glaser:2016qso, Schluter:2021egm}, the maximum radio emission position $X^{\text{radio}}_{\text{max}}$ is expected to occur earlier than the particle maximum,  $X^{\text{particle}}_{\text{max}}$. 
Therefore, $X^{\text{radio}}_{\text{max}}>X_{\text{threshold}}$ should provide a stringent rejection of charged CRs.
Since the upper limit of the UHE $\gamma$ flux is strongly constrained \cite{PierreAuger:2022aty}, we will not treat it as a neutrino background in this work.

For neutrinos, the first interaction $X_1$ can occur anywhere.
Therefore, the absolute position $X^{\text{radio}}_{\text{max}}$ can be much larger than $X_{\text{threshold}}$, which can be used to identify neutrinos in the background caused by charged CRs.
The key criterion for identifying neutrinos is $X^{\text{radio}}_{\text{max}} - X_{\text{threshold}} > 3\sigma_{X^{\text{radio}}_{\text{max}}}$.

Therefore, it is essential to reconstruct the emission point $X^{\text{radio}}_{\text{max}}$ and the shower geometry, $\theta$, in order to identify neutrinos. 
Linking the $X^{\text{radio}}_{\text{max}}$ geometry point and the core of the geomagnetic radiation footprint allows us to determine the shower axis.

\begin{figure}[bth]
    \centering
    \includegraphics[width=0.49\linewidth]{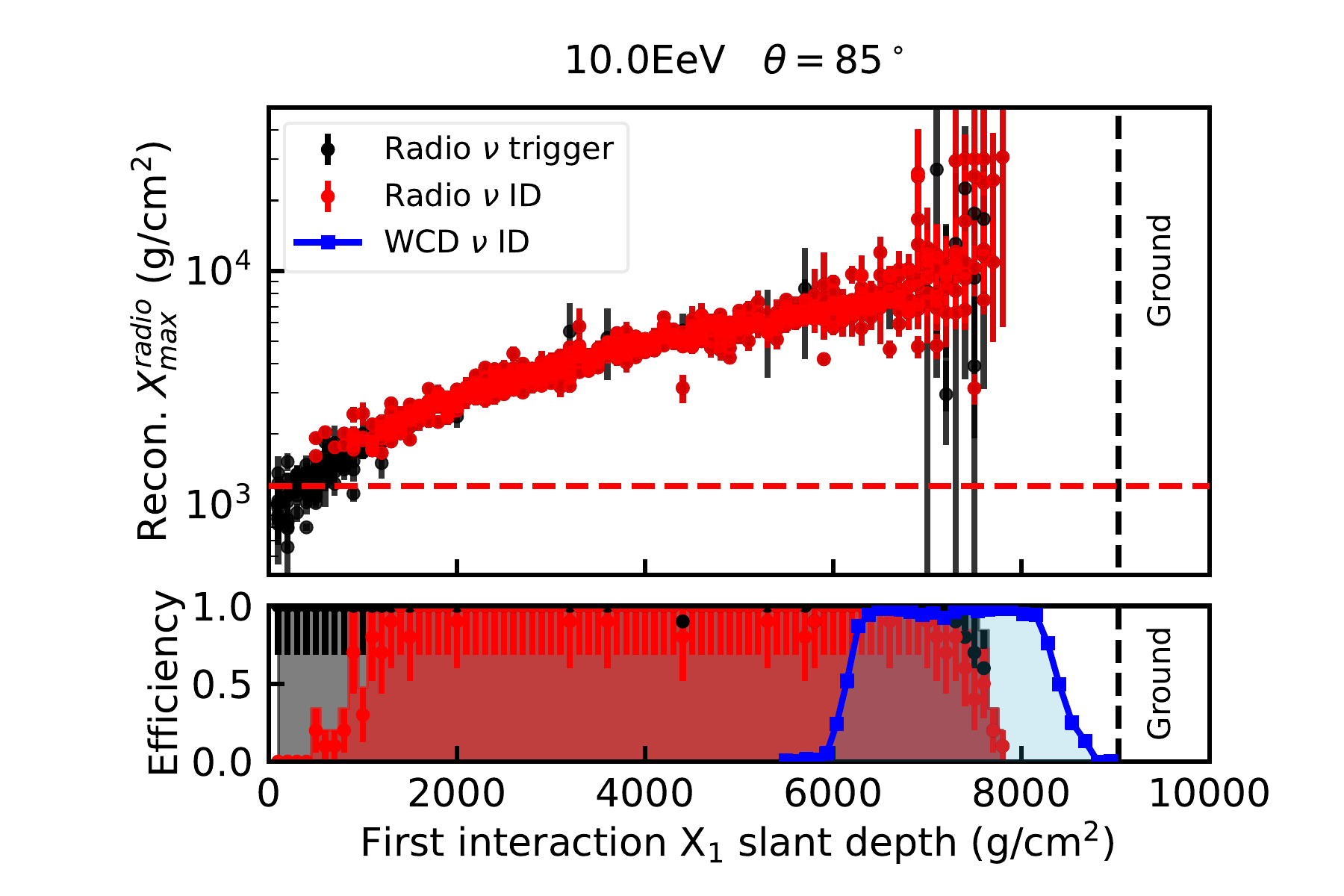}
    \includegraphics[width=0.485\linewidth]{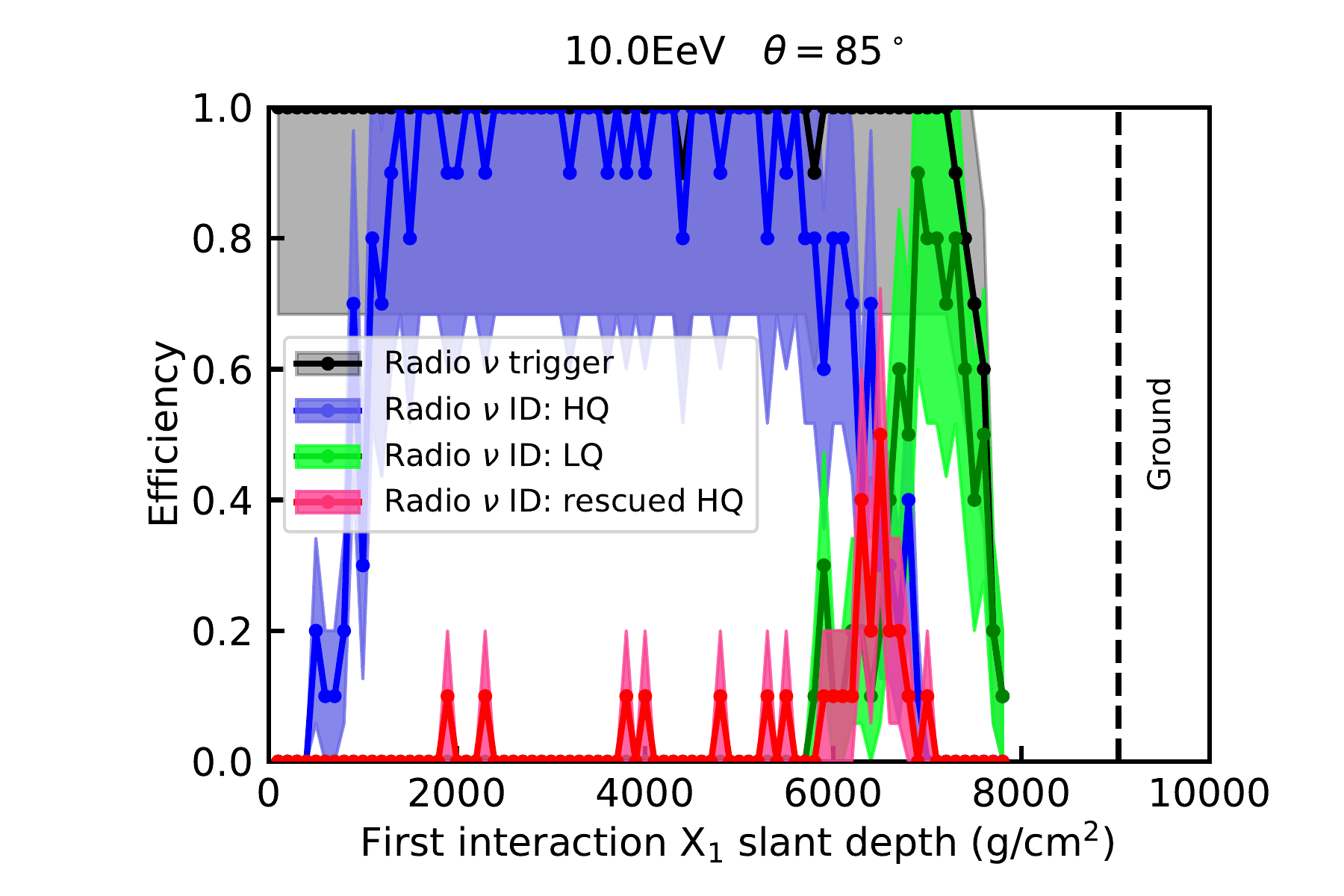}
    \caption{Left: $X_{\rm max}^{\rm radio}$ reconstruction and neutrino identification efficiency at $E_{\nu} = 10$\,EeV and $\theta=85^\circ$ as a function of the slant depth $X_1$ for $\nu_e$-CC interactions for the case of $T_\mathrm{trigger}^{\rm SNR}=10$. 
    Right: radio trigger efficiency and neutrino identification efficiencies for different components: high quality (HQ) events, low quality (LQ) events, and rescued HQ events.}
    \label{fig:XmaxRecon_NeutrinoID}
\end{figure}

The upper graph on the left in  Fig.\,\ref{fig:XmaxRecon_NeutrinoID} shows the reconstructed shower maximum of the radio emission for $E_{\nu_e} = 10$\,EeV and $\theta=85^\circ$.
The lower graph compares the trigger and neutrino identification efficiencies of the radio antenna array compared to the additional neutrino identification efficiency of the WCDs from Fig.\ 7.3 in \cite{Tiffenberg:2011thesis}.
As shown in the figure, the radio antenna array extends the neutrino detection capabilities to shallower showers, and the constant $X^{\text{radio}}_{\text{max}}$ threshold effectively identifies neutrinos.
The right plot in Fig.~\ref{fig:XmaxRecon_NeutrinoID} shows the neutrino identification efficiencies from the various components.
We divided the simulation events again into HQ and LQ samples (see Sec.\,\ref{sec:ShowerAxisRecon}).
The flow chart in Fig.\,\ref{fig:LogicDiagram} shows that failed HQ events can be rescued using a reconstruction method designed for LQ events.
The HQ component dominates the neutrino detections at $E_{\nu} = 10$\,EeV and $\theta=85^\circ$ and for shower maxima far from the ground, as shown in Fig.~\ref{fig:XmaxRecon_NeutrinoID}.
LQ neutrino detections are dominated by relatively young showers close to the ground.
The rescued HQ component that failed the radio footprint fit for HQ events (see Fig.\,\ref{fig:LogicDiagram})
can be treated as LQ events and reconstructed using an alternative footprint core option.
These events are concentrated within the transition range from ``old'' to ``young'' showers, complementing the HQ and LQ categories.




Reconstructing $X^{\text{radio}}_{\text{max}}$ is shown to be a powerful technique for identifying neutrinos.
However, there is still room for improvements in the reconstruction method.
For example, interferometric mapping \cite{LOPES:2005ipv,Schoorlemmer:2020low} can improve the reconstruction of the shower axis, resulting in more precise $X^{\text{radio}}_{\text{max}}$ estimations.
Additionally, more precise GPS time synchronization of radio antenna arrays can improve the $X^{\text{radio}}_{\text{max}}$ reconstruction, as demonstrated in \cite{Schoorlemmer:2020low}.
In general, the radio technique can provide more information about showers induced by neutrinos than particle detectors can.
Investigating more such features of neutrino-induced showers will benefit neutrino searches in real data.







\subsubsection{Neutrino flavor and interaction type identifications}
After identifying neutrinos, we aim to investigate their flavors or types of interaction.
Identifying the flavor of neutrinos is more difficult than identifying the atmospheric showers they induce.
In principle, we can only identify $\nu_e$-CC and $\nu_\tau$-CC induced air showers for downward-going neutrinos.
$\nu_\mu$-CC interactions have features similar to $\nu$-NC interactions because the outgoing primary muon will not decay in the EeV regime and will escape the atmosphere.
The remaining shower is induced by its hadronic jets.
Therefore, the difference between a $\nu_\mu$-CC and $\nu$-NC induced shower cannot be determined if the primary muon is invisible to the detector array on the ground.
A $\nu_\tau$-CC interaction could produce two showers in the air.
The first is caused by its hadronic jets and
the second is a $\tau$-decay induced shower, provided that the $\tau$ does not escape the atmosphere.
If the $\tau$ penetrates the ground, the second shower is invisible to the detector array. This results in a shower that resembles a $\nu$-NC shower.
Conversely, if double showers are identified, they are most likely caused by a $\nu_\tau$-CC interaction.
The $\nu_e$-CC induced air showers have a greater electromagnetic component than the $\nu$-NC (or $\nu_\mu$-CC and $\nu_\tau$-CC showers without $\tau$ decay) induced showers because the primary electromagnetic energy is transferred from the primary electron or positron at the first interaction of the shower.
However, techniques for distinguishing $\nu_e$-CC interactions from hadronic jets ($\nu$-NC like shower) still need to be investigated using radio antennas. Once we can determine the neutrino flavors, we can better estimate the primary energy.

\subsection{The $\nu_e$-CC detection effective area for transient events}

The overall neutrino detection efficiency, $\epsilon_{\nu}(X_1)$, is the product of three efficiencies: the trigger efficiency ($\epsilon_t(X_1)$), the reconstruction efficiency (referring to the data quality) ($\epsilon_q(X_1)$), and the neutrino identification efficiency ($\epsilon_{id}(X_1)$), i.e.
\begin{equation}
    \epsilon_\nu(X_1) = \epsilon_t(X_1) \cdot \epsilon_q(X_1) \cdot \epsilon_{id}(X_1)\,.
\end{equation}



\begin{figure}[ht]
    \centering
    \includegraphics[width=0.45\linewidth]{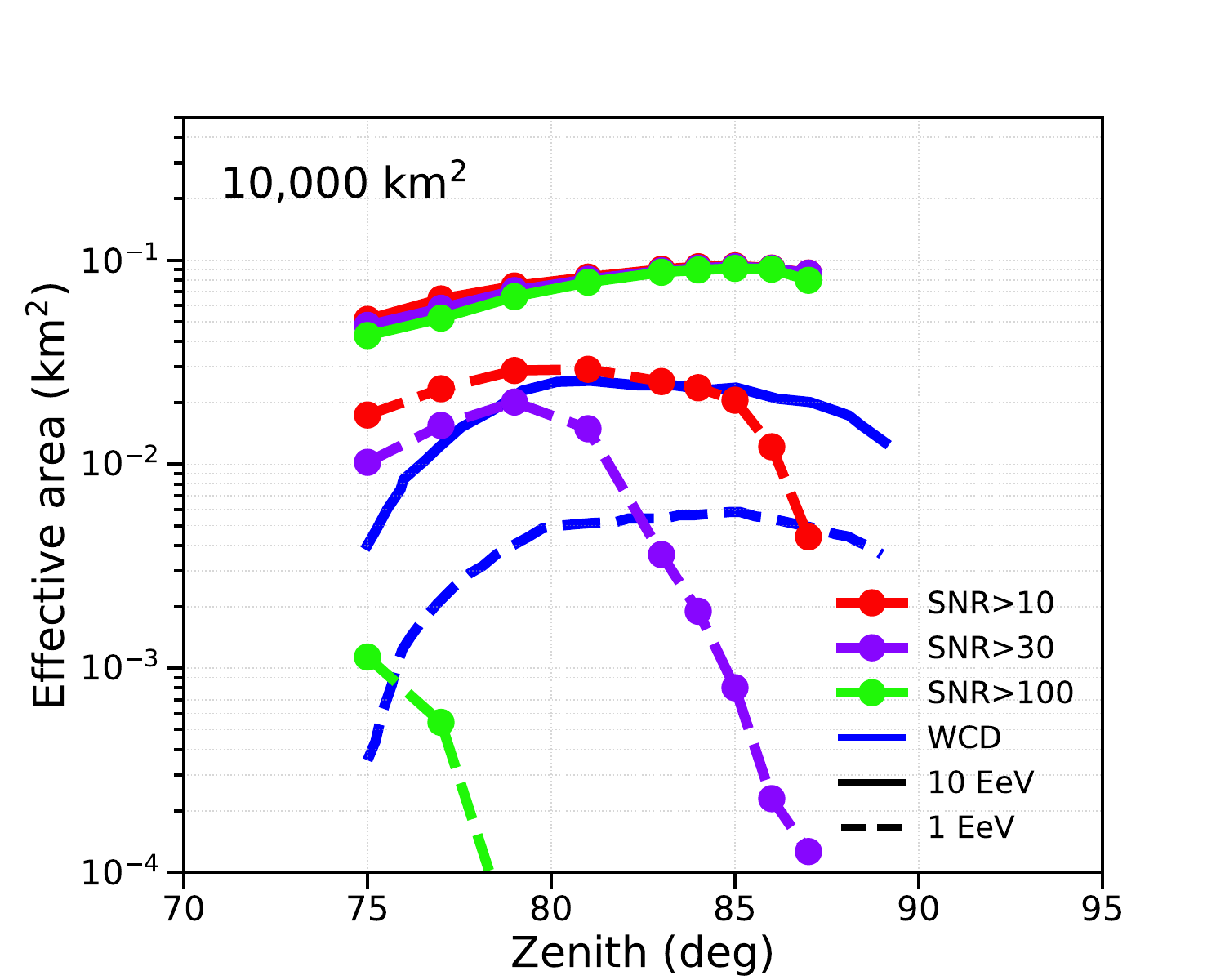}
    \includegraphics[width=0.45\linewidth]{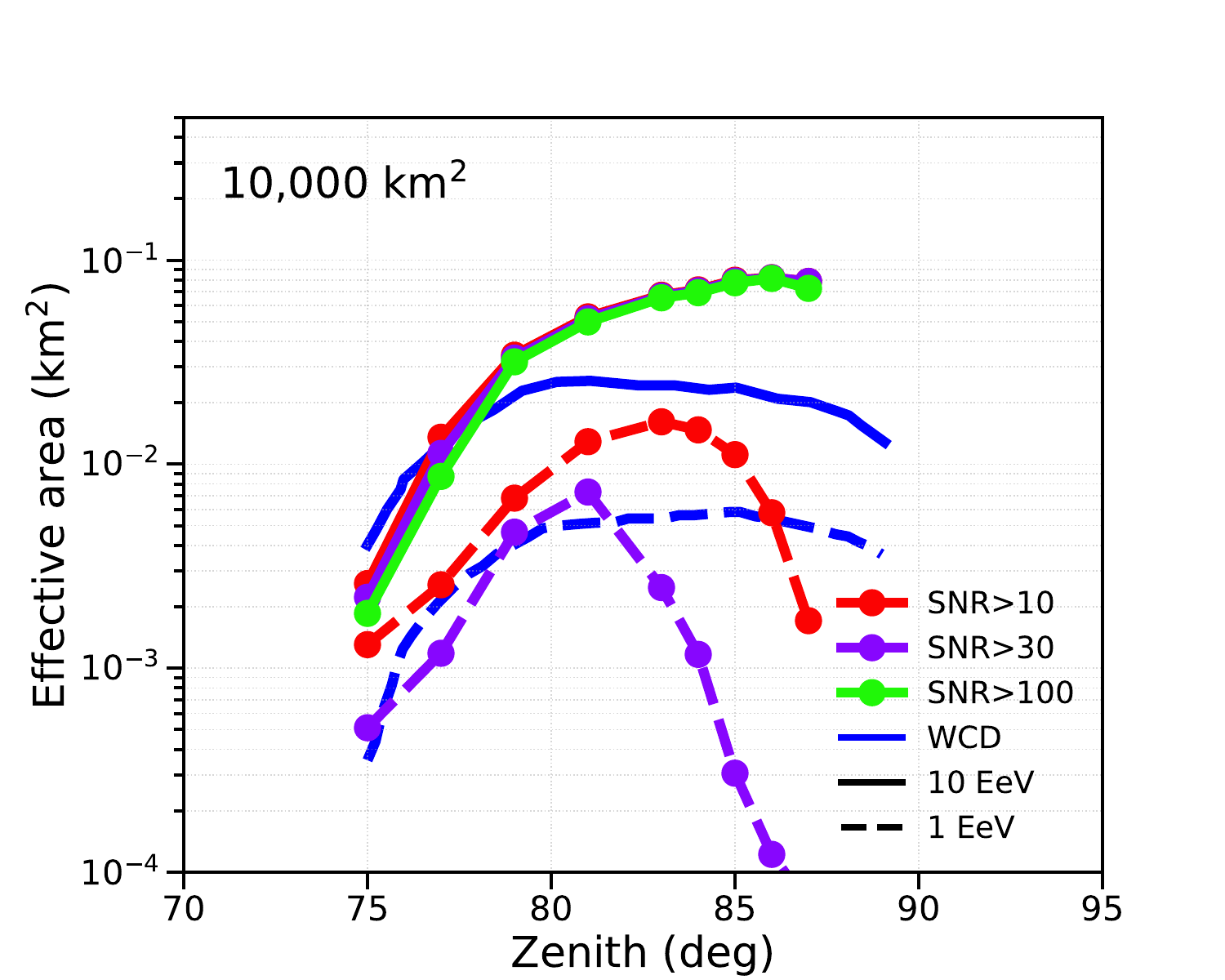}
    \caption{Left: $\nu_e$-CC effective area with $\epsilon_{id}(X_1)$=100\% as a function of primary zenith angle. Right: $\nu_e$-CC effective area with $\epsilon_{id}(X_1)$ obtained from Sec.\,\ref{sec:constant_threshold}.
    The WCD effective areas are scaled from Ref.~\cite{PierreAuger:2019azx}.
     }
    \label{fig:EffectiveAreaByTheta}
\end{figure}

\begin{figure}[ht]
    \centering
    \includegraphics[width=0.45\linewidth]{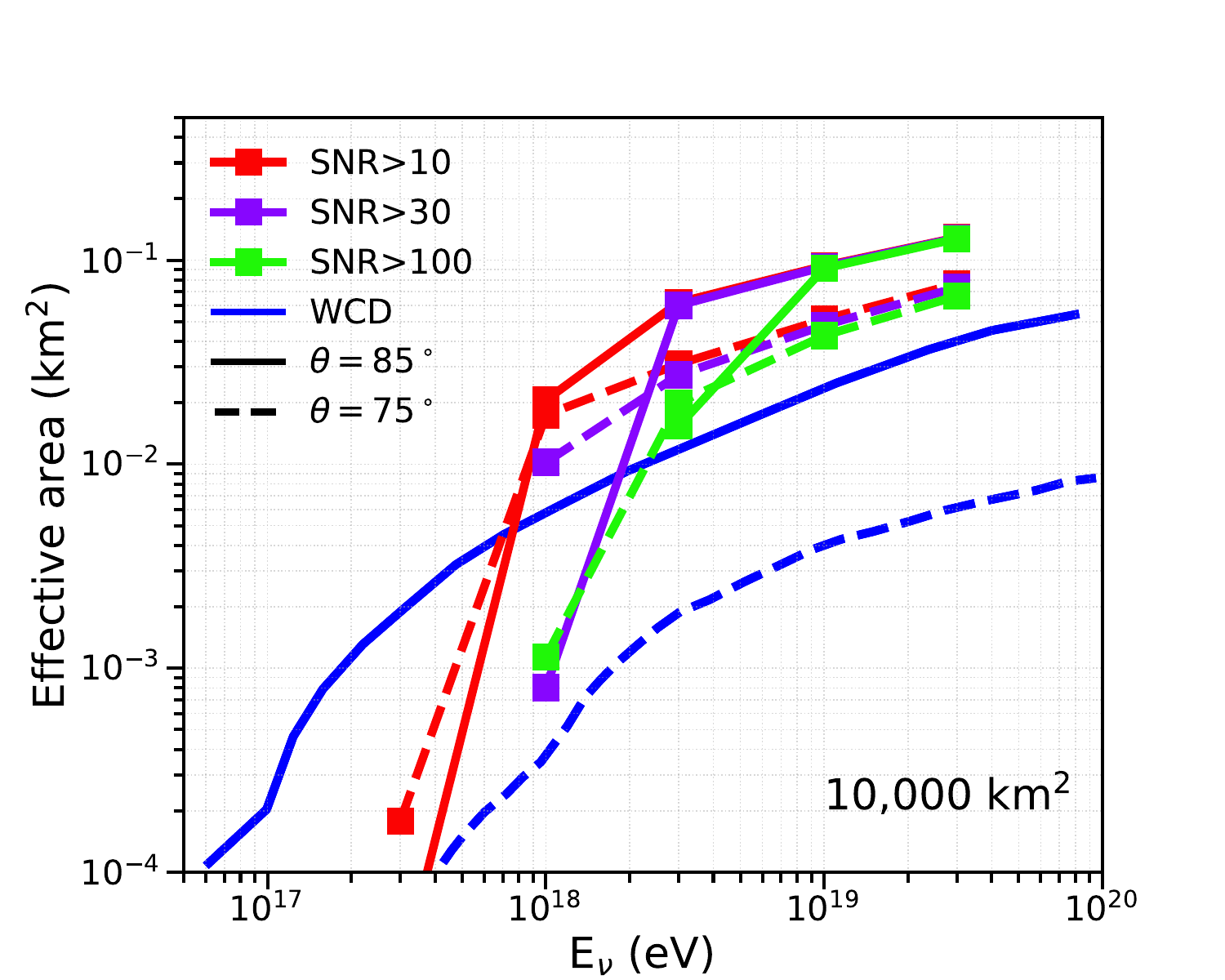}
    \includegraphics[width=0.45\linewidth]{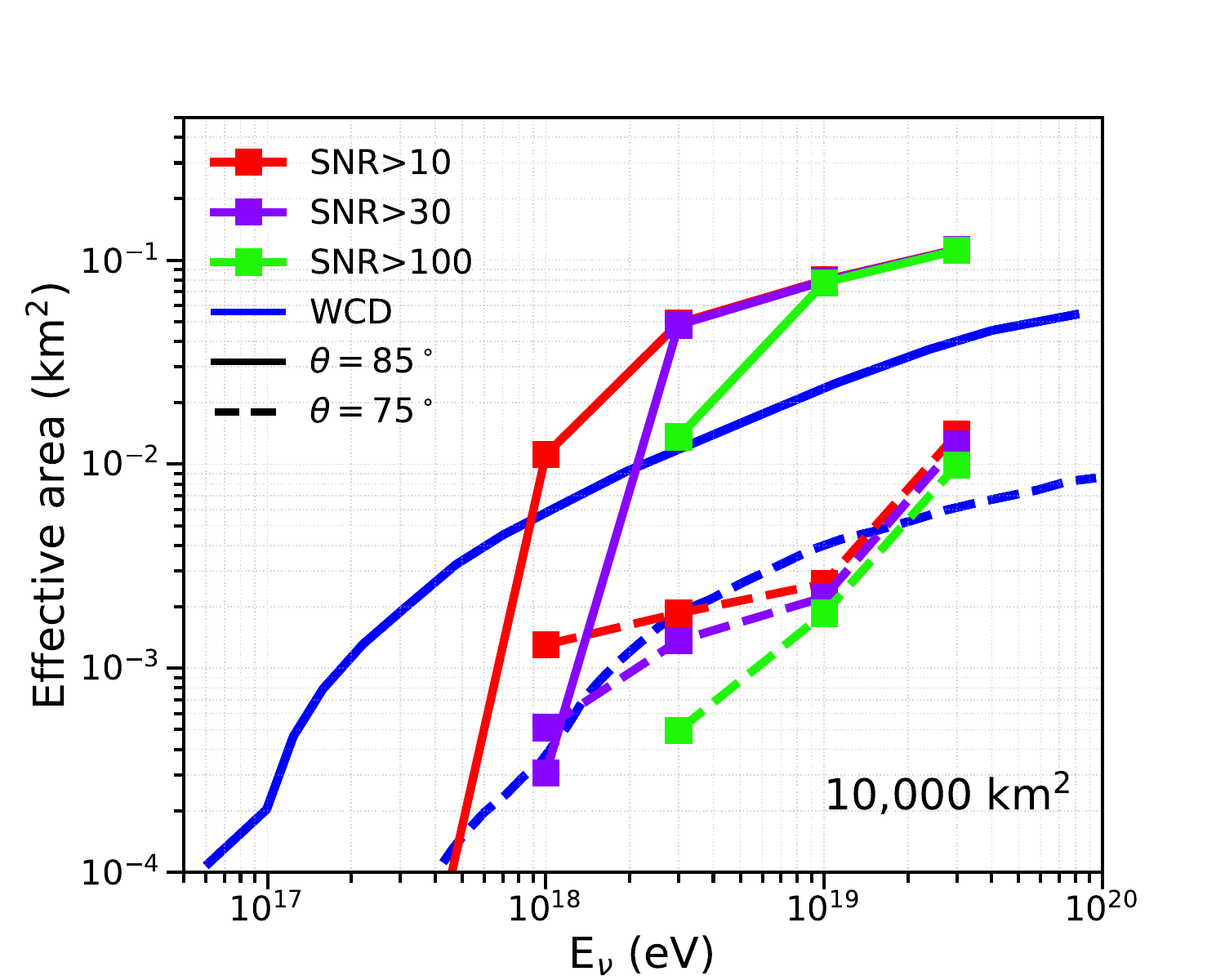}
    \caption{Left: $\nu_e$-CC effective area with $\epsilon_{id}(X_1)$=100\% as a function of primary neutrino energy. Right: $\nu_e$-CC effective area with $\epsilon_{id}(X_1)$ obtained from Sec.\,\ref{sec:constant_threshold}.
    The WCD effective areas are scaled from Ref.~\cite{PierreAuger:2019azx}.}
    \label{fig:EffectiveAreaByEnergy}
\end{figure}

\noindent The effective area of downward-going $\nu_e$-CC detection is computed as follows:
\begin{equation}
    A_{\text{eff}} = \iint \cos\theta \, \epsilon_\nu(X_1) \, \sigma_{\nu_e}^\text{CC} \, m_p^{-1} \, dX_1 \, dA \,,
\end{equation}
where $\sigma_{\nu_e}^\text{CC}$ is the cross-section of $\nu_e$-CC interactions \cite{Cooper-Sarkar:2011jtt}, $m_p$ is the proton mass, and $A$ is the area covered by the radio antenna array ($A=10,000$\,km$^2$ for the reference Observatory).

The neutrino detection efficiency, $\epsilon_\nu(X_1)$, is based on the method described in Sec.\,\ref{sec:constant_threshold}.
Assuming the neutrino identification efficiency $\epsilon_{id}(X_1)$ is 100\%, the effective area is maximum. 
However, if we use the method proposed in Sec.\,\ref{sec:constant_threshold} to estimate $\epsilon_{id}(X_1)$, the effective area will be a conservative estimate since this method can be improved.
The left and right plots in Fig.\,\ref{fig:EffectiveAreaByTheta} depict the ideal and conservative cases, respectively.
Additionally, there are three self-trigger cases: SNR>10, 30, and 100 with $N_{\rm trigger}\geq 3$.
Figure~\ref{fig:EffectiveAreaByTheta} shows two cases: 1 EeV and 10 EeV.
The WCD effective areas are scaled from Ref.~\cite{PierreAuger:2019azx} and are provided as reference.

In the ideal case, a significant enhancement in neutrino detection with a radio antenna array exists even at 1 EeV and SNR>100.
For 10 EeV, the effective radio detection area is always greater than that of the WCDs within the shown zenith angle range.
Additionally, the 1 EeV and 10 EeV cases demonstrate that, as the zenith angle increases, the antenna response weakens rapidly and the area decreases quickly.

In the conservative case (plot on the right in Fig.~\ref{fig:EffectiveAreaByTheta}), the enhancement by radio detection only occurs when the zenith angle is greater than $80^\circ$.
Since the total slant depth is minimal when the zenith angle is small, and since the discrimination method in Sec.\,\ref{sec:constant_threshold} removes all events with a reconstructed $X_{\rm max}^{\rm radio}$ less than 1200\,g\,cm$^{-2}$, the detection of neutrinos is greatly suppressed.
Therefore, methods are necessary to identify downward-going neutrino-induced showers with $X_{\rm max}^{\rm radio}<1200$\,g\,cm$^{-2}$ to boost the neutrino detection enhancement by radio antenna arrays. 

Figure\,\ref{fig:EffectiveAreaByEnergy} presents the effective area as a function of the primary neutrino energy, again for the ideal (left) and conservative (right) cases.
In both cases, the zenith angles are fixed at $75^\circ$ and $85^\circ$.
In the ideal case, the radio detection enhances the effective area if the neutrino energy is greater than 3 EeV (10 EeV) at an angle of $75^\circ$ ($85^\circ$) and even with a very high trigger threshold (SNR>100).
In the conservative case, enhancement is expected when the neutrino energy exceeds 30 EeV (10 EeV) for all the listed trigger thresholds, regardless of the zenith angle.





\subsection{The $\nu_e$-CC detection aperture for diffuse neutrino searches}

\begin{figure}
    \centering
    \includegraphics[width=0.48\linewidth]{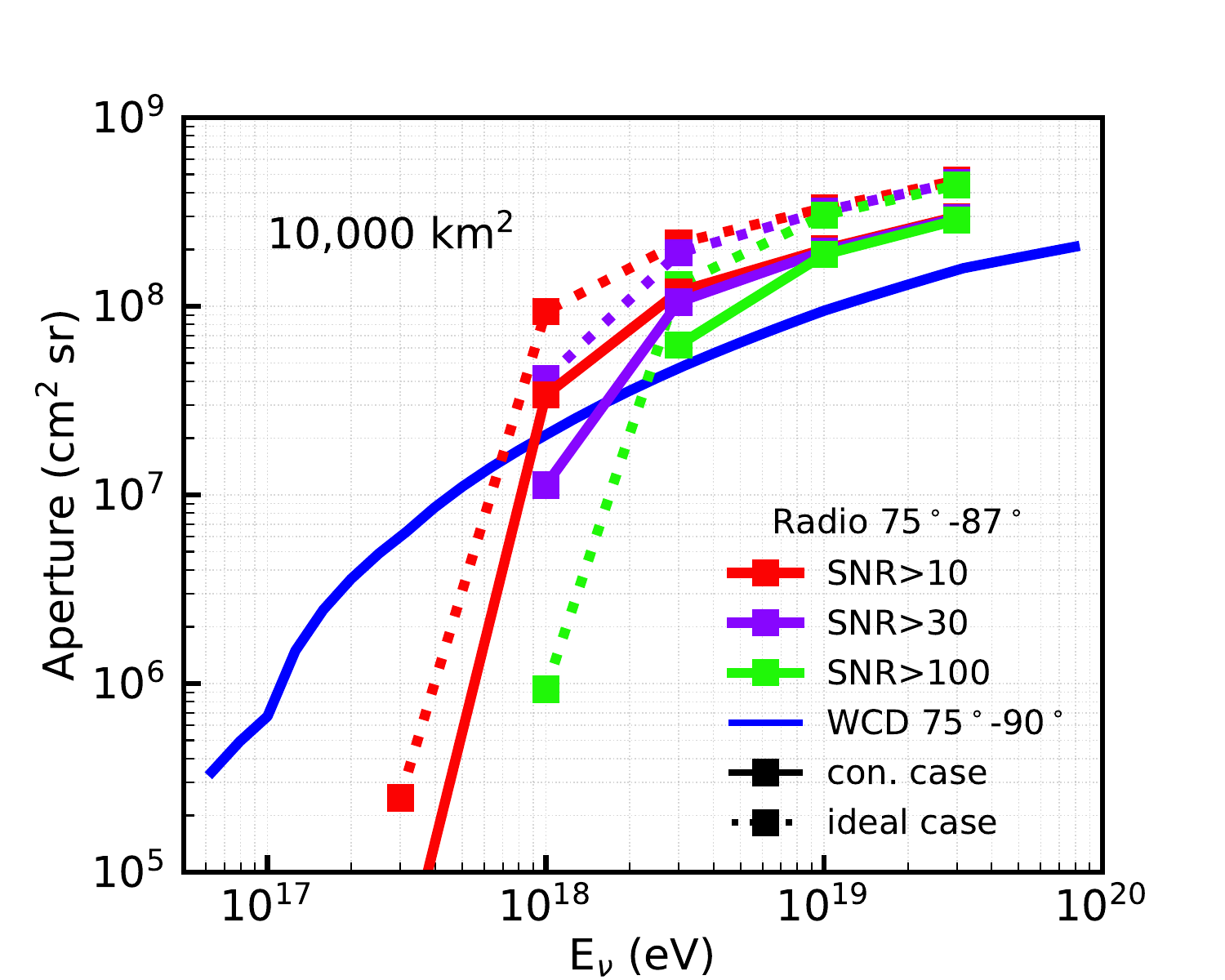}
    \includegraphics[width=0.48\linewidth]{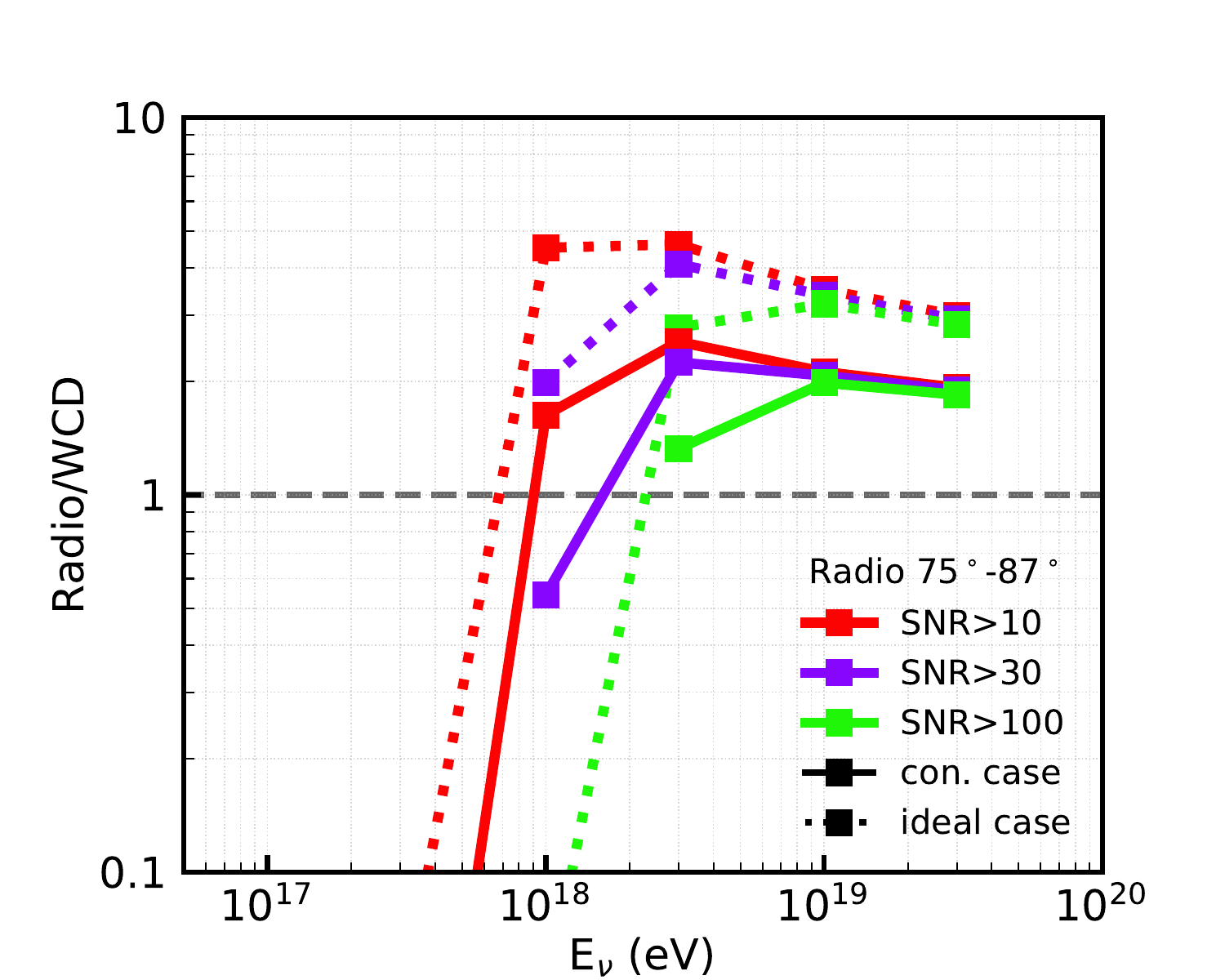}
    \caption{Left: the radio detection aperture for $\nu_e$-CC interactions compared with WCDs in the same zenith angle range $75^\circ < \theta \leq 85^\circ$, for the ideal and conservative cases, as described in the text. Right: aperture ratio.}
    \label{fig:exposure}
\end{figure}

For diffuse neutrino searches, we calculate the expected aperture for radio detection.
The expected aperture can be written as:
\begin{equation}
     \mathcal{E}_{radio}(E_\nu) = \iint \sin\theta \, A_{eff}(E_\nu,\theta) \, d\theta \,d\phi \, .
\end{equation}
For the neutrino simulation with radio antennas, the zenith angle ranges from $\theta=75^\circ$ to $87^\circ$.
The left plot in Fig.~\ref{fig:exposure} presents the expected aperture for radio detection compared to WCDs scaled from Ref.~\cite{PierreAuger:2019azx}. 
The ideal case assumes an identification efficiency $\epsilon_{id}(X_1)$=100\,\%.
The conservative case considers the neutrino identification method described in Sec.~\ref{sec:constant_threshold}.
The right plot in Fig.~\ref{fig:exposure} shows the aperture ratio of radio detection to WCDs for different cases.
For neutrino energies of $E_\nu\leq 1$ EeV, only the naive trigger case (SNR>10) provides compatible exposures for radio detection and WCDs.
For $E_\nu= 3$ EeV, radio detection is enhanced in both the ideal and conservative cases, even for SNR>100.
For $E_\nu \geq 10$\,EeV, radio detection provides at least about 100\% more aperture than WCDs in all cases.

\section{Summary and Outlook}

The detection of UHE neutrinos is one of the major frontiers in astroparticle physics. In this study, we have demonstrated the feasibility of using ground-based radio antennas to detect inclined neutrino-induced air showers. Specifically, we introduced an algorithm to reconstruct the maximum of the radio emission ($X^{\text{radio}}_{\text{max}}$), enabling us to efficiently distinguish between EAS induced by neutrinos and those induced by cosmic-rays. 

Using a reference radio array similar to the Pierre Auger Observatory, we demonstrated that radio arrays can enhance the effective area for detecting downward-going $\nu_e$-CC induced air showers at energies above 1~EeV and zenith angles greater than $75^\circ$. Their sensitivity to neutrinos complements that of existing surface detectors.
In this study, we proved that the $X^{\text{radio}}_{\text{max}}$ observable is a powerful tool for rejecting cosmic-ray backgrounds. 
%
The reconstruction and identification techniques developed in this work can be applied directly to large-scale instruments, such as GRAND ~\cite{GRAND:2018iaj}, or adapted for them.

Future work can build on this method by incorporating neutral-current (NC) and other flavor interactions, and developing more robust approaches for handling young showers close to the ground, as well as old showers that overlap with the background of UHECRs and possibly UHE photons. All this will contribute to paving the way for detecting EeV neutrinos and identifying their astrophysical sources using large-scale radio antenna arrays.

\acknowledgments
We would like to thank our colleagues in the radio group of the Pierre Auger Collaboration for their valuable suggestions and fruitful discussions. 
We are also grateful to the Pierre Auger Collaboration for providing simulation resources with $\nu_e$-CC interactions. 
The calculations for this work were performed on the Pleiades Cluster at the University of Wuppertal.
We would like to thank Felix Schlüter for his valuable suggestions regarding the effects of the antenna response and Mohit Saharan for valuable cross-checks on the efficiency of the radio trigger and for contributing to neutrino simulations in Auger.
This work was supported by the BMFTR Verbundforschung Astroteilchenphysik and by the Deutsche Forschungsgemeinschaft (DFG).

\appendix
\section{Radio antenna response modeling}
\label{AntennaModeling}
There are two antenna responses. One is for the east-west (EW) arm, and the other is for the north-south (NS) arm. 
Since they have an identical design, their simulations and responses are quite similar.
Therefore, we will only present results for the EW arm. The NS arm is modeled in the same way. 

The response patterns of ${H}^{EW}(\Theta, \Phi, f)$ can be decomposed into ${H}^{EW}_\Theta(\Theta, \Phi, f)$ and ${H}^{EW}_\Phi(\Theta, \Phi, f)$.
${H}^{EW}_\Theta$ and ${H}^{EW}_\Phi$ are complex values.
${H}^{EW}_\Theta$ can be simplified as a function of azimuth angle $\Phi$ for different frequency $f$ and zenith angle $\Theta$:
\begin{equation}
\label{eq:HEWTheta}
    H^{EW}_\Theta(\Phi) = \left(\mathcal{A}_{\Theta}^{EW}  + i \mathcal{B}_{\Theta}^{EW} \right) \cdot \cos(\Phi-\Phi_{max}^{EW}) \,,
\end{equation}
where $\mathcal{A}_{\Theta}^{EW}$ and $\mathcal{B}_{\Theta}^{EW}$ are the amplitudes of the real and imaginary parts, and $\Phi_{max}^{EW}$ is the angle representing the maximum antenna response of $H^{EW}_\Theta(\Phi)$. 
In addition, $\mathcal{A}_{\Theta}^{EW}$ and $\mathcal{B}_{\Theta}^{EW}$ can be positive and negative.
${H}^{EW}_\Phi$ can also be simplified as a function of azimuth angle $\Phi$ for a given zenith $\Theta$:
\begin{equation}
\label{eq:HEWPhi}
    H^{EW}_\Phi(\Phi) = \left( \mathcal{A}_{\Phi}^{EW}  + i \mathcal{B}_{\Phi}^{EW}\right) \cdot \sin(\Phi-\Phi_{max}^{EW}) \,,
\end{equation}
where $\mathcal{A}_{\Phi}^{EW}$ and $\mathcal{B}_{\Phi}^{EW}$ are the amplitudes of the real and imaginary parts, $\Phi_{max}^{EW}$ is the same angle in Eq.~\eqref{eq:HEWTheta} representing the zero antenna response of $ H^{EW}_\Phi(\Phi)$.
With this modeling, we get the VEL amplitudes as functions of $\Theta$ and $f$: $\mathcal{A}_{\Theta}^{EW}(\Theta, f)$, $\mathcal{B}_{\Theta}^{EW}(\Theta, f)$, $\mathcal{A}_{\Phi}^{EW}(\Theta, f)$ and $\mathcal{B}_{\Phi}^{EW}(\Theta, f)$.




The antenna response VEL is simulated with discrete values of the frequency $f$ and the electric field zenith and azimuth angles $\Theta$ and $\Phi$, respectively.
We build models at \eqref{eq:HEWTheta} and \eqref{eq:HEWPhi} to describe the VEL as a function of the azimuth angle $\Phi$ for a given frequency $f$ and zenith $\Theta$.

First, we fit the models to the simulated VEL at the available (discrete) combinations of frequency $f$ and zenith angle $\Theta$, obtaining the corresponding model parameters (e.g., $\mathcal{A}_{\Theta}^{EW}(\Theta, f)$, $\mathcal{B}_{\Theta}^{EW}(\Theta, f)$, $\mathcal{A}_{\Phi}^{EW}(\Theta, f)$, and $\mathcal{B}_{\Phi}^{EW}(\Theta, f)$) for each simulation point. 
Then, we linearly interpolate these fitted parameters in $(\Theta, f)$ to evaluate $\mathcal{A}_{\Theta}^{EW}(\Theta, f)$, $\mathcal{B}_{\Theta}^{EW}(\Theta, f)$, $\mathcal{A}_{\Phi}^{EW}(\Theta, f)$, and $\mathcal{B}_{\Phi}^{EW}(\Theta, f)$ at arbitrary frequencies $f$ and zenith angles $\Theta$. With this, we obtain the VEL for a given $(\Theta, f)$ by using Eqs.~\eqref{eq:HEWTheta} and \eqref{eq:HEWPhi}.

According to \cite{Glaser2017AugerRadioCalibration} interpolating the vector effective length in terms of its real and imaginary components yields results that are better than, or at least comparable to, alternative approaches across all three dimensions: frequency, zenith angle, and azimuth angle.
Therefore, we interpolate only the real parts of \eqref{eq:HEWTheta} and \eqref{eq:HEWPhi} and the imaginary parts of \eqref{eq:HEWTheta} and \eqref{eq:HEWPhi} between discrete values of the frequency $f$ and zenith $\Theta$.

\begin{figure}[htb]
    \centering
    \includegraphics[width=0.32\linewidth]{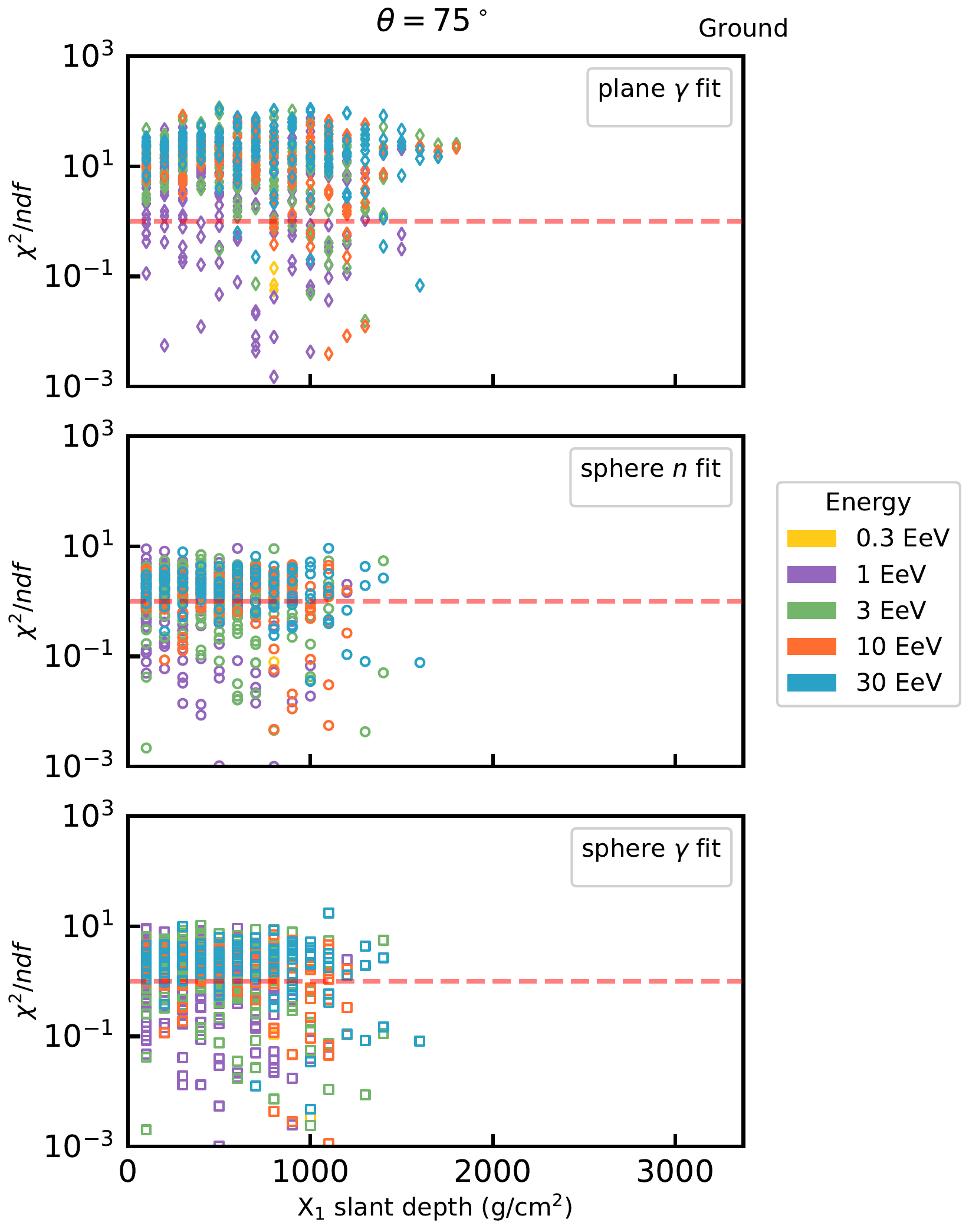}
    \includegraphics[width=0.32\linewidth]{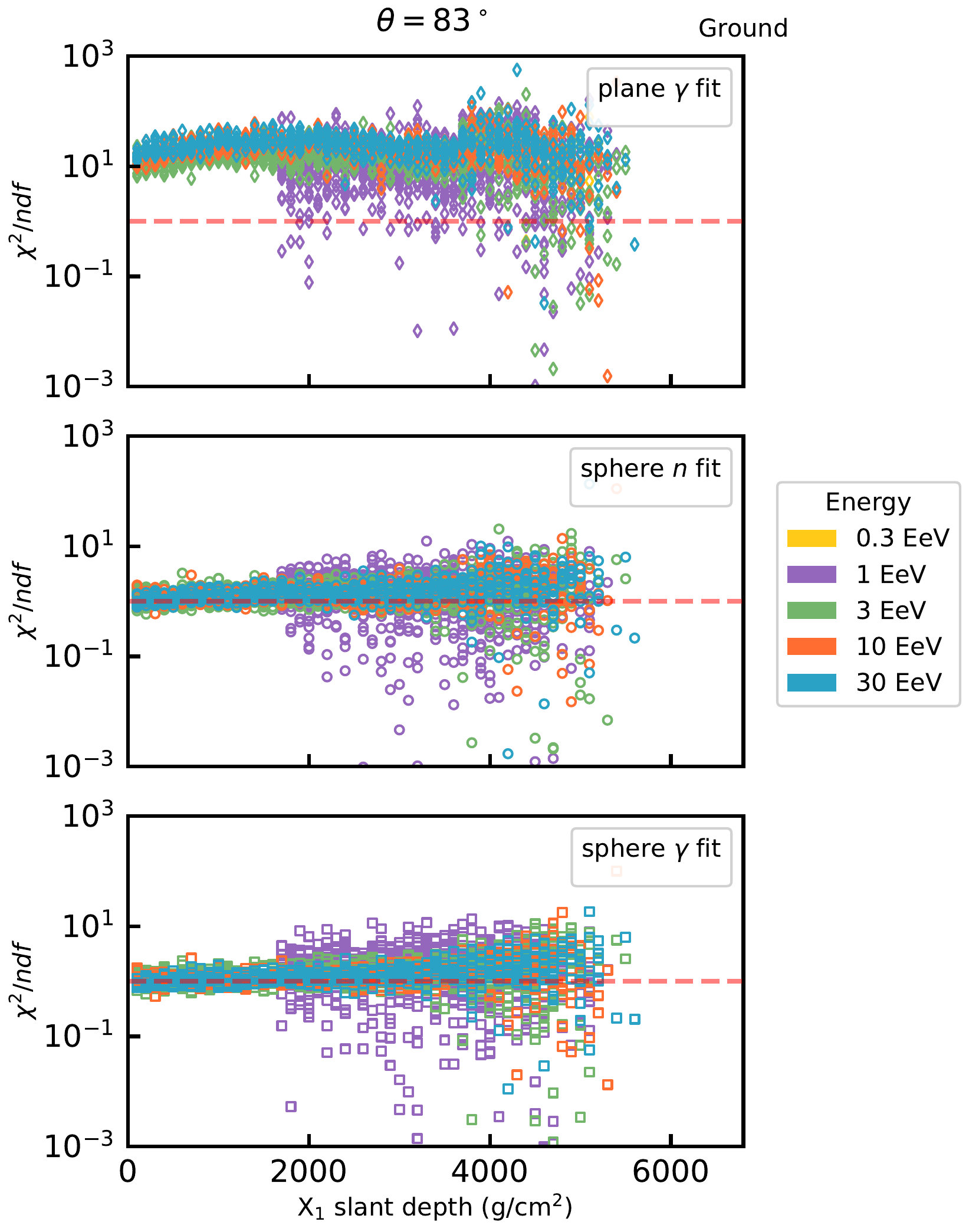}
    \includegraphics[width=0.32\linewidth]{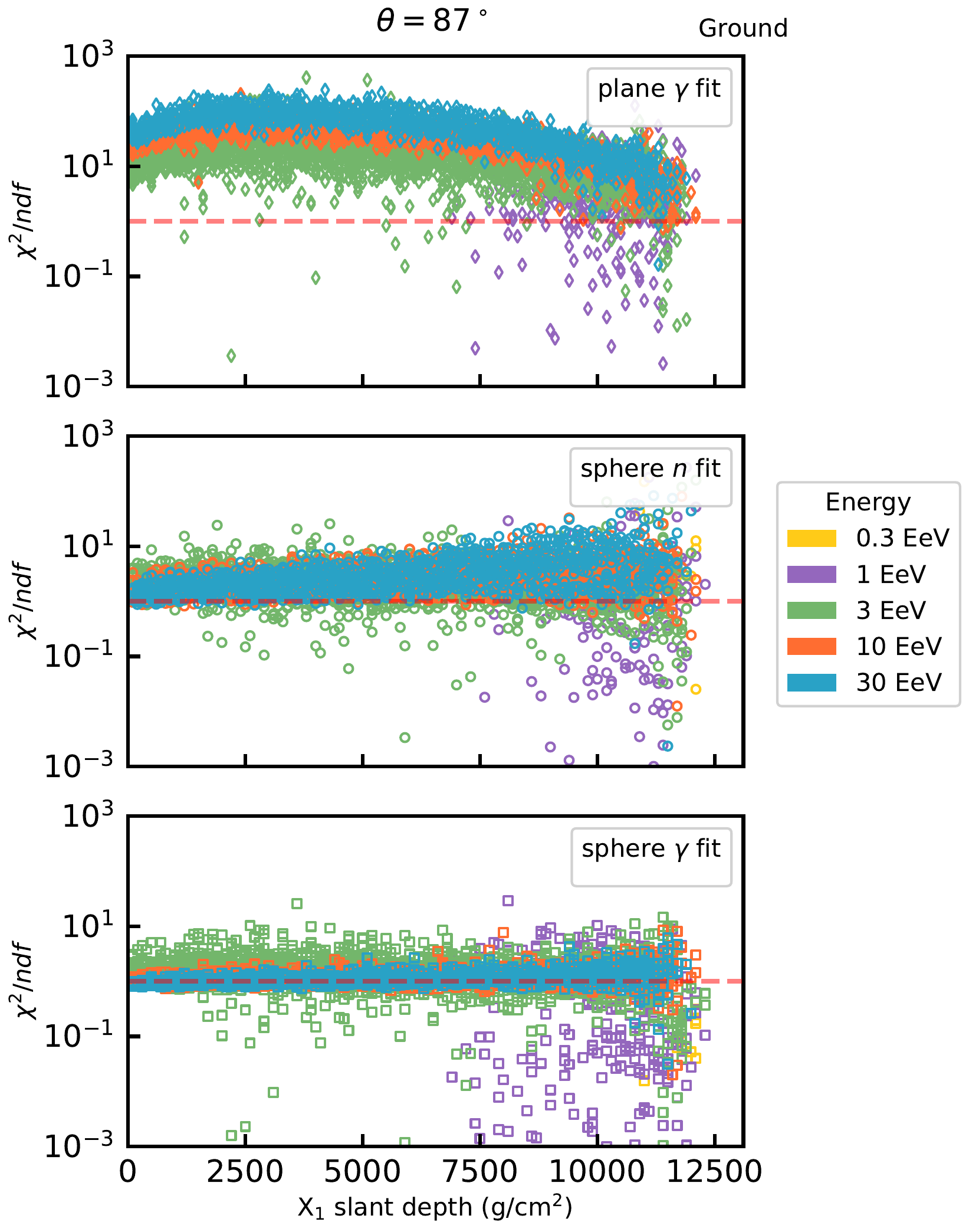}
    \caption{Three cases with different zenith angles (75$^\circ$, 83$^\circ$, and 87$^\circ$) for the performances of three different the wavefront models on the simulations with 30-80\,MHz bandpass, white noise, antenna response and GPS timing uncertainty (5 ns).}
    \label{fig:wavefront}
\end{figure}

\section{Radio wavefront}
\label{sec:Wavefront}
In this section, we test three wavefront models for the neutrino induced horizontal showers.
These three models are: the planar wavefront with the $\gamma_{n}$-factor, the spherical wavefront with the average refractive index $\bar{n}$, and the spherical wavefront with the $\gamma_{n}$-factor.
\paragraph{The planar wavefront with the $\gamma_{n}$-factor}
For the planar wavefront, the arrival time different between two antennas ($i$ and $j$) can be written as
\begin{equation}
    \Delta t_{ij} = \frac{\vec{n}(\theta, \phi)\cdot (\vec{r}_i - \vec{r}_j)}{\gamma_n\cdot c}\,,
\end{equation}
where $\vec{n}(\theta, \phi)$ is the unit vector of the shower axis and $\gamma_n$ can mimic the effect from the refractive index $n$.
\paragraph{The spherical wavefront with the average refractive index $\bar{n}$}
The time difference between the maximum radio emission time $t_{max}$ and trigger time $t_i$ of each antenna can be expressed as
\begin{equation}
    \Delta t_i =  \frac{|\vec{r}_\text{max}-\vec{r}_i| \cdot \bar{n}}{c}\,,
\end{equation}
where $\bar{n}$ is the average refractive index from the maximum radio emission point to each antenna.
More details can be found in Ref.~\cite{Schluter:2021egm}.
\paragraph{The spherical wavefront with the $\gamma_{n}$-factor}
The time difference is
\begin{equation}
    \Delta t_i =  \frac{|\vec{r}_\text{max}-\vec{r}_i| }{\gamma_n \cdot c}\,.
\end{equation}

Fig.~\ref{fig:wavefront} presents the performances of different wavefront models mentioned above with different zenith angles (75$^\circ$, 83$^\circ$, and 87$^\circ$).
The planar wavefront does not work well for the most of the cases. 
The spherical wavefront model with the average refractive index $\bar{n}$ works well as same as the spherical wavefront model with the $\gamma_{n}$-factor when $\theta = 75^\circ$. 
When the zenith angle is above $83^\circ$, the performance of the spherical wavefront model with the average refractive index $\bar{n}$ is not good.
However, the model of the spherical wavefront with the $\gamma_{n}$-factor works very well even for $87^\circ$.
We think, for the case of horizontal showers above $83^\circ$, the wavefront deviates from the spherical model and $|\vec{r}_\text{max}-\vec{r}_i|$ is not the real pathway of the radio propagation due to the refraction in the air. 
Hence, the model of the spherical wavefront with the average refractive index $\bar{n}$ can not describe the time difference well.
However, the $\gamma_{n}$-factor can effectively compensate it.
Therefore, the model of the spherical wavefront with the $\gamma_{n}$-factor is the best among these three models for the horizontal showers.

Furthermore, the better wavefront model, like the hyperbolic wavefront \cite{Corstanje:2014waa}, and the ray tracing are needed at each antenna to calculate the time difference for the accurate reconstruction of the maximum radio emission induced by the horizontal showers. 

\section{Shower axis reconstructions}
\label{sec:ShowerAxisReconWithDifferentCores}
\begin{figure}
    \centering
    \includegraphics[width=0.49\linewidth]{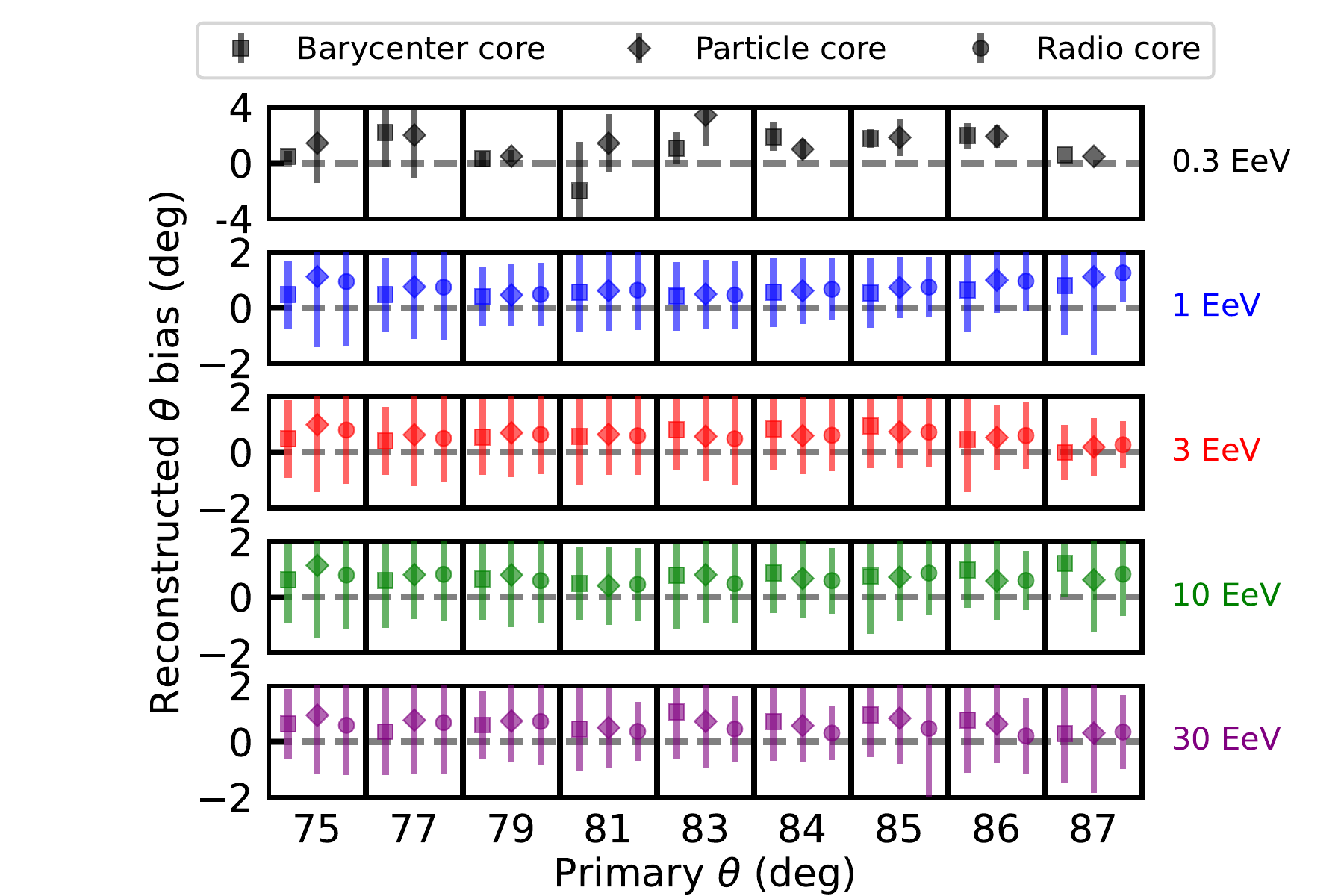}
    \includegraphics[width=0.49\linewidth]{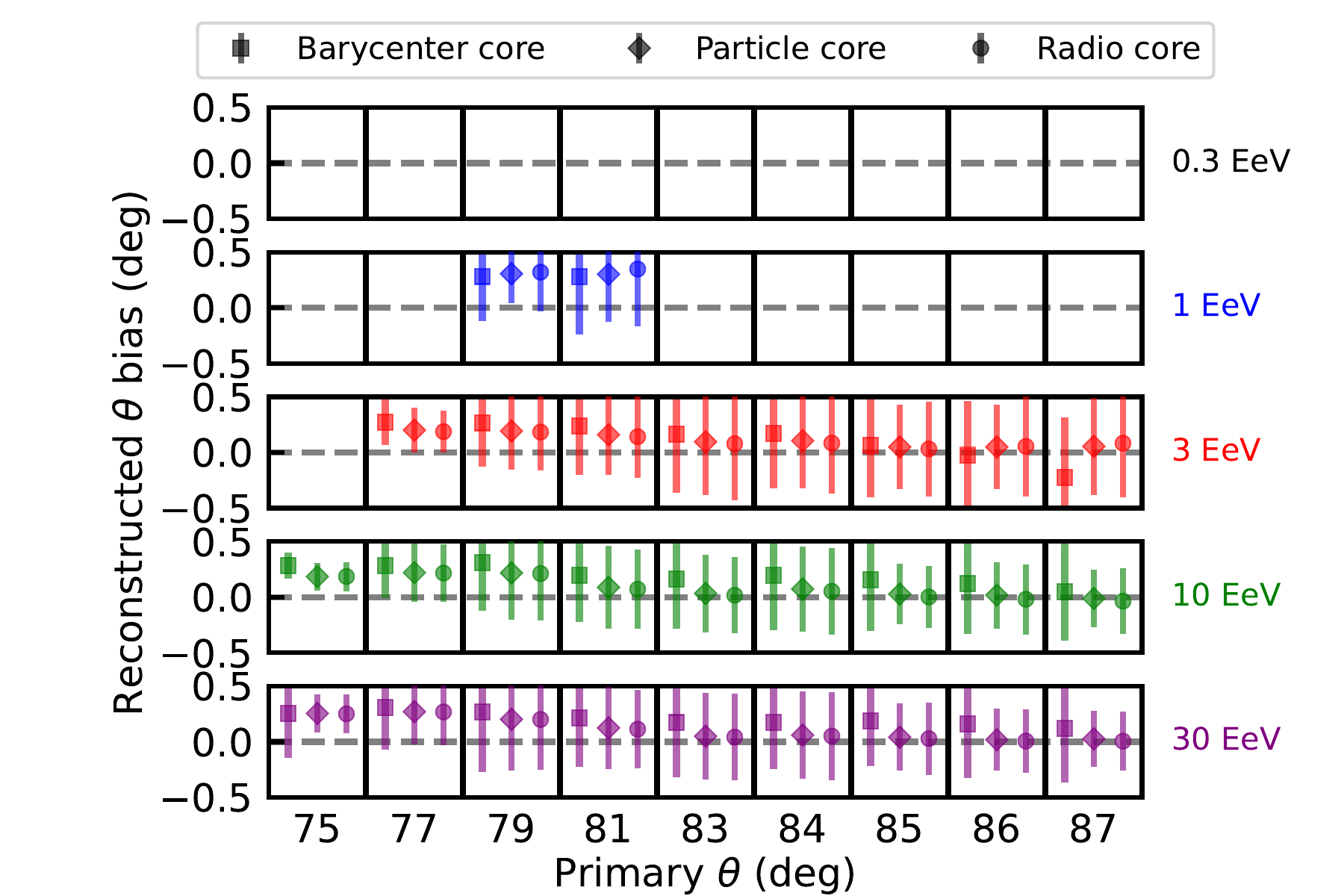}
    \caption{The performances of the shower axis reconstruction with three different shower cores: the barycenter position $\vec{r}_{b}$, the particle footprint core position $\vec{r}_{p}$, and the radio footprint core position $\vec{r}_{\rm core}$. All simulations have 30-80\,MHz bandpass, white noise, antenna response and GPS timing uncertainty (5 ns). Left: low quality events with additional fit quality cuts. Right: high quality events with additional fit quality cuts.}
    \label{fig:ShowerAxisReconWithDifferentShowerCore}
\end{figure}

In this section, we compare the shower axis reconstruction with three different shower cores, the barycenter position $\vec{r}_{b}$, the particle footprint core position $\vec{r}_{p}$, and the radio footprint core position $\vec{r}_{\rm core}$, as introduced in the section~\ref{sec:ShowerAxisRecon}.
Fig.~\ref{fig:ShowerAxisReconWithDifferentShowerCore} presents the reconstruction performance with different shower cores for both low quality events and high quality events with additional fit quality cuts: $\vec{r}_{max}$ fit quality cut ($\chi^2/ndf<10$) for all cases and radio footprint quality cut ($\chi^2/ndf<10$) for the radio core $\vec{r}_{\rm core}$ case.
For the case in the left plot, three cases have similar performances: the bias and the deviation of the reconstructed $\theta$ can be about $1^\circ$ and $2^\circ$, respectively.
For the case in the right plot, the bias is less than $0.5^\circ$ and the deviation is less than $0.5^\circ$ for all the shower core cases. 
Furthermore, the cases of the particle footprint core position $\vec{r}_{p}$ and the radio footprint core position $\vec{r}_{\rm core}$ present similar reconstruction performance.

\section{Radiation energy}

\subsection{Footprint feature}
Based on Sec.\,\ref{sec:Footprint}, we study the parametrization of the geomagnetic footprint.
We found patterns that could be universally applicable not only to $\nu_e$-CC interactions, but also to charged cosmic rays at Auger.

After reconstructing the maximum radio emission point, $\vec{r}_{\rm max}$, as described in Sec.\,\ref{sec:RadioPositionRecon}, the air density $\rho$ and the refractive index $n$ at this point can be determined according to the known atmospheric model. With this information, 
the Cherenkov angle, $\theta_{\rm C}=\arccos(1/n)$, and
the Cherenkov radius in the shower plane at the ground can be predicted as
\begin{equation}
    R_{\rm C}= \tan(\theta_{\rm C})\cdot d_{\rm max}\,,
\end{equation}
where $d_{\rm max}$ is the absolute distance between $\vec{r}_{\rm max}$ and $\vec{r}_\text{core}$, i.e.\ the distance from the shower maximum to the ground, $|\vec{v}|$.

\begin{figure}[htb]
    \centering
    \includegraphics[width=0.48\linewidth]{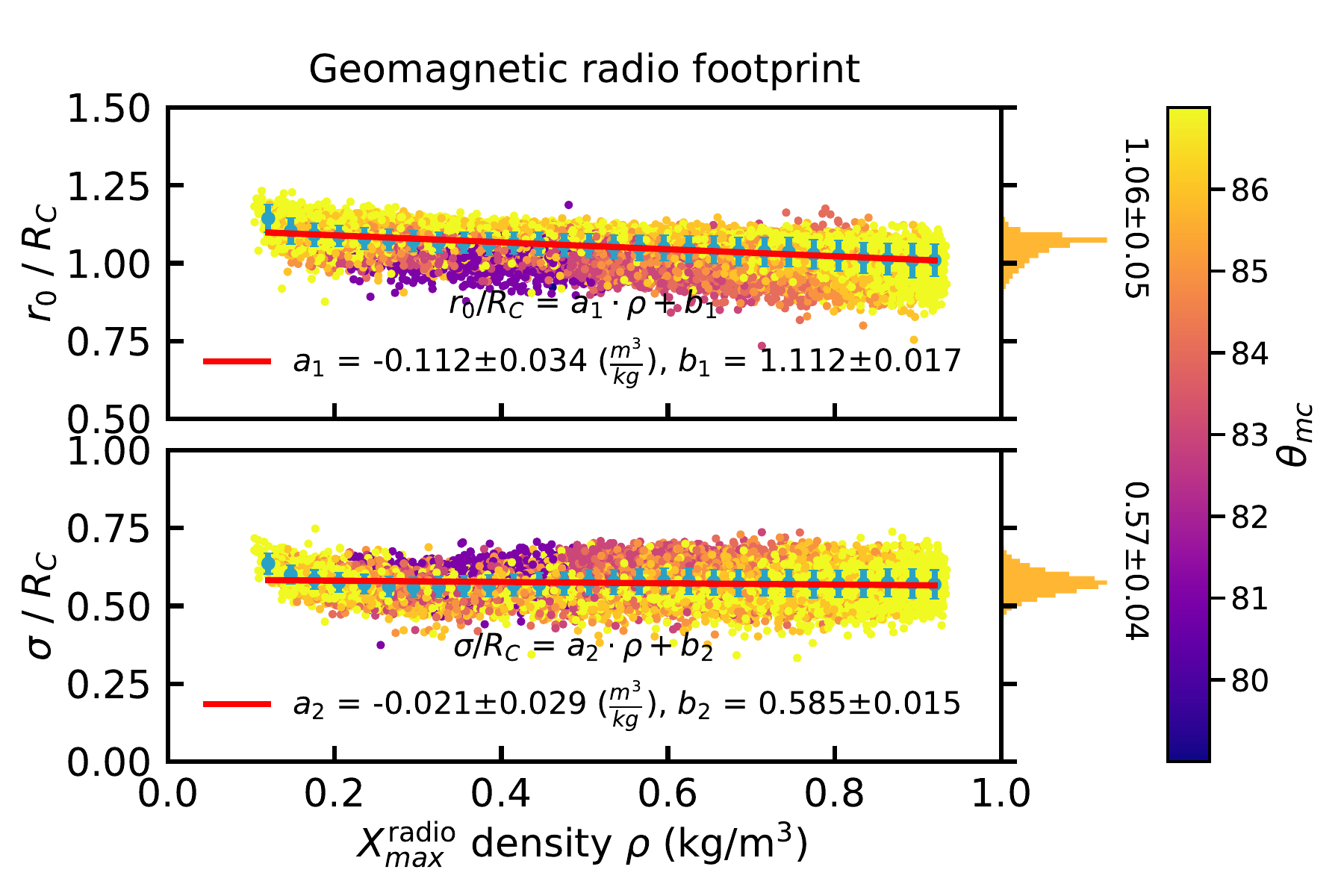}
    \includegraphics[width=0.48\linewidth]{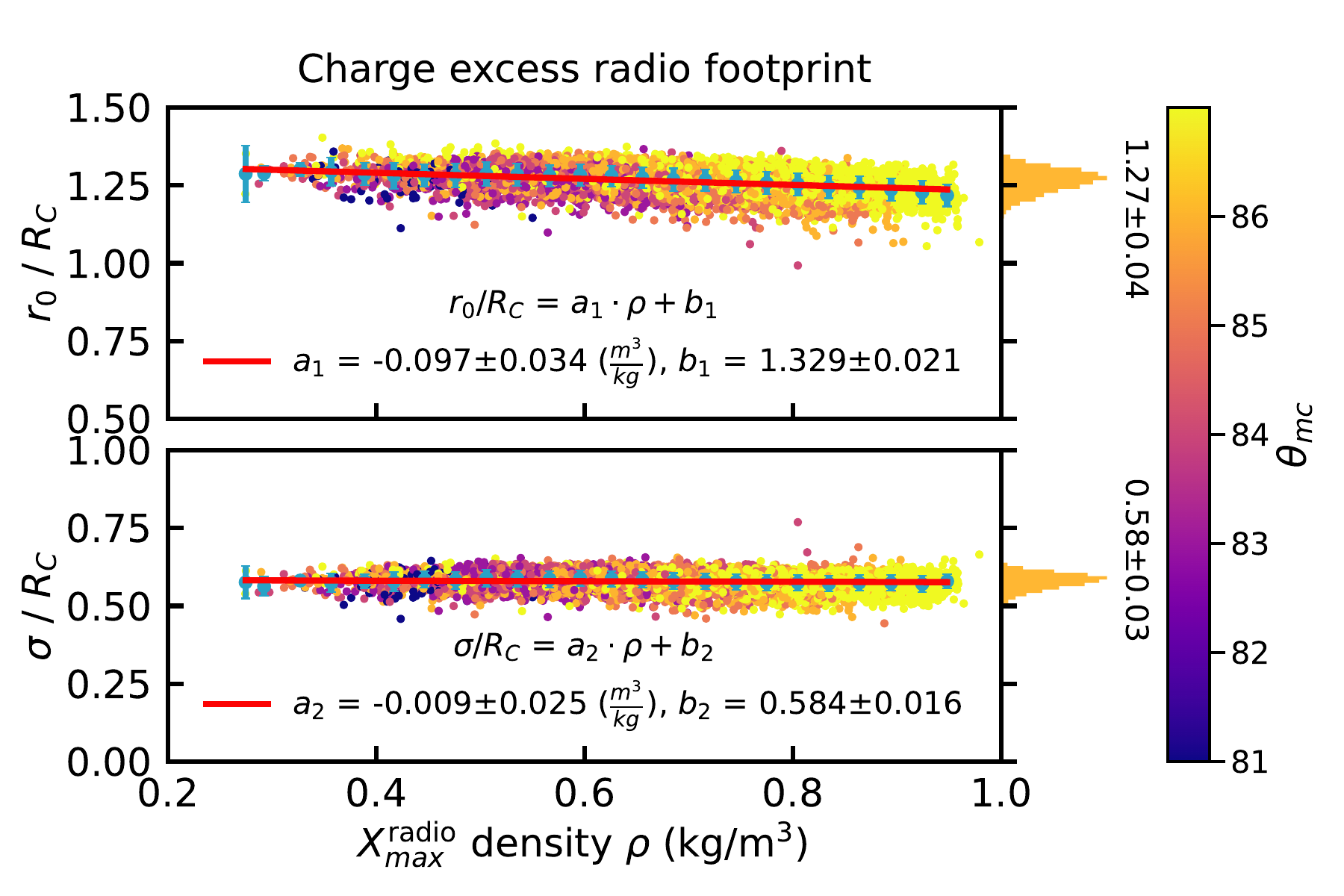}
    \caption{Left: features of the geomagnetic radiation footprint. Right: features of the charge excess radiation footprint.}
    \label{fig:FootprintPrediction}
\end{figure}

The maximum radius $r_0$ of the Gaussian distribution in Eq.~\eqref{eq:fgeo2} and Eq.~\eqref{eq:fce} can be correlated with the predicted Cherenkov angle $\theta_{\rm C}$.
The two upper plots in Fig.~\ref{fig:FootprintPrediction} show the ratio $r_0/R_{\rm C}$ as a function of the air density at $\vec{r}_{\rm max}$. 
Additionally, the width $\sigma$ of the Gaussian distribution can be predicted and be compared to the size of the Cherenkov cone, as shown in the two lower plots of Fig.~\ref{fig:FootprintPrediction}.
These linear correlations can be used for the footprint fit once $\vec{r}_{\rm max}$ is estimated.

\subsection{The noise influence on the footprint fit}
\begin{figure}[ht]
    \centering    
    \includegraphics[width=0.48\linewidth]{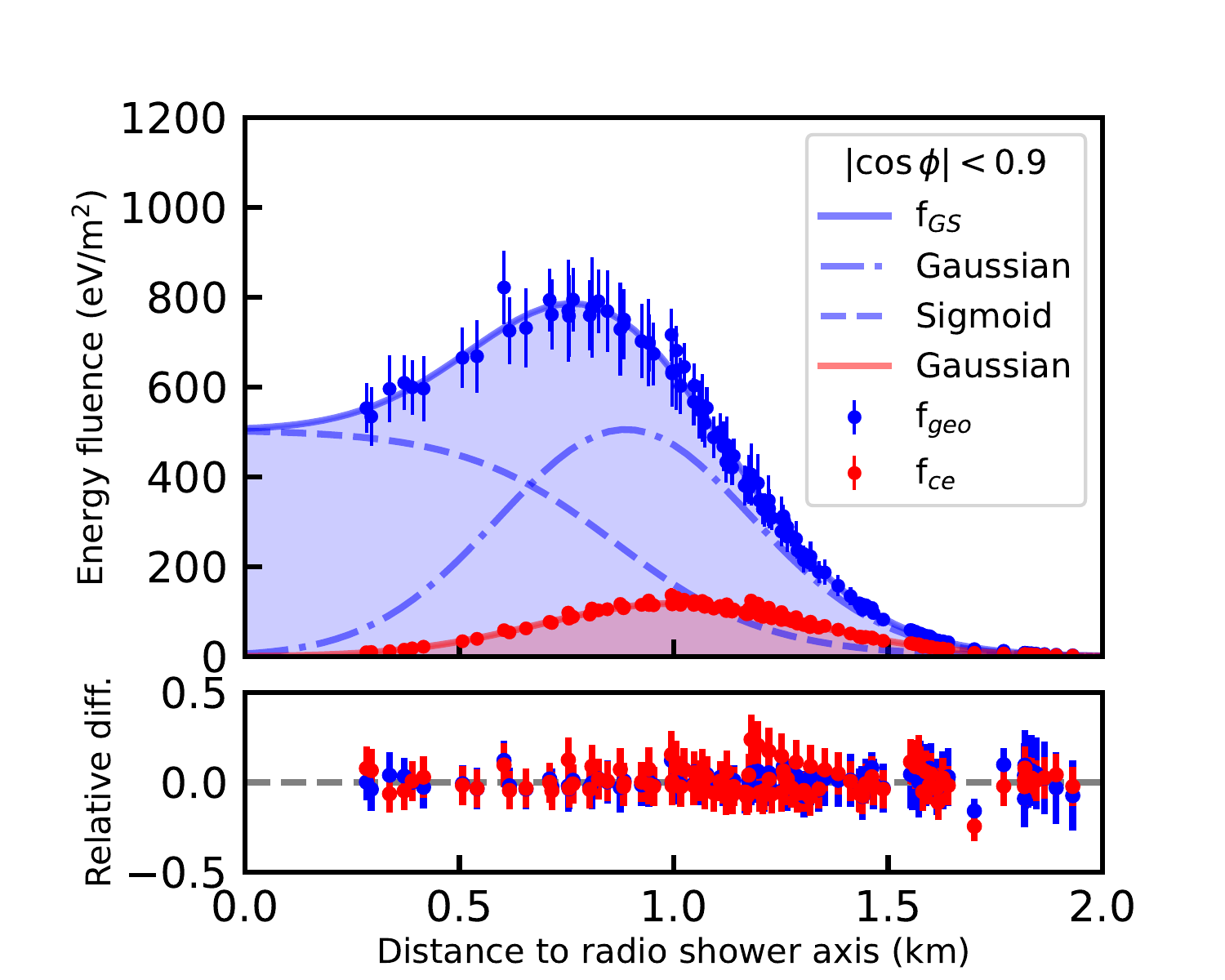}
   \includegraphics[width=0.48\linewidth]{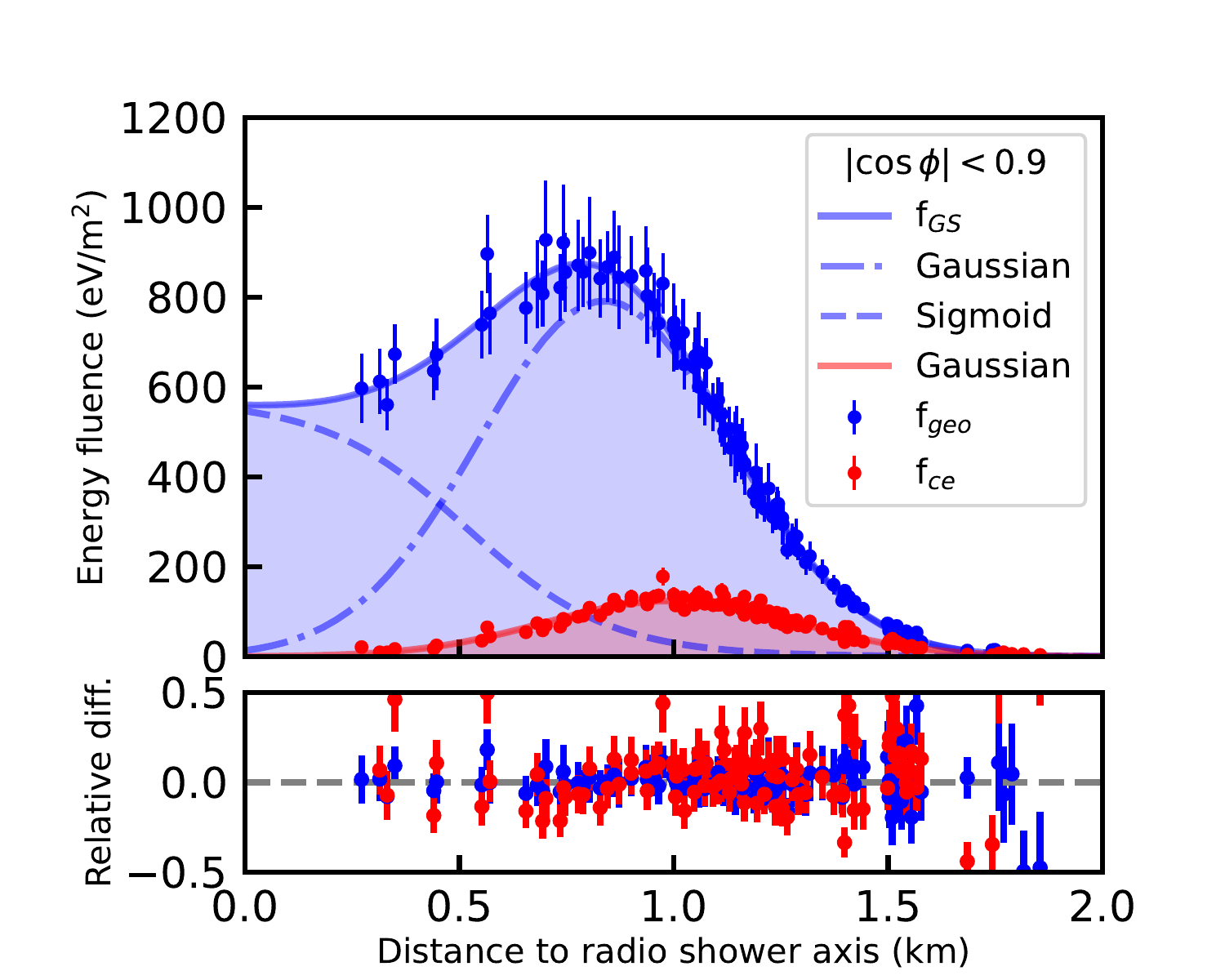}
    \caption{The performance of the footprint fits for a $\nu_e$-CC simulation of $E_{\nu_e}=30$ EeV, $\theta=87^\circ$, and $X_1=8000$\,g\,cm$^{-2}$ with 30-80\,MHz bandpass, and GPS timing uncertainty. There is no radio antenna response.
    Left: without noise. Right: with noise.
    }
    \label{fig:NoisyFootprint}
\end{figure}

Noise can severely impact the accuracy of radio footprint fits. Especially, if the estimated energy fluence is biased by the noise, the footprint fit will be inaccurate.
In this study, we use the method for estimating the energy fluence described in Ref.\,\cite{PierreAuger:2025aik}.
Figure~\ref{fig:NoisyFootprint} shows two simulations of a $\nu_e$-CC interaction at $E_{\nu_e}=30$ EeV, $\theta=87^\circ$, and $X_1=8000$\,g\,cm$^{-2}$, and with a bandpass of 30-80\,MHz.
The simulation on the left is without noise and antenna response.
Its footprint fit is very smooth.
However, after adding noise, the footprint becomes rough, as shown in the right plot of Fig.~\ref{fig:NoisyFootprint}.
Adding the antenna response further deteriorates the fit, as shown in Fig.~\ref{fig:Footprint}.
Better methods for estimating the energy fluence with less bias, especially when the SNR is low, were proposed in Ref.\,\cite{Martinelli:2024bzg, Ravn:2025puy}. 

\subsection{Radiation energy for the air density of the particle maximum}
\label{sec:ParticleMaxModeling}

We also studied the radiation energy as a function of the air density at the position of the particle shower maximum instead of that at the radio shower maximum. The two differ slightly, as also noted in Ref.\,\cite{Glaser:2016qso}.
Table~\ref{tab:ParticleAirDensityEgeo} shows the results for the geomagnetic radiation energy parameterizations for $\nu_e$-CC interactions compared with hadron showers \cite{Schluter:2022mhq}.
For comparison purposes with Ref.\,\cite{Schluter:2022mhq}, the geomagnetic radiation energy is not scaled by 111\% in this section. 
The reference air density $\rho_{\rm ref}$ is $\langle\rho\rangle=0.3$ kg\,m$^{-3}$.

\begin{table*}[!ht]
\centering
{\small
\begin{tabular}{ lccc}
 \toprule
  & $\nu_e$-CC Sibyll-2.3d & $\nu_e$-CC  EPOS-LHC & Hadron \cite{Schluter:2022mhq} \\
\midrule\midrule
$p_0$                   & 0.5490 $\pm$ 0.0262  & 0.5524 $\pm$ 0.0233 & 0.5045\\
$p_1$  (m$^{3}$/kg)     & -2.8673 $\pm$ 0.3008 & -2.8317 $\pm$ 0.2766 & -2.7083\\
$S_{19}$ (GeV)          & 3.2873 $\pm$ 0.0956 &  3.2791 $\pm$ 0.0946 & 3.1461\\
$\gamma$                & 1.9700 $\pm$ 0.0109 & 1.9697 $\pm$ 0.0092 & 1.9997\\
\bottomrule
\end{tabular}
}
\caption{Results of the combined fits with the air density taken at the position of the particle shower maximum $X_{\rm max}^{\rm particle}$ based on Eq.\,\eqref{eq:EgeoFraction} and \eqref{eq:EgeoEem}. The reference air density $\rho_{\rm ref}$ is $\langle\rho\rangle=0.3$ kg\,m$^{-3}$.}
\label{tab:ParticleAirDensityEgeo}
\end{table*}

Table~\ref{tab:ParticleAirDensityEceFraction} shows the corresponding results for the charge excess radiation fraction for $\nu_e$-CC interactions compared with the hadron showers \cite{Glaser:2016qso}.
The reference air density $\rho_{\rm ref}$ is $\langle\rho\rangle=0.65$ kg\,m$^{-3}$ as was used in \cite{Glaser:2016qso}.

\begin{table*}[!ht]
\centering
{\small
\begin{tabular}{ lccc }
 \toprule
 &  $\nu_e$-CC Sibyll-2.3d &  $\nu_e$-CC EPOS-LHC & Hadron \cite{Glaser:2016qso}\\
\midrule\midrule
$q_0$                   & 0.47580 $\pm$ 0.00437  & 0.46782 $\pm$ 0.00398 &  0.43\\
$q_1$  (m$^{3}$/kg)     & 1.05904 $\pm$ 0.00997 & 1.07982 $\pm$ 0.00951 &  1.11\\
$q_2$                   & -0.26432 $\pm$ 0.00433 & -0.25682 $\pm$ 0.00394 &  -0.24\\
\bottomrule
\end{tabular}
}
\caption{The fit result with the air density of the particle maximum $X_{max}^{particle}$ based on Eq.\eqref{eq:charge_excess_a} for the charge excess radiation energy fraction. 
The reference air density $\rho_{ref}$ is $\langle\rho\rangle=0.65$ kg\,m$^{-3}$.}
\label{tab:ParticleAirDensityEceFraction}
\end{table*}

\section{Electromagnetic energy reconstruction}
\label{sec:EemRecon}

Figure \ref{fig:EemRecon} shows a bias of $-2.7$\% in the reconstructed electromagnetic energy $E_{\rm geo}$.
The reconstruction bias of $E_{\rm geo}$ can be caused by all other parameters in Fig.\,\ref{fig:EemRecon}.
To quantify the bias propagation, we evaluated the median reconstruction biases of $E_{\rm geo}$, $\rho_{\rm max}^{\rm radio}$, and $\alpha$, and propagated them to the reconstructed $E_{\rm em}$ using Eqs.\,\eqref{eq:EgeoFraction} and \eqref{eq:EgeoEem}. 
We find that the dominant contribution comes from the bias in $E_{\rm geo}$ ($-0.8$\%), followed by the bias in $\rho_{\rm max}^{\rm radio}$ ($-0.5$\%), whereas the contribution from $\alpha$ is negligible. Altogether, these three propagated biases result in a median bias in $E_{\rm em}$ of about $-1.3$\%. 
However, this is still substantially smaller than the total median bias of about $-2.7$\% observed in the $E_{\rm em}$ reconstruction with respect to the MC truth. 
Hence, these propagated biases alone cannot explain the full reconstruction bias, indicating that additional bias sources are present in the reconstruction chain.

\begin{figure}[t]
    \centering
    \includegraphics[width=1\linewidth]{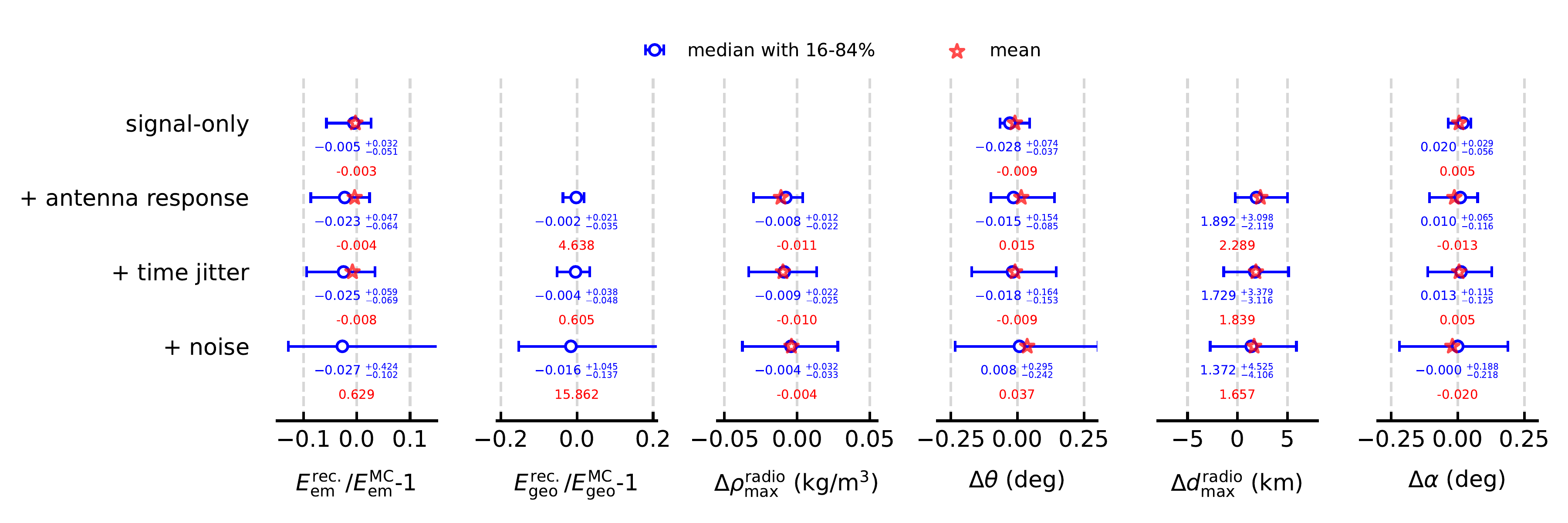}
    \centering
    \caption{
Step-by-step comparison of the reconstruction biases of the main observables relevant for the electromagnetic-energy reconstruction, obtained by progressively increasing the realism of the simulated radio signals for the HQ events (Sibyll-2.3d). 
Starting with the ``signal-only'' case, in which pure simulated signals are filtered with only 30-80~MHz rectangular bandpass filter, we successively include the antenna response, a $5$~ns time jitter, and Galactic noise. 
The six panels show the biases of $E_{\rm em}$, $E_{\rm geo}$, $\rho_{\rm max}^{\rm radio}$, $\theta$, $d_{\rm max}^{\rm radio}$, and $\alpha$. 
In each panel, the circle and horizontal bar indicate the median bias, along with the $16\%$--$84\%$ interval in blue. The star marks the mean value in red. 
CORSIKA 7 is used for the true values of $E_{\rm em}$, $\theta$, and $\alpha$. 
In contrast, the reference values of $E_{\rm geo}$, $\rho_{\rm max}^{\rm radio}$, and $d_{\rm max}^{\rm radio}$ cannot be directly read out from CORSIKA 7 or CoREAS. Therefore, they are obtained by reconstructing the ``signal-only'' simulations using the method presented in this work. 
For this reason, the ``signal-only'' entries are omitted from the panels of $E_{\rm geo}$, $\rho_{\rm max}^{\rm radio}$, and $d_{\rm max}^{\rm radio}$, since their biases are zero by construction. 
As shown in the figure, the dominant bias in the reconstructed $E_{\rm em}$ is introduced when the antenna response is included. The subsequent addition of time jitter and Galactic noise produces only comparatively small further changes. 
At this stage, significant biases appear in $\rho_{\rm max}^{\rm radio}$, $\theta$, and $d_{\rm max}^{\rm radio}$, while the bias in $E_{\rm geo}$ remains small. 
This suggests that the remaining $E_{\rm em}$ bias is not simply the linear sum of the median biases of the individual reconstructed parameters. Instead, it points to coupled effects induced by the antenna response, which likely involve nonlinear dependencies and correlations among the reconstructed observables.
}
    \label{fig:BiasStepByStep}
\end{figure}

To qualitatively investigate the origin of the remaining bias, we performed a step-by-step study in which we reconstructed the simulated signals with increasing levels of realism.
Starting with pure simulated radio signals filtered through a 30-80~MHz rectangular bandpass filter in the ``signal-only'' simulations, we successively included the antenna response, time jitter (5~ns), and Galactic noise. 
For each step, we evaluated the reconstruction biases of the main observables relevant to the $E_{\rm em}$ reconstruction, namely $E_{\rm em}$, $E_{\rm geo}$, $\rho_{\rm max}^{\rm radio}$, $\theta$, $d_{\rm max}^{\rm radio}$, and $\alpha$. 
Figure\,\ref{fig:BiasStepByStep} summarizes the median bias together with the 16\%--84\% interval and the mean value for each of these quantities for each reconstruction stage.

This step-by-step comparison shows that the dominant $E_{\rm em}$ reconstruction bias is introduced when the antenna response is included. 
The subsequent addition of time jitter (5~ns) and Galactic noise leads only to comparatively minor changes. 
Although biases are also observed in other reconstructed quantities, they do not fully explain the bias observed in $E_{\rm em}$. 
This suggests that the effect is not simply the linear propagation of the median biases from individual parameters. 
Since the $E_{\rm em}$ reconstruction depends nonlinearly on several correlated quantities, transmitting median biases is itself nontrivial. Correlations among the reconstructed parameters can further modify the final bias. 
Therefore, we conclude that the bias induced by the antenna response requires its own dedicated study, which is beyond the scope of this article. 
One possible explanation is a frequency-dependent group delay of the antenna response, which could introduce a systematic distortion of the reconstructed signals and thereby bias the $E_{\rm em}$ reconstruction. However, this remains to be studied in detail.

Furthermore, the bias in the $E_{\rm geo}$ reconstruction primarily originates from the noise subtraction method in Eq.~\eqref{eq:energyfluence}, as discussed in \cite{Martinelli:2024bzg, Ravn:2025puy}. 
Therefore, an unbiased estimation of energy fluence is essential for reconstructing both geomagnetic radiation energy $E_{\rm geo}$ and electromagnetic energy $E_{\rm em}$.
While we expect the methods proposed in \cite{Martinelli:2024bzg, Ravn:2025puy} to reduce the bias in $E_{\rm em}$ reconstruction, further investigation is needed.

Additionally, Fig.~\ref{fig:EemReconCorrelation} shows that geometry reconstruction is crucial for determining the primary electromagnetic energy, $E_{\rm em}$, since the correlation coefficients are significant. 
In particular, the calculation of the air density at the shower maximum $\rho_{\rm max}^{\rm radio}$, which is used to estimate the geomagnetic radiation energy correction via Eq.~\eqref{eq:EgeoFraction}, is based on the reconstruction of the shower maximum position and the knowledge of the atmospheric model.
Thus, to improve the reconstruction of $E_{\rm em}$, better geometry reconstructions are needed, including wavefront modeling and ray tracing for very inclined air showers, as well as atmospheric monitoring.

\section{Electromagnetic energy fraction}
\label{sec:EemOverEnu}
To derive the ratio of electromagnetic energy of the $\nu_e$-CC induced shower to the primary neutrino energy, $E_{\rm em}/E_{\nu_e}$, we employ again simulations.
To model this ratio, we use a model consisting of two Gaussians and one sigmoid cutoff function:
\begin{equation}
\label{eq:EemOverEnu}
f(x) \;=\; \Big[ A_1\cdot G(\mu=1,\sigma_1)\;+\;A_2 \cdot G (\mu=1,\sigma_2) \Big]\cdot S(x_c,w)  \,,
\end{equation}
where $G(\mu, \sigma)$ is a Gaussian distribution with a fixed $\mu=1$, and $S(x_c, w)$ is the sigmoid function.
%
The sigmoid function is expressed by
\begin{equation}
S(x_c,w)\;=\;\frac{1}{1+\exp\!\left(\frac{x-x_c}{w}\right)} \,.
\end{equation}

\begin{figure}[htb]
    \centering
    \includegraphics[width=0.5\linewidth]{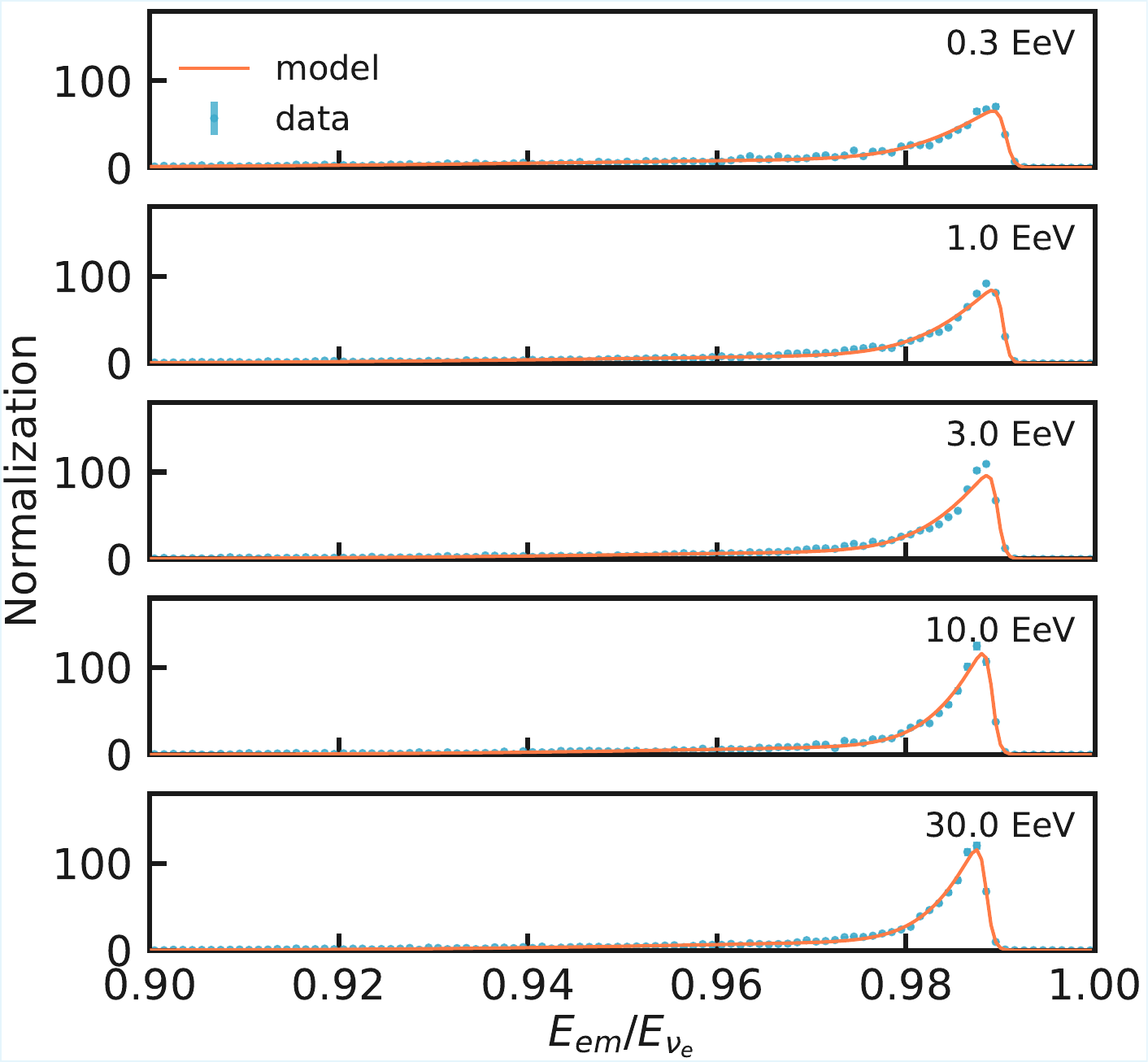}
    \caption{Distributions of the electromagnetic energy fraction $E_{\rm em}/E_{\nu_e}$ based on fits to Sibyll-2.3d simulations of different primary energies.}
    \label{fig:EemOverEnu}
\end{figure}

Figure\,\ref{fig:EemOverEnu} presents the fitting of the ratio using Sibyll-2.3d simulations and Eq.\eqref{eq:EemOverEnu} for different neutrino energies.
Table~\ref{tab:emratio_fit_params_noerr_sibyll} and \ref{tab:emratio_fit_params_noerr_EPOS} show the best fits for Sibyll-2.3d and EPOS-LHC models separately.

\begin{table}[!ht]
\centering
\begin{tabular}{r cccccc}
 \toprule
$E$ (EeV) &
$A_{1}$ & $\sigma_1$ &
$A_{2}$ & $\sigma_2$ &
$x_c$ & $w$ \\
\midrule\midrule
0.3  &
$1357.91$ &
$1.04\times 10^{-2}$ &
$586.45$ &
$5.08\times 10^{-2}$ &
0.9906 &
$3.58\times 10^{-4}$ \\
1.0  &
$1951.69$ &
$9.15\times 10^{-3}$ &
$561.57$ &
$4.24\times 10^{-2}$ &
0.9900 &
$3.91\times 10^{-4}$ \\
3.0  &
$2435.43$ &
$8.72\times 10^{-3}$ &
$541.42$ &
$3.83\times 10^{-2}$ &
0.9896 &
$3.73\times 10^{-4}$ \\
10.0 &
$3121.98$ &
$8.46\times 10^{-3}$ &
$499.46$ &
$3.57\times 10^{-2}$ &
0.9891 &
$3.30\times 10^{-4}$ \\
30.0 &
$4616.24$ &
$7.56\times 10^{-3}$ &
$575.74$ &
$3.44\times 10^{-2}$ &
0.9883 &
$3.85\times 10^{-4}$ \\
\bottomrule
\end{tabular}
\caption[]{Fit results for the $E_{\rm em}/E_{\nu_e}$ ratio to Eq.\,\ref{eq:EemOverEnu} using the Sibyll-2.3d model. Only $A_1$ and $A_2$ are rescaled by the normalization factor such that the fitted model integrates to unity over the fitted range; all other parameters are direct fit outputs.\label{tab:emratio_fit_params_noerr_sibyll}
}
\end{table}

\begin{table}[!ht]
\centering
\begin{tabular}{r cccccc}
 \toprule
$E$ (EeV) &
$A_{1}$ & $\sigma_1$ &
$A_{2}$ & $\sigma_2$ &
$x_c$ & $w$ \\
\midrule\midrule
0.3  &
$1279.38$ &
$9.86\times 10^{-3}$ &
$671.56$ &
$4.82\times 10^{-2}$ &
0.9905 &
$3.99\times 10^{-4}$ \\
1.0  &
$1773.67$ &
$9.37\times 10^{-3}$ &
$573.07$ &
$4.32\times 10^{-2}$ &
0.9902 &
$3.26\times 10^{-4}$ \\
3.0  &
$2279.77$ &
$9.03\times 10^{-3}$ &
$525.17$ &
$4.07\times 10^{-2}$ &
0.9897 &
$3.52\times 10^{-4}$ \\
10.0 &
$3332.78$ &
$8.16\times 10^{-3}$ &
$513.21$ &
$3.59\times 10^{-2}$ &
0.9891 &
$3.50\times 10^{-4}$ \\
30.0 &
$3798.15$ &
$8.07\times 10^{-3}$ &
$555.88$ &
$3.62\times 10^{-2}$ &
0.9884 &
$3.71\times 10^{-4}$ \\
\bottomrule
\end{tabular}
\caption[]{
Same as Table\,\ref{tab:emratio_fit_params_noerr_sibyll}, but for the \textsc{EPOS-LHC} model. \label{tab:emratio_fit_params_noerr_EPOS}
}
\end{table}

\section{Details about the neutrino trigger efficiency}
\label{sec:TriggerEff}
Since neutrinos can interact with the atmosphere anywhere, radio radiation coherence may be enhanced or weakened by different atmospheric conditions as the shower develops. 
\begin{figure}[t]
    \centering
    \includegraphics[width=0.8\linewidth, trim=0 45mm 0 60mm, clip]{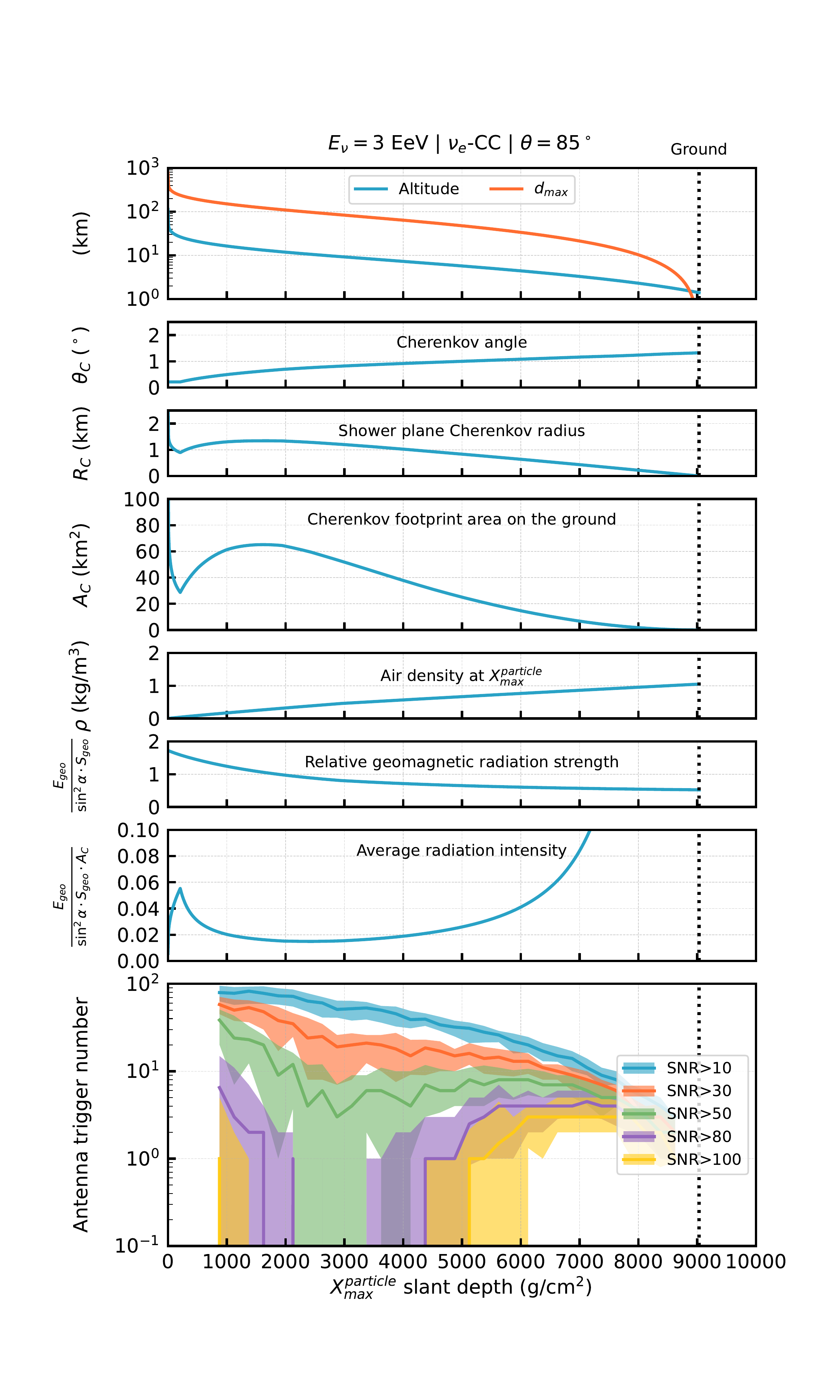}
    \caption{Nonlinear variation in the number of antenna triggers caused by multiple factors.}
    \label{fig:MagicNeutrinoTriggerNumber}
\end{figure}
%
Figure~\ref{fig:MagicNeutrinoTriggerNumber} summarizes the geometrical and atmospheric factors that determine the size of the Cherenkov footprint on the ground and, consequently, the trigger multiplicity for different SNR thresholds as a function of the true particle-maximum slant depth, $X_{\max}^{\rm particle}$. The vertical dotted line marks the ground level (1.4 km), i.e., the maximum traversed slant depth for the chosen zenith angle of $85^\circ$.

\paragraph{Panel-by-panel description.}
\begin{itemize}
    \item \textbf{Altitude and propagation distance (top panel).}
    We show the emission altitude $h(X)$ (in kilometers) corresponding to a given slant depth, $X$, as well as the maximum propagation distance to the detector plane, $d_{\max}(X)$. $h(X)$ is obtained from the atmospheric model and $d_{\max}(X)$ is the line-of-sight distance from the emission point to the ground level, calculated using the shower geometry at a zenith angle of $\theta$.

    \item \textbf{Cherenkov angle (second panel).}
    The Cherenkov angle $\theta_{\rm C}(X)$ (in degrees) is computed from the refractive index $n(h)$ as follows: $\theta_{\rm C}(X) = \arccos\!\left(n(h(X)^{-1})\right)$.
    Since $n(h)$ increases toward lower altitudes, the Cherenkov angle $\theta_{\rm C}$ generally grows as the emission point approaches the ground.

    \item \textbf{Cherenkov ring radius (third panel).}
    The characteristic Cherenkov radius in the shower plane is
        $R_{\rm C}(X) \;=\; d_{\max}(X)\cdot\tan\theta_{\rm C}(X)$.
    This equation combines the shrinking propagation distance, $d_{\max}$, and the increasing angle, $\theta_{\rm C}$.

    \item \textbf{Cherenkov footprint area on the ground (fourth panel).}
    We define the Cherenkov footprint area as the projected area of the Cherenkov ring on the ground, calculated by
    \begin{equation}
        A_{\rm C}(X) = \frac{\pi R_{\rm C}^2(X)}{\cos\theta} .
        \label{eq:Ac_def}
    \end{equation}
    The factor $1/\cos\theta$ accounts for the projection from the shower plane to the horizontal detector plane. In our geometry, $A_{\rm C}(X)$ exhibits a broad maximum around $X \simeq 2000$\,g\,cm$^{-2}$.

    \item \textbf{Atmospheric density (fifth panel).}
    The local mass density, $\rho(X)$, (given in kg\,m$^{-3}$) is evaluated at altitude $h(X)$ using the atmospheric profile $\rho(h)$.

    \item \textbf{Geomagnetic-radiation energy (sixth panel).}
    The quantity $E_{\rm geo}/(\sin\alpha \cdot S_{\rm geo})$ is shown as a function of $X_{\max}^{\rm radio}$, where $E_{\rm geo}$ denotes the geomagnetic contribution to the radiated energy, $\alpha$ is the angle between the shower axis and the geomagnetic field, and $S_{\rm geo}$ is the geomagnetic scaling factor used in the energy reconstruction/normalization. As this panel shows, the total radiated energy by the geomagnetic effect decreases smoothly as $X_{\max}^{\rm particle}$ increases.

    \item \textbf{Average radiation intensity (seventh panel).}
    To isolate the effect of the footprint size, we plot an average radiation intensity at ground level:
    \begin{equation}
        I_{\rm avg}(X) \;\propto\; \frac{E_{\rm geo}}{\sin\alpha \cdot S_{\rm geo}\cdot A_C(X)} \: .
        \label{eq:Iavg_def}
    \end{equation}
    Because the peak of $A_{\rm C}(X)$ occurs near $X \simeq 2000$\,g\,cm$^{-2}$, the radiated energy is distributed over the largest ground area, resulting in a pronounced minimum in $I_{\rm avg}(X)$.

    \item \textbf{Trigger multiplicity for different SNR thresholds (bottom panel).}
    The shaded bands and central curves illustrate the number of triggered antennas (assuming a 1.5\,km triangular grid size) as a function of $X_{\max}^{\rm particle}$ for several SNR cuts. The band represents the spread within each $X_{\max}^{\rm particle}$ bin, while the central line corresponds to the median of the trigger multiplicity.
\end{itemize}

\paragraph{Origin of the minimum in number of triggered antennas at high SNR.}
For high SNR thresholds, triggering becomes sensitive primarily to the amplitude of the signal at individual antennas, which scales more closely with an intensity-like quantity than with the total radiated energy. Consequently, the geometrical modulation of the footprint area, $A_{\rm C}(X)$, plays a dominant role. At around $X \simeq 2000$\,g\,cm$^{-2}$, the footprint area reaches its maximum (Eq.~\eqref{eq:Ac_def}). Thus, the available radiation is spread over the largest ground area, which reduces the average intensity $I_{\rm avg}(X)$ (see Eq.~\eqref{eq:Iavg_def}). Consequently, fewer antennas exceed the very high SNR thresholds in this region, resulting in the observed minimum in the trigger number near $X \sim 2000$\,g\,cm$^{-2}$. At larger $X_{\max}^{\rm radio}$ (i.e.\ emission closer to the ground), $A_{\rm C}(X)$ decreases, the intensity increases, and the high-SNR trigger multiplicity rises again.





\bibliographystyle{JHEP}
\bibliography{reference}

\end{document}